\documentclass{article}

     \usepackage[nonatbib,preprint]{neurips_2021}

\usepackage[utf8]{inputenc} 
\usepackage[T1]{fontenc}    
\usepackage{hyperref}       
\usepackage{url}            
\usepackage{booktabs}       
\usepackage{amsfonts}       
\usepackage{nicefrac}       
\usepackage{microtype}      
\usepackage{xcolor}         
\usepackage{graphicx}
\usepackage{amsmath}
\usepackage{siunitx}
\usepackage{xcolor}

\title{Modelling Neuronal Behaviour with Time Series Regression: Recurrent Neural Networks on \textit{C. Elegans} Data}

\author{%
  Gon\c{c}alo Mestre
    \\
  INESC-ID / IST Tecnico Lisboa\\
  Universidade de Lisboa\\
  Rua Alves Redol 9, 1000-029 Lisboa\\
  \texttt{goncalo.mestre@tecnico.ulisboa.pt} \\ 
  \And
  Ruxandra Barbulescu \\
  INESC-ID \\
  Rua Alves Redol 9, 1000-029 Lisboa \\
  \texttt{ruxi@inesc-id.pt} \\
  \AND
  Arlindo L. Oliveira \\
  INESC-ID / IST Tecnico Lisboa \\
  Universidade de Lisboa \\
  Rua Alves Redol 9, 1000-029 Lisboa \\
  \texttt{aml@inesc-id.pt} \\
  \And
  L. Miguel Silveira \\
  INESC-ID / IST Tecnico Lisboa \\
  Universidade de Lisboa \\
  Rua Alves Redol 9, 1000-029 Lisboa \\
  \texttt{lms@inesc-id.pt} \\
}

\begin{document}

\maketitle

\begin{abstract}
Given the inner complexity of the human nervous system, insight into the dynamics of brain activity can be gained from understanding smaller and simpler organisms, such as the nematode \textit{C. Elegans}. The behavioural and structural biology of these organisms is well-known, making them prime candidates for benchmarking modelling and simulation techniques. In these complex neuronal collections, classical,
white-box modelling techniques based on intrinsic structural or behavioural information are either unable to capture the profound nonlinearities of the neuronal response to different stimuli or generate extremely complex models, which are computationally intractable. In this paper we show how the nervous system of \textit{C. Elegans} can be modelled and simulated with data-driven models using different neural network architectures. Specifically, we target the use of state of the art recurrent neural networks architectures such as LSTMs and GRUs and 
compare these architectures in terms of their properties and their accuracy as well as the complexity of the resulting models. 
We show that GRU models with a hidden layer size of 4 units are able to accurately reproduce with high accuracy the system's response to very different stimuli.


\end{abstract}

\section{Introduction}
\label{sec:introduction}

The study of the human brain is probably one of the greatest challenges in the field of neuroscience.
Recent developments in experimental neuroscience have considerably increased the availability of novel recordings and reconstructions shedding further light into the structure and function of the brain as well as many other systems. But understanding the complexities behind the relations between structure and  function as well as the behaviour of such systems across multiple scales in these neuronal collections is constrained by the methods available to study them.
This challenge has raised interest in many related fields, such as electrophysiological analysis, imaging techniques, brain-related medicine, computational modelling and simulation, model reduction.

One research direction is the study of smaller and simpler nervous systems, for which the underlying principles of network organization and information processing are easier to postulate. These organisms can become useful models to gain insight into the fundaments of neuronal dynamics and whole brain organization, to validate hypotheses, to develop and test modelling methods, simulation instruments and model reduction techniques. 

\textit{Caenorhabditis Elegans} (\textit{C. Elegans}) belongs to this category of organisms and is quickly becoming one of the benchmarks in whole brain organization studies. \textit{C. Elegans} is a nematode (roundworm) of about 1 mm in length with a compact nervous system consisting of less than 1000 cells across all sexes and around 15000 connections~\cite{cook2019whole}. This rather small nervous system allows the worm to solve basic problems such as feeding, predator avoidance and mate-finding. Moreover, at least the cell-lineage and the anatomy of \text{C. Elegans} are invariant, in the sense that every individual possesses the same number of neurons and they occupy fixed positions in the organism; the invariance of the synaptic connections is still under debate~\cite{brittin2020beyond}.

The relatively small size of the \textit{C. Elegans} nervous system allowed for its almost complete description from different perspectives and scales, from its genetics and genomics to its molecular biology, structural anatomy, neuronal function, circuits and behaviour. This information is available in comprehensive databases of genetics and genomics~\cite{wormbase}, \cite{wormgeneexp}, \cite{hunt2007high}, electron micrographs and associated data~\cite{wormimage}, online books~\cite{wormbook}, \cite{jackson2014use} and atlases~\cite{wormatlas} of the neurobiology, structural and behavioural anatomy.
%
However, creating a realistic model that encapsulates all this information is not a trivial task. 
Open-source databases of digitally reconstructed neurons~\cite{opensourcebrain}, \cite{gleeson2019open}, computational models~\cite{openworm}, \cite{szigeti2014openworm} and collaborative solutions~\cite{geppetto}, \cite{cantarelli2018geppetto} are opening the door for more flexible, multi-scale and multi-algorithm simulation environments for \textit{C. Elegans} and other complex biological systems. 

The underlying models are based on the connectome (''connectivity graph'' or ''wiring diagram''), which is the map of the neuronal connections in the brain. Also called a neuronal network, the connectome can be described as a graph where the nodes are the neurons and the edges represent the synapses. 
The complete connectome of \textit{C. Elegans} contains 302 neurons for the adult hermaphodite \cite{varshney2011structural} and 385 neurons for the male \cite{cook2019whole}, but for the latter the respective 3D reconstructions are not yet published \cite{wormwiring}. Digital reconstructions for the male are only available for the posterior nervous system of 144 neurons~\cite{jarrell2012connectome}.


The more complex the organism, the more complicated the resulting model, which implies more computationally demanding, potentially intractable, simulations of its dynamic behaviour. This increased complexity stems from the detailed modelling of the internal structure whereas often one is really only interested in the peripheral, or input-output behaviour.
In this work we propose a methodology for generating a model of the neuronal behaviour of organisms using only peripheral information.
We use \textit{C. Elegans} as a proxy for our study.
Realistic models of \textit{C. Elegans}, which take into account spatial distribution and biophysical properties of neuronal compartments have been reported in the literature~\cite{gleeson2018c302}.  We 
start with a similar model created in-house.
We assume no prior knowledge of the original system's structure and equations, creating a completely data-driven model using neural networks trained on datasets representing the system's response to different input signals. The ultimate goal is to generate a reduced model to replace the high-fidelity one. This reduced model should be able to reproduce with reasonable accuracy the behaviour of the realistic model while having fewer
degrees of freedom. In this work we focus on the issue of accuracy, showing that we can produce sufficiently accurate models for analysing the behaviour of the \textit{C. Elegans} nervous system using neural networks.

\section{Related work and context}
\label{sec:related_work}

The connectome-based models mentioned above
are often termed white-box
models, as they are based on
direct knowledge and access to the internal structure and parameters' values of the modelled system. These are distributed models, where each neuron has a 3D description and position in space and the synapses are associated with neuronal sections.
Such models enable highly accurate simulation of the dynamic behaviour of organisms, but easily become extremely complex as they incorporate detailed structural and functional information of the system.

While the white-box approach ensures access to and evaluation of inner parameters during simulation, it has been shown that the activity of complex networks of neurons can often be described by relatively few distinct
patterns, which evolve on low-dimensional subspaces~\cite{karasozen2020model}. This knowledge, together with the ever-present need to avoid potential numerical intractability in large-scale networks with many degrees of freedom, has generated renewed interest
in applying model reduction, often also referred to as model compression, to these neuronal networks, including techniques such as Dynamic Mode Decomposition (DMD)~\cite{brunton2016extracting}, Proper Orthogonal Decomposition (POD)~\cite{kellems2009low} and Discrete Empirical Interpolation (DEIM)~\cite{lehtimaki2019projection}.
Depending on the level of morphological accuracy of the underlying models, model reduction techniques can have any shade of grey from white-box to black-box, the latter assuming no preliminary knowledge of the system structure and building the model solely out of knowledge of its input-output behaviour. 

Black-box approaches are often built upon
data-driven models, sometimes learning-based, which have the ability to grasp more naturally and more efficiently the complexity induced by the profound nonlinearities in the neuronal transmission of information. Machine-learning techniques are used to extract data-driven reduced order models for systems arising from 
differential equations describing the intrinsic dynamics~\cite{regazzoni2019machine} and even to extract the governing equations of the estimated model~\cite{sun2020neupde}.
It is therefore quite natural to consider using state of the art learning methods for developing reduced models of neuronal behaviour using data obtained from available recordings or even simulations obtained with more complex models.

Dynamical models generate multivariate time series, for which neural networks are mostly used for forecasting in industrial applications~\cite{lai2018modeling},~\cite{jin2020prediction}. Especially designed to capture temporal dynamic behaviour, Recurrent Neural Networks (RNNs), in their various architectures such as Long Short-Term Memory (LSTMs) and Gated Recurrent Units (GRUs), have been extensively and successfully used for forecasting or detecting faults in industrial multivariate time series data~\cite{massaoudi2019novel},~\cite{gallicchio2018comparison},~\cite{yuan2020using},~\cite{filonov2016multivariate}.
Bidirectional LSTMs were also used to model genome data~\cite{tavakoli2019modeling}. A combination of CNNs and LSTMs generates a model for epileptic seizure recognition using EEG signal analysis~\cite{xu2020one}. An attempt to model the human brain activity based on fMRI using RNNs (LSTMs and GRUs) is reported in~\cite{gucclu2017modeling}.

\section{Methods}
\label{sec:methods}

In the previous sections
we stated that the data modelling of the system will be done using time series data obtained from simulations of a more complex, connectome-based model.
From that viewpoint, we can consider that the modeling task is akin to a sequence to sequence conversion for which
the most suitable neural network models are sequential ones.

In this work we analyze the suitability of
three of the most 
commonly used architectures for recurrent neural networks. We start with the least complex unit, the simple RNN, which was originally proposed in the 1980's for modelling sequence data~\cite{rumelhart1986},~\cite{werbos1988} and~\cite{elman1990}. The second model used for the recurrent layer is the LSTM unit, developed by Hochreiter and Schmidhuber \cite{lstms1997original} and later improved
with the introduction of the forget gate to adaptively release internal resources when necessary \cite{gers1999}. Finally we analyze its sibling, the GRU \cite{cho2014gru}.

\subsection{Recurrent neural networks}
\label{sec:methods-rnn}

RNNs \cite{rumelhart1986},~\cite{werbos1988},~\cite{elman1990} are a family of neural networks used for processing sequential data,
particularly adept to
processing a sequence of values \(\textbf{x}^{(1)},...,\textbf{x}^{(t)}\), and in most cases
capable to process sequences of variable length. RNNs appear from the relaxation of the condition on Feedforward Neural Networks (FFNNs) that neurons in a given layer do not form connections among themselves.

\begin{figure}
    \centering
    \includegraphics[scale=0.25]{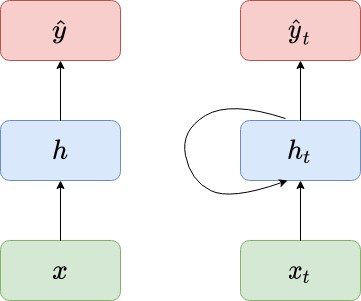}
    \caption{A simple version of a FFNN graph vs. a rolled RNN graph.}
    \label{fig:ffnn_vs_rnn}
\end{figure}

Figure~\ref{fig:ffnn_vs_rnn} illustrates both types of networks described mathematically by the following set of equations:
\noindent\begin{minipage}{.5\linewidth}
\begin{flalign}
    \mathbf{h} &= \phi(\mathbf{Vx} + \mathbf{c}) \label{eq:ffnn1} \\
    \mathbf{\hat{y}} &= \mathbf{Wh} + \mathbf{b} \label{eq:ffnn2}
\end{flalign}
\end{minipage}%
\begin{minipage}{.5\linewidth}
\begin{flalign}
  \mathbf{h}_t &= \phi(\mathbf{Vx_t} + \mathbf{Uh_{t-1}} + \mathbf{c}) \label{eq:rnn1} \\
    \mathbf{\hat{y}}_t &= \mathbf{Wh}_t + \mathbf{b} \label{eq:rnn2}.
\end{flalign}
\end{minipage}
\newline

 Equations~(\ref{eq:ffnn1}) and~(\ref{eq:ffnn2}) describe the FFNN, while (\ref{eq:rnn1}) and~(\ref{eq:rnn2}) describe the RNN. In both cases \(\mathbf{x}\) is the input vector, \(\mathbf{x_t}\) is the input point \(t\) of the sequence, \(\mathbf{V}\), \(\mathbf{W}\) and \(\mathbf{U}\) are weights, \(\mathbf{b}\) and \(\mathbf{c}\) are biases and \(\phi(\cdots)\) is an activation function, usually the hyperbolic tangent. \(\mathbf{h}\) and \(\mathbf{h_t}\) are the outputs of the hidden layer in each case, while \(\mathbf{\hat{y}}\) is the output of the feedforward neural network and \(\mathbf{\hat{y}_t}\) is the output point \(t\) of the sequence produced by the recurrent neural network.

As Figure~\ref{fig:ffnn_vs_rnn} illustrates, the overall structure is similar with equations~(\ref{eq:ffnn1}) and~(\ref{eq:rnn1}) computing the output of the hidden layer, while equations~(\ref{eq:ffnn2}) and~(\ref{eq:rnn2}) compute the output of the model, for the FFNN and the RNN, respectively.
However, comparing (\ref{eq:ffnn1}) and~(\ref{eq:ffnn2}) for the FFNN and (\ref{eq:rnn1}) and~(\ref{eq:rnn2}) for the RNN~\cite{elman1990}, both with one hidden layer, as seen in Figure~\ref{fig:ffnn_vs_rnn}, it is apparent
that the difference between these two types of networks 
lies in the fact that the RNN also takes into account values from previous units in the same layer, thereby encoding the idea of sequence that is present in our time series data.

Although RNNs, which are trained using Backpropagation Through Time (BPTT)~\cite{werbos1990}, seem to be a good model for sequential tasks, they are known to suffer mainly from two problems, vanishing and exploding gradients~\cite{bengio1994}. Exploding gradients, as described in~\cite{bengio1994}, refer to a large increase in the norm of the gradient during training, which appears due to the explosion of long term components that can grow exponentially more than short term ones. This is the less common of the two problems
and 
there are known solutions to handle it, such as the clipping gradient technique \cite{pascanu2013}.

A
harder to solve problem is the vanishing gradient issue~\cite{bengio1994}, which refers to when long term components go exponentially fast to norm $0$, making it impossible for the model to learn the correlation between temporally distant events.
Unfortunately in order to
faithfully reproduce the dynamics of our system, the simulations used for generating the datasets require the use of fine time steps, leading to long data sequences.
This in turn implies that the response at a given time will depend on values which are far back in the sequence.  This situation, however unavoidable, may lead the RNN to experience difficulties in learning our data with the desired accuracy.


\subsection{Long short-term memory}
\label{sec:methods-lstm}

In order to tackle the vanishing gradient problem~\cite{bengio1994}, Hochreiter and Schmidhuber proposed a Long Short-Term Memory Unit~\cite{lstms1997original}. This architecture was later improved with the inclusion of the forget gate, by Felix Gers~\cite{gers1999}, a student of Schmidhuber, who went on to further improve the unit~\cite{gers2000}. A LSTM unit consists of three main gates, the input gate (\ref{eq:inputgate}) that controls whether the cell state is updated, the forget gate (\ref{eq:forgetgate}) that  defines how the previous memory cell affects the current one and the output gate (\ref{eq:outputgate}), which controls how the hidden state is updated. Note that LSTM units exhibit a major difference from RNN simple units, since besides the hidden state they also output a cell state to the next LSTM unit, as can be seen in Figure~\ref{fig:rnn-lstm-gru}. The LSTM mechanism is described by the following:
%
\noindent\begin{minipage}{.5\linewidth}
\begin{flalign}
    \mathbf{i_t} &= \sigma(\mathbf{W_ix_t} + \mathbf{U_ih_{t-1}} + \mathbf{b_i}) \label{eq:inputgate} \\
    \mathbf{f_t} &= \sigma(\mathbf{W_fx_t} + \mathbf{U_fh_{t-1}} + \mathbf{b_f}) \label{eq:forgetgate} \\
    \mathbf{o_t} &= \sigma(\mathbf{W_ox_t} + \mathbf{U_oh_{t-1}} + \mathbf{b_o}) \label{eq:outputgate}
\end{flalign}
\end{minipage}%
\begin{minipage}{.5\linewidth}
\begin{flalign}
  \mathbf{\Tilde{c}_t} &= \phi(\mathbf{W_cx_t} + \mathbf{U_ch_{t-1}} + \mathbf{b_c}) \label{eq:candidatecelllstm} \\
    \mathbf{c_t} &= \mathbf{f_t} \circ \mathbf{c_{t-1}} + \mathbf{i_t} \circ \mathbf{\Tilde{c}_t} \label{eq:cellstatelstm} \\
    \mathbf{h_t} &= \mathbf{o_t} \circ \phi(\mathbf{c}_t), \label{eq:hiddenstatelstm}
\end{flalign}
\end{minipage}
\newline

where \(\mathbf{W_i}\), \(\mathbf{U_i}\), \(\mathbf{W_f}\), \(\mathbf{U_f}\), \(\mathbf{W_o}\), \(\mathbf{U_o}\), \(\mathbf{W_c}\) and \(\mathbf{U_c}\), are the weights and \(\mathbf{b_i}\), \(\mathbf{b_f}\), \(\mathbf{b_o}\) and \(\mathbf{b_c}\) are the biases. 
All these 12 parameters are learned by the model, while \(\sigma(\cdots)\) and \(\phi(\cdots)\) are the logistic sigmoid and the hyperbolic tangent activation functions, respectively.
The outputs of the LSTM unit, the hidden state and the cell state, are computed using~(\ref{eq:cellstatelstm}) and (\ref{eq:hiddenstatelstm}), respectively. The computation of the cell state requires the candidate cell state, obtained through~(\ref{eq:candidatecelllstm}).
%
%

\subsection{Gated recurrent units}
\label{sec:methods-gru}

The use of LSTM units in recurrent neural networks already produced models that were able to learn very distant dependencies~\cite{lstms1997original},~\cite{gers1999},~\cite{gers2000}, but these units are complex structures composed of three gates. For that reason, in 2014 a new type of unit, the GRU~\cite{cho2014gru}, was suggested, described as follows:
\noindent\begin{minipage}{.5\linewidth}
\begin{flalign}
    \mathbf{z}_t &= \sigma(\mathbf{W_zx}_t + \mathbf{U}_z\mathbf{h}_{t-1} + \mathbf{b}_z) \label{eq:updategate} \\
    \mathbf{r}_t &= \sigma(\mathbf{W_rx_t} + \mathbf{U_rh_{t-1}} + \mathbf{b_r}) \label{eq:resetgate}
\end{flalign}
\end{minipage}%
\begin{minipage}{.5\linewidth}
\begin{flalign}
  \mathbf{\hat{h}_t} &= \phi(\mathbf{W_hx_t} + \mathbf{U_h}(\mathbf{r_t} \circ \mathbf{h_{t-1}}  + \mathbf{b_h}) \label{eq:candidategru} \\
    \mathbf{h}_t &= (\mathbf{1} - \mathbf{z_t}) \circ \mathbf{h_t} + \mathbf{z_t} \circ \mathbf{\hat{h}_t}. \label{eq:hiddengru}
\end{flalign}
\end{minipage}
\newline


Note that in~(\ref{eq:updategate}),~(\ref{eq:resetgate}) and~(\ref{eq:candidategru}), the weights, \(\mathbf{W_z}\), \(\mathbf{U_z}\), \(\mathbf{W_r}\), \(\mathbf{U_r}\), \(\mathbf{W_h}\), \(\mathbf{U_h}\) and the biases \(\mathbf{b_z}\), \(\mathbf{b_r}\), \(\mathbf{b_h}\) are the parameters that the model should learn. \(\sigma(\cdots)\) and \(\phi(\cdots)\) are, again, the logistic sigmoid and the hyperbolic tangent activation functions, respectively.

\begin{figure}
    \centering
    \includegraphics[width=0.95\textwidth]{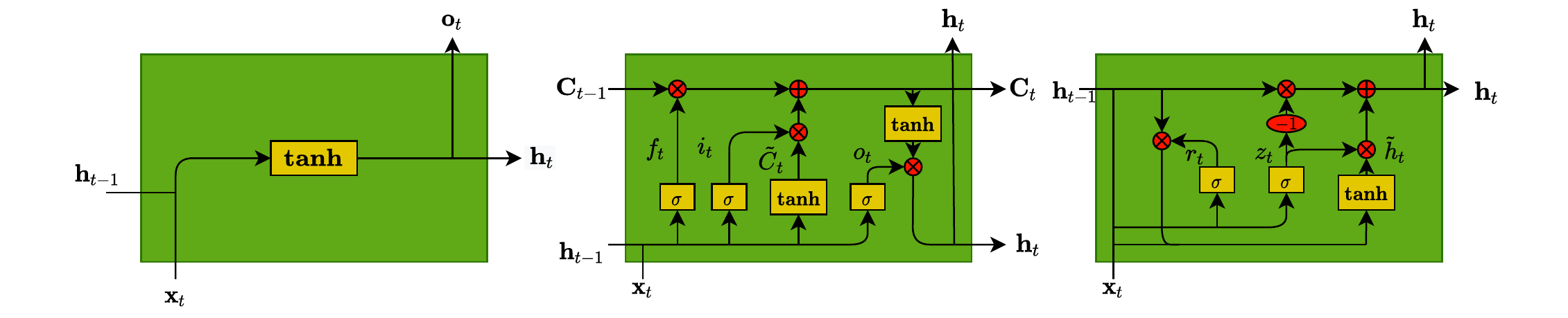}
    \caption{Comparison between the three different units, RNN, LSTM and GRU, respectively.}
    \label{fig:rnn-lstm-gru}
\end{figure}

The GRU, as shown in Figure~\ref{fig:rnn-lstm-gru}, is only composed of two gates, the update gate~(\ref{eq:updategate}) and the reset gate~(\ref{eq:resetgate}). The GRU only outputs the hidden state~(\ref{eq:hiddengru}), computed based on the candidate hidden state~(\ref{eq:candidategru}). The update gate controls how much of the past information needs to be passed along to the future, while the reset gate is used to decide how much information the model should forget.

\section{Experimental setting}

\subsection{Data}


The high-fidelity model is based on the connectome of the adult hermaphodite of \textit{C. Elegans} with 302 neurons \cite{varshney2011structural}, described in the NEURON simulator~\cite{carnevale2006neuron}. We conducted multiple simulations with varying input sequences applied to a set of four chosen neurons and recorded the responses in four additional neurons.
%
The data obtained is comprised of two datasets of 40 data files, one for each input setting, with four input time series measuring the input current in each of the four input neurons, and four output time series which measure the output voltage on the four output neurons, herein referred to as "DB1", "LUAL", "PVR" and "VB1". For the two datasets, each file corresponds to a simulation of \(500~\si{\milli\second}\), with a time step of \(0.5~\si{\milli\second}\) for the first and \(0.1~\si{\milli\second}\) for the second.

\begin{figure}
  \centering
  \includegraphics[width=0.32\columnwidth]{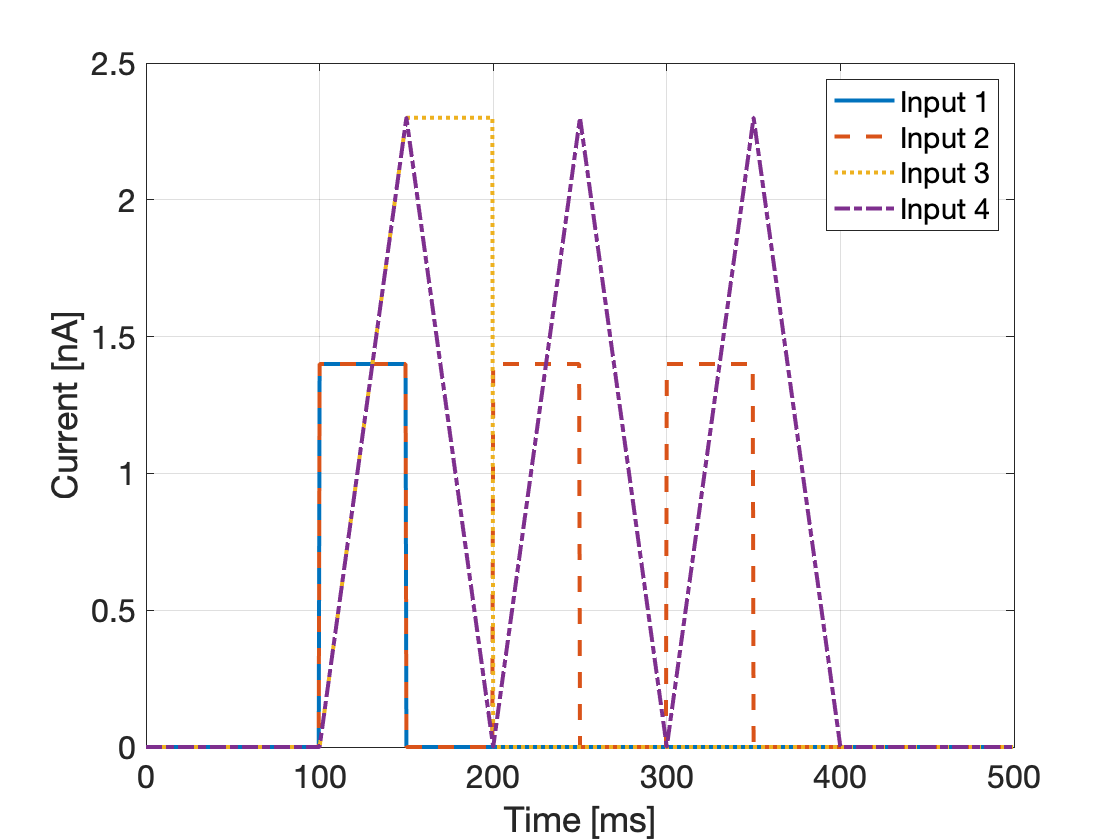}
  \includegraphics[width=0.32\columnwidth]{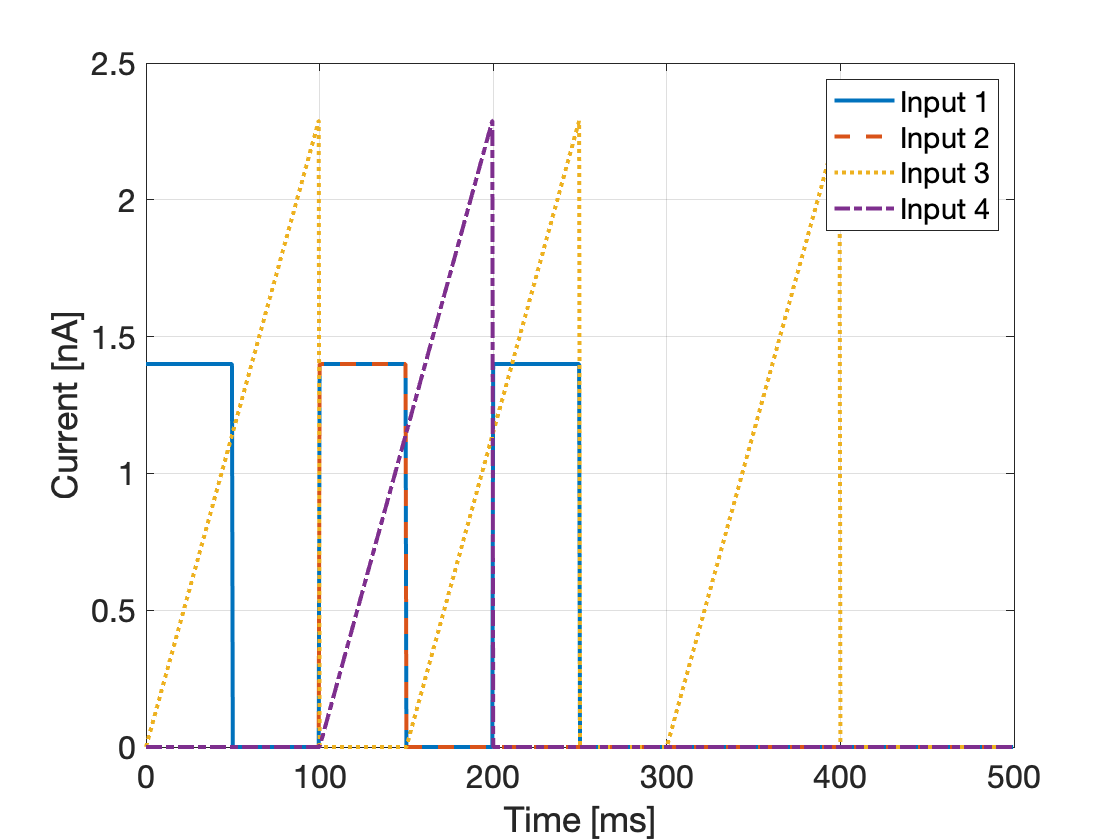}
  \includegraphics[width=0.32\columnwidth]{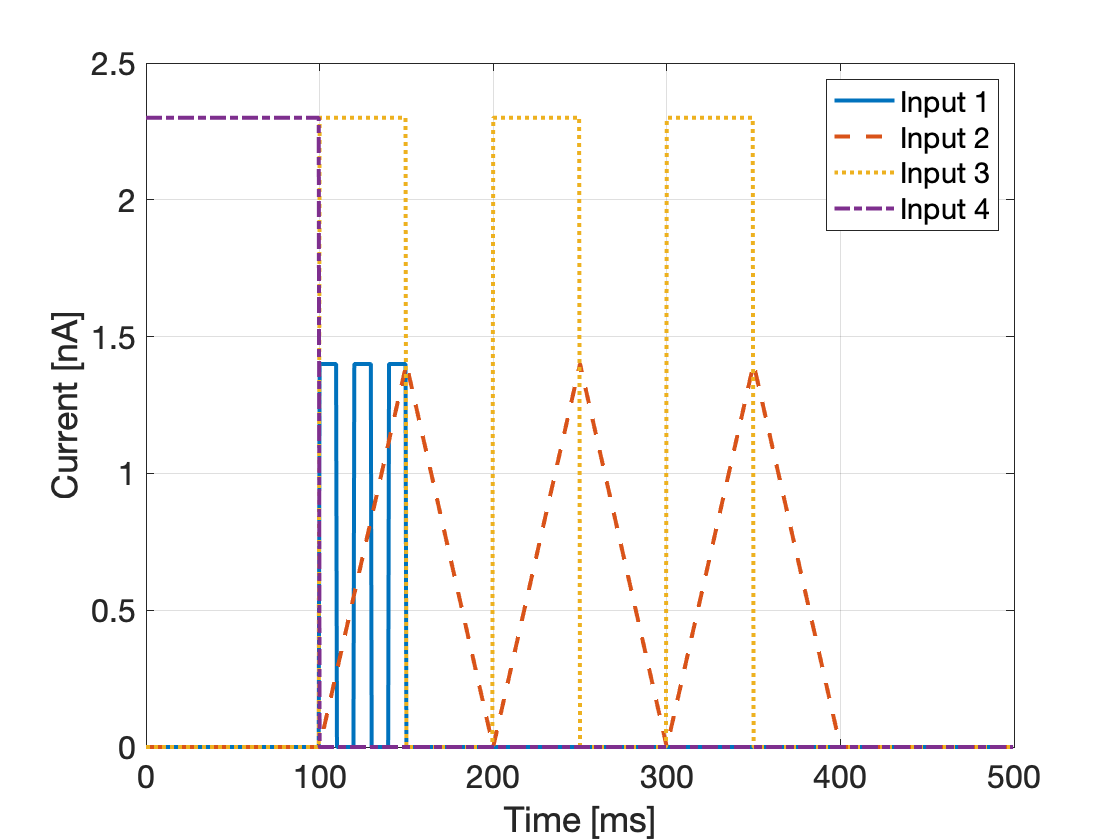}
  \includegraphics[width=0.32\columnwidth]{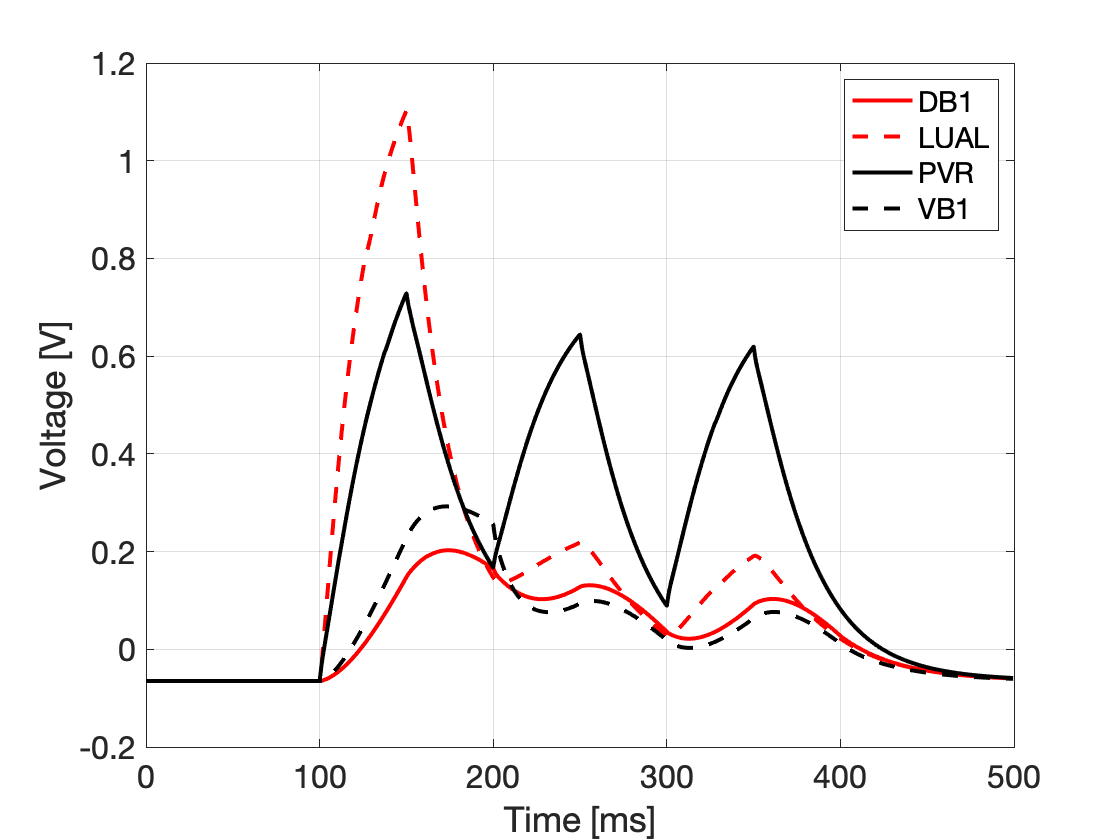}
  \includegraphics[width=0.32\columnwidth]{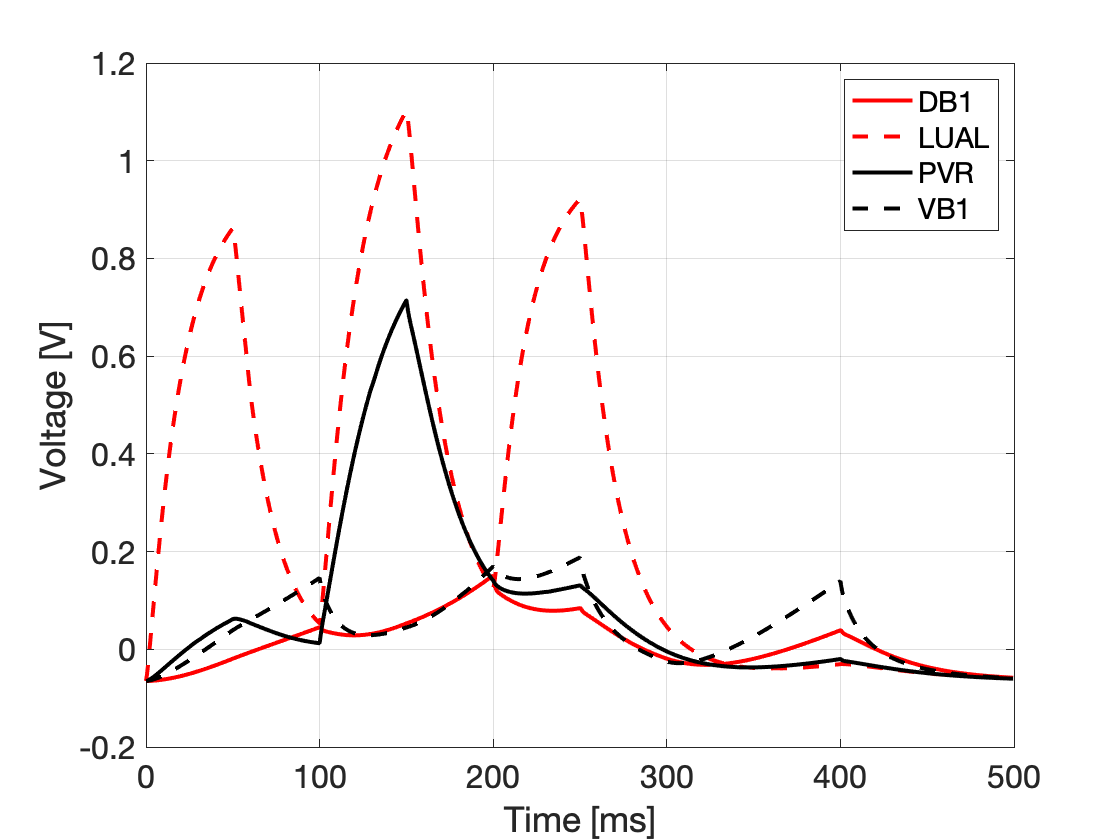}
  \includegraphics[width=0.32\columnwidth]{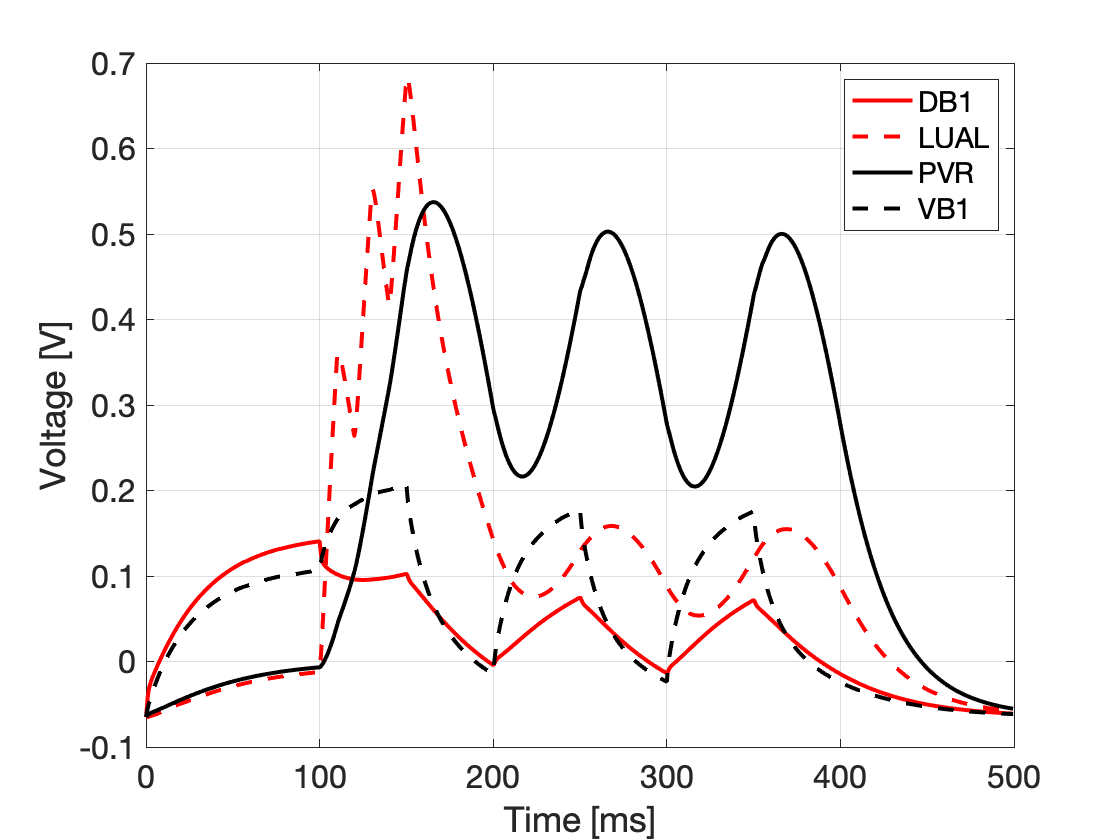}
  \caption{Example input (top row) and output (bottom row) time sequences.}
  \label{fig:example-io}
\end{figure}

To train and tune the models' hyperparameters, learning rate and batch size, the data  was divided into three sets: training, validation and test. 
The separation of data is done as follows: training set \(50\%\) of the data, validation set \(25\%\) of the data and for test the remaining \(25\%\). The separation is done by hand, so that validation and test sets are as diverse and demanding as possible for the models. Three examples of the diverse set of inputs and outputs are shown in Figure~\ref{fig:example-io}.


\subsection{Modelling}

The models are developed in Python~\cite{python}, using the Keras~\cite{keras} and Tensorflow~\cite{tensorflow} libraries. Details on the code and dependencies to run the experiments are listed in a Readme file available together with the code in the Supplemental Material.

Each model tested consists of a recurrent layer described in section~\ref{sec:methods} followed by a dense layer. The dense layer performs a simple linear transformation for each sequence point to convert the output of the recurrent layer, of size "hidden size", into the four outputs:
\begin{flalign}
    \mathbf{y_t} = \mathbf{x_t}\mathbf{A^T} + \mathbf{b}, \label{eq:dense_layer}
\end{flalign}

where \(\mathbf{A^T}\) are weights and \(\mathbf{b}\) are biases, both fitted by the model.

For a consistent comparison of the models, we fixed the optimizer to Adam~\cite{adam} and the loss function to the mean squared error.
The other hyperparameters, learning rate and batch size, were tuned through experimentation (further info available in the Supplemental Material). The experiments on each of the two hyperparameters are conducted using the LSTM recurrent layer with a fixed hidden size of \(64\) units, and the other hyperparameter fixed. Each model is trained for \(5000\) epochs, with the final model chosen as the best iteration on the validation set. After some experimentation we chose to use a learning rate of \(0.001\) and a batch size of \(20\) as that seemed to provide the best results. 



\section{Experiments and results}
\label{sec:results}

To test the ability of these models to reproduce the \textit{C. Elegans} neuronal data with high accuracy while maintaining simplicity, we conduct three experiments. In Experiment 1 (Section~\ref{sec:experiment1}) we compare the performance of the three types of layers under investigation, RNN, LSTM and GRU. Experiment 2 (Section~\ref{sec:experiment2}) includes a comparison between different sizes of the recurrent layer in order to determine the optimal size given some accuracy constraints. Finally, Experiment 3 (Section~\ref{sec:experiment3}), is an investigation upon the ability of the models to reproduce results for data resulting from simulations with a finer time step, therefore involving longer sequences with more data points.

\subsection{Experiment 1} \label{sec:experiment1}


%
%
In this experiment, the performances of the three types of units described in Section~\ref{sec:methods} are compared on the collected data of the dataset corresponding to the coarser time step (\(0.5~\si{\milli\second}\)). To test the ability of each unit to reproduce the outputs using different sizes of the recurrent layer, the three units are tested with a hidden layer of \(16\) and \(64\) units.

\begin{figure}[h]
    \centering
    \begin{minipage}{0.45\textwidth}
        \centering
        \includegraphics[width=0.8\textwidth]{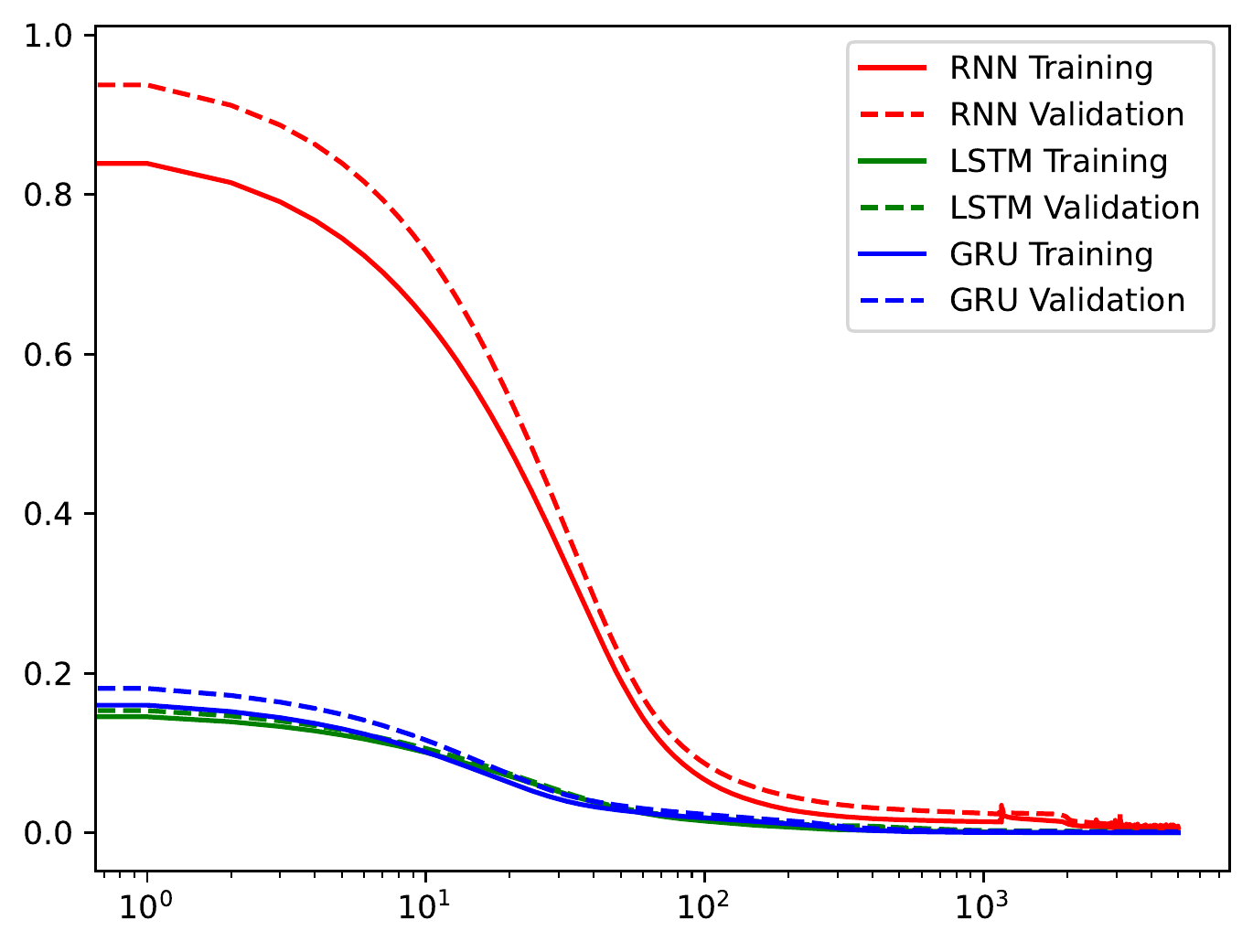}
    \end{minipage}\hfill
    \begin{minipage}{0.45\textwidth}
        \centering
        \includegraphics[width=0.8\textwidth]{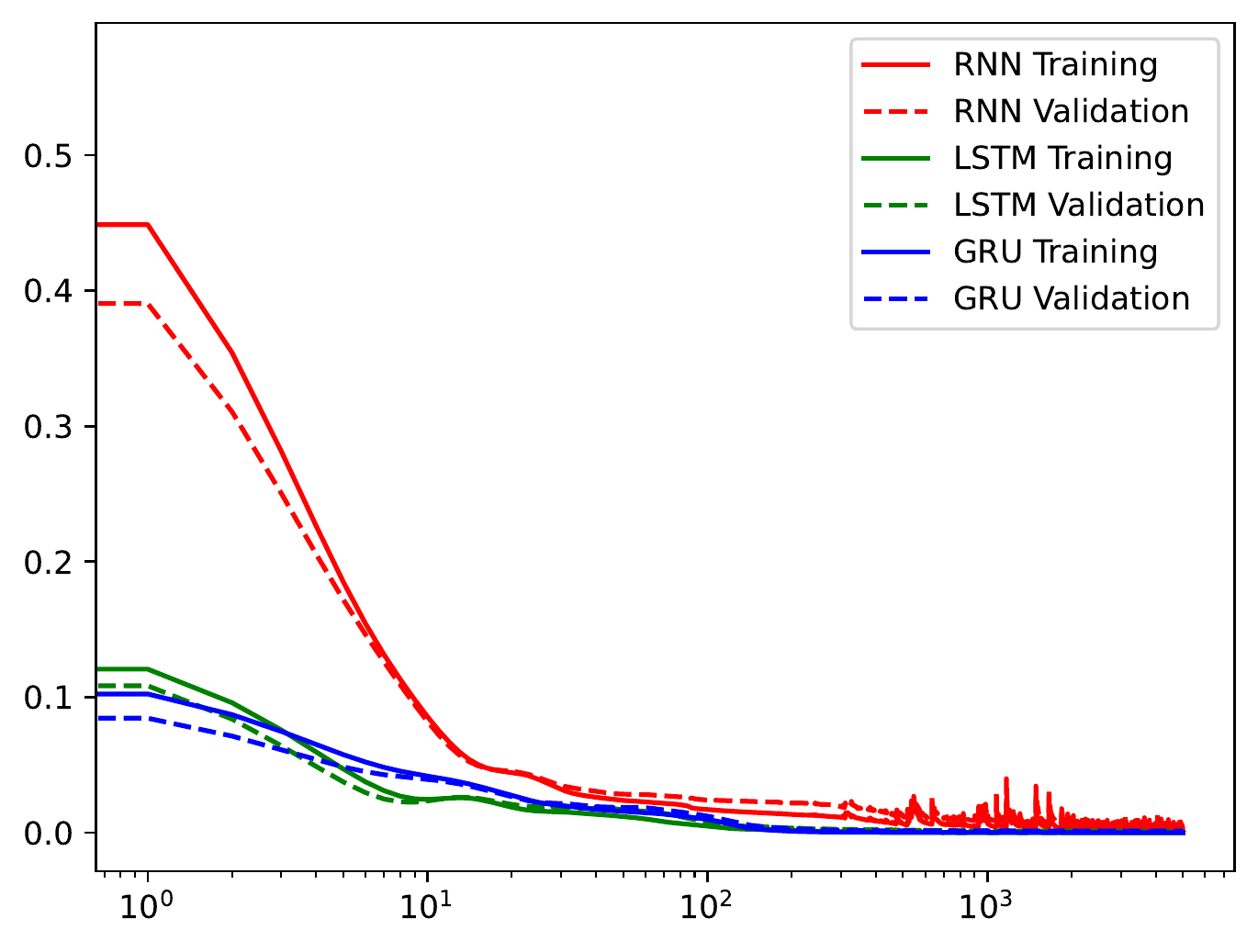} 
    \end{minipage}
    \caption{Training and validation loss for the three architectures, with recurrent layer size of 16 (left) and 64 (right) units.}
    \label{fig:exp1loss}
\end{figure}

\begin{figure}
 \setlength\tabcolsep{2pt}
  \centering
  \hspace*{-0.25cm}\begin{tabular}{cccccc}
    \textbf{RNN-16} & \textbf{LSTM-16} & \textbf{GRU-16} & \textbf{RNN-64} & \textbf{LSTM-64} & \textbf{GRU-64} \\
		\includegraphics[width=0.16\textwidth]{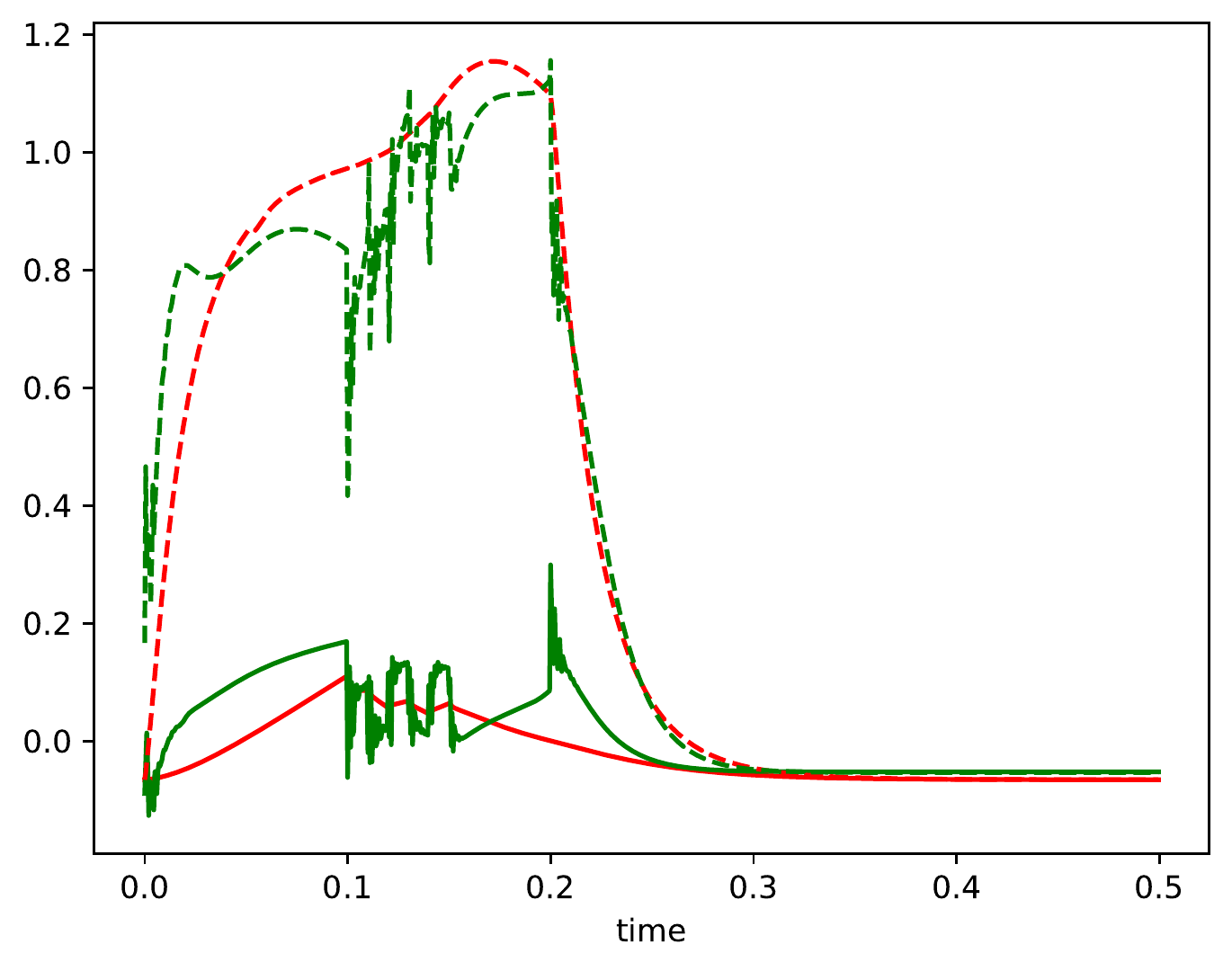} & \includegraphics[width=0.16\textwidth]{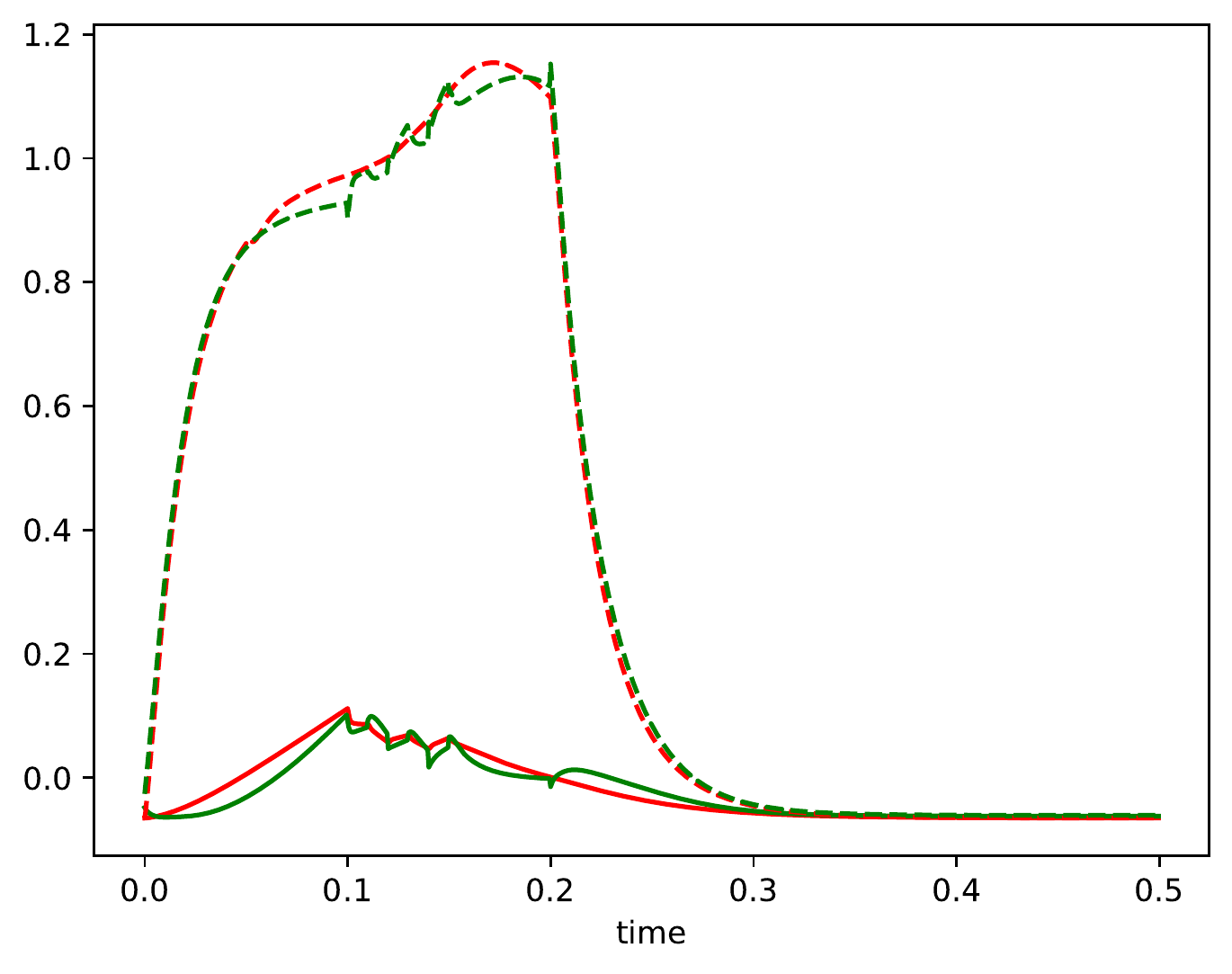} & 
		\includegraphics[width=0.16\textwidth]{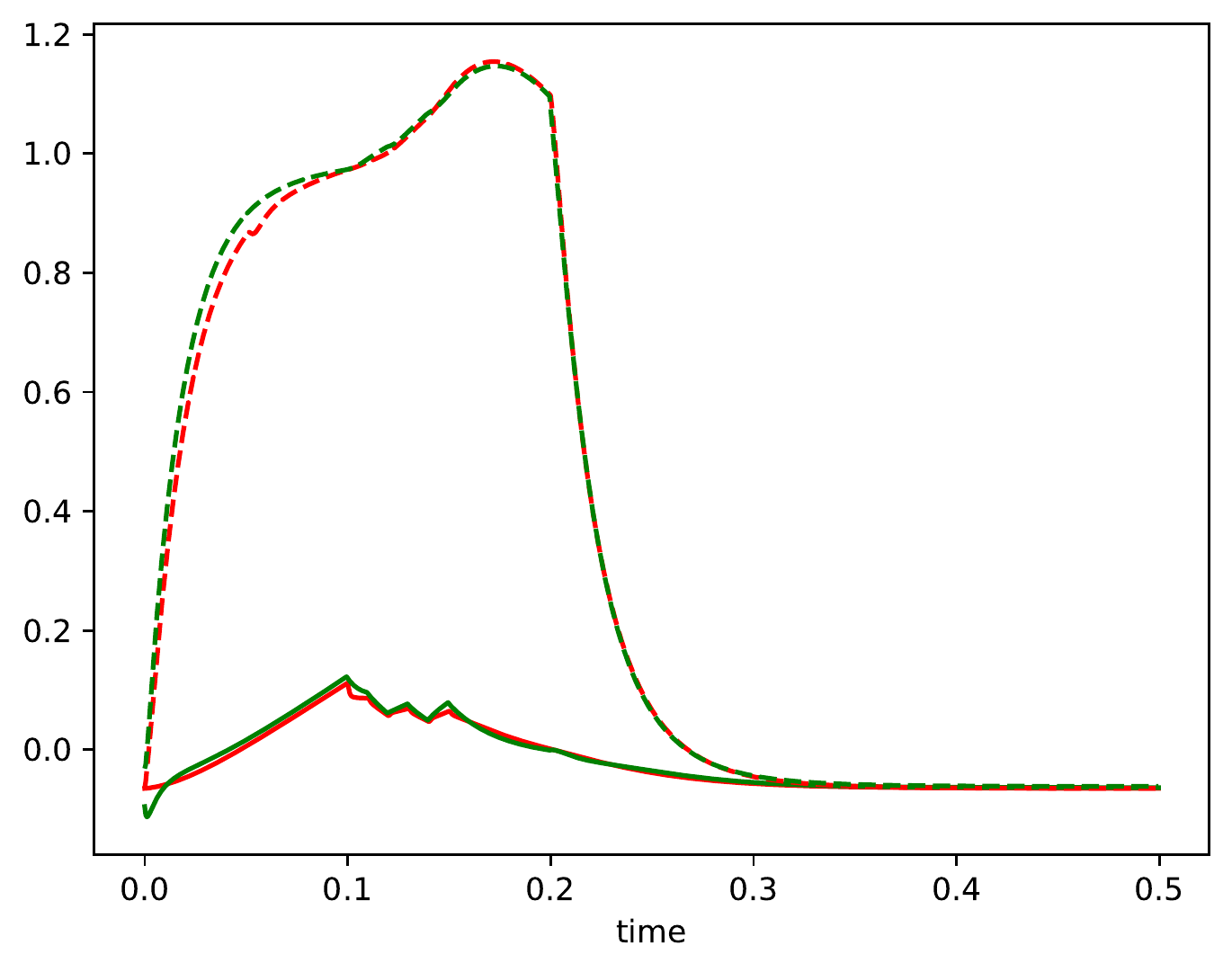} & 
		\includegraphics[width=0.16\textwidth]{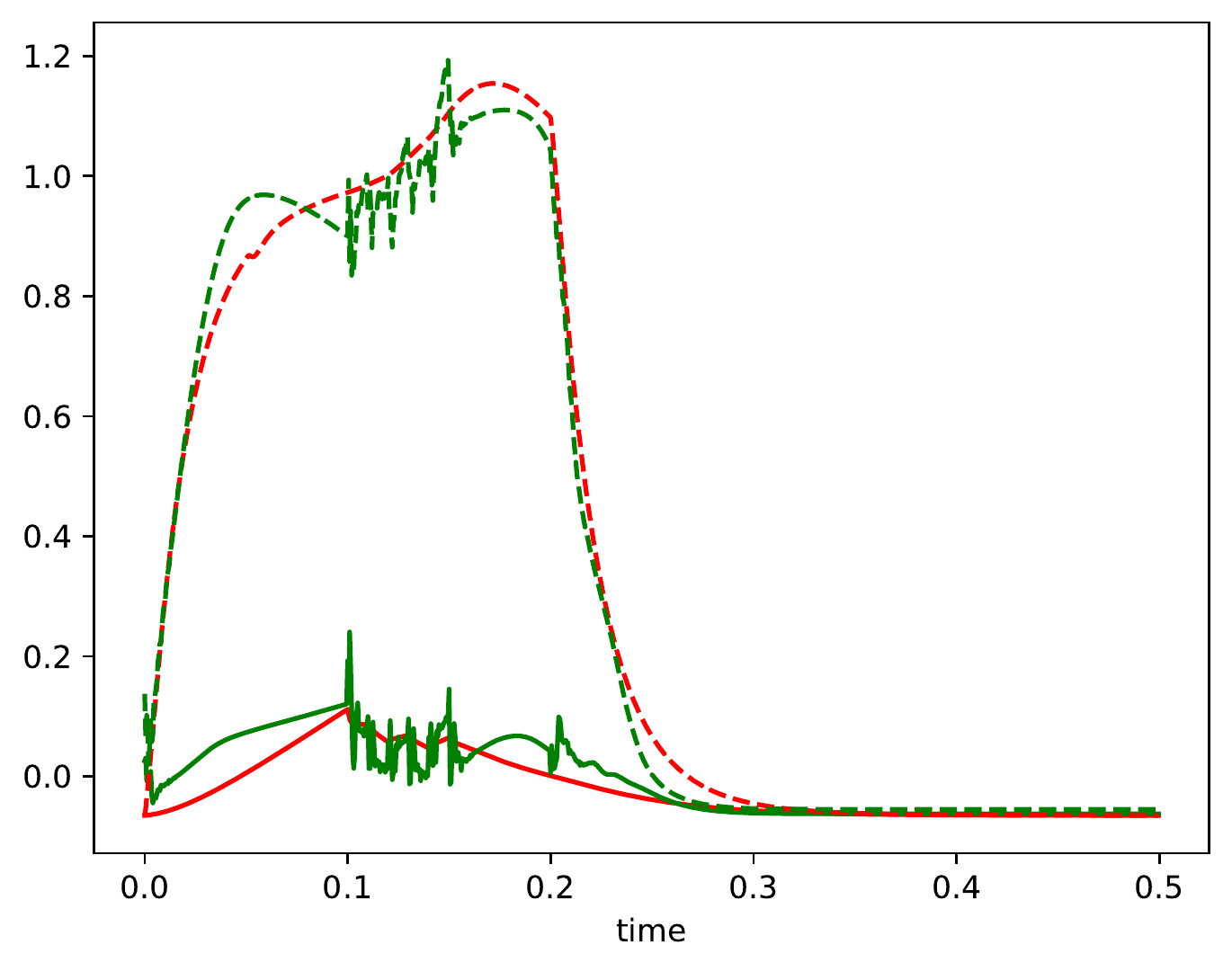} & \includegraphics[width=0.16\textwidth]{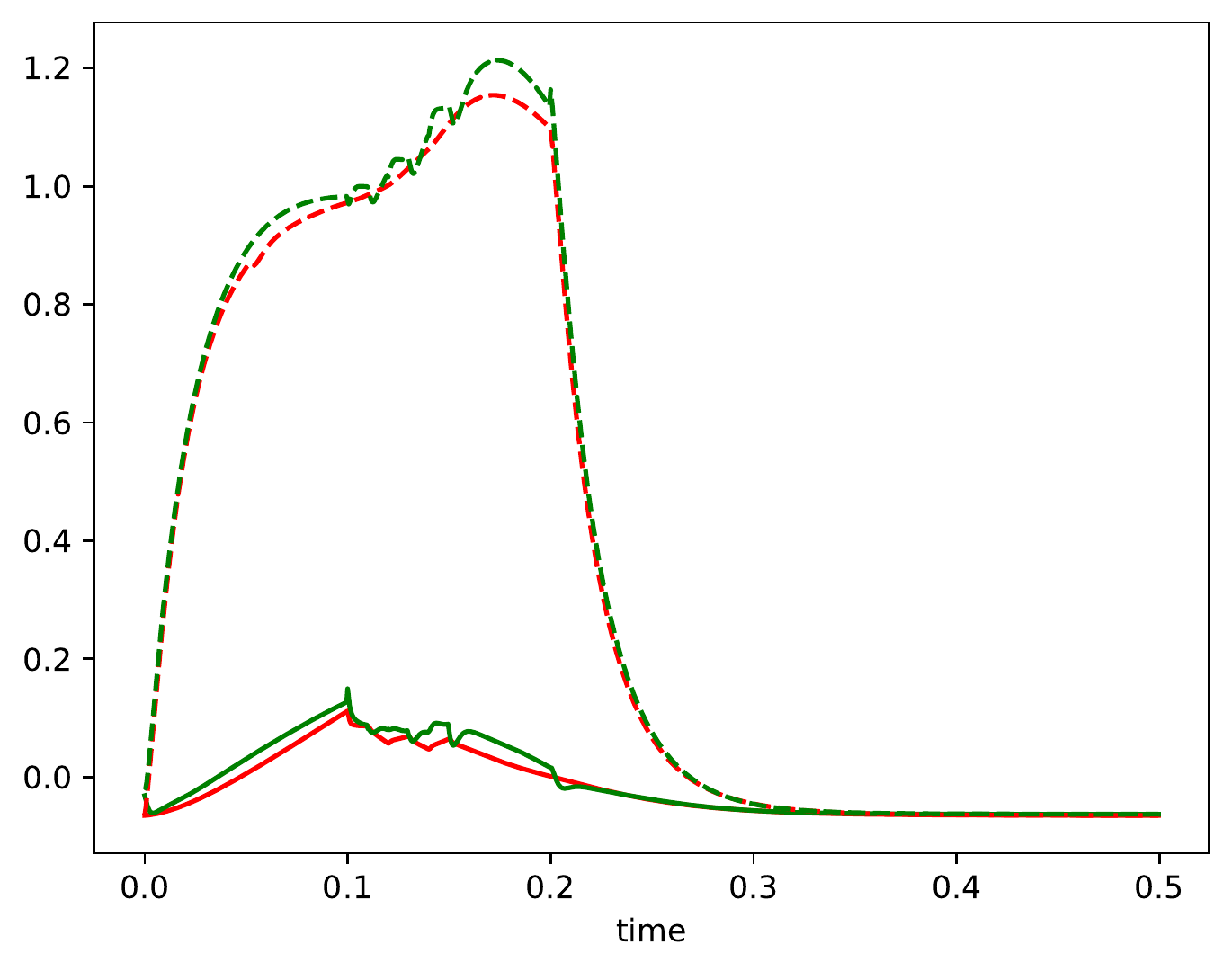} & 
		\includegraphics[width=0.16\textwidth]{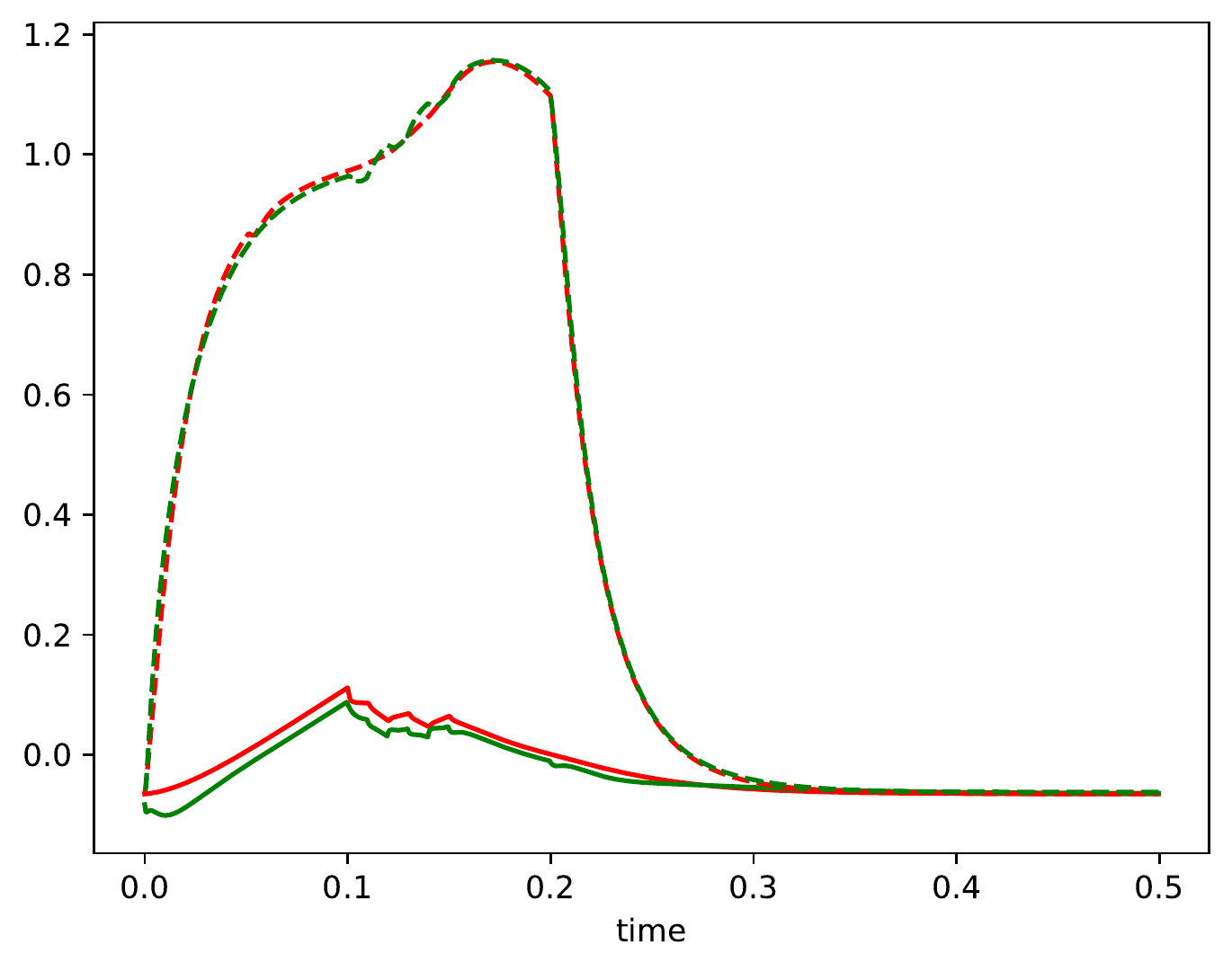} 
	\\
		\includegraphics[width=0.16\textwidth]{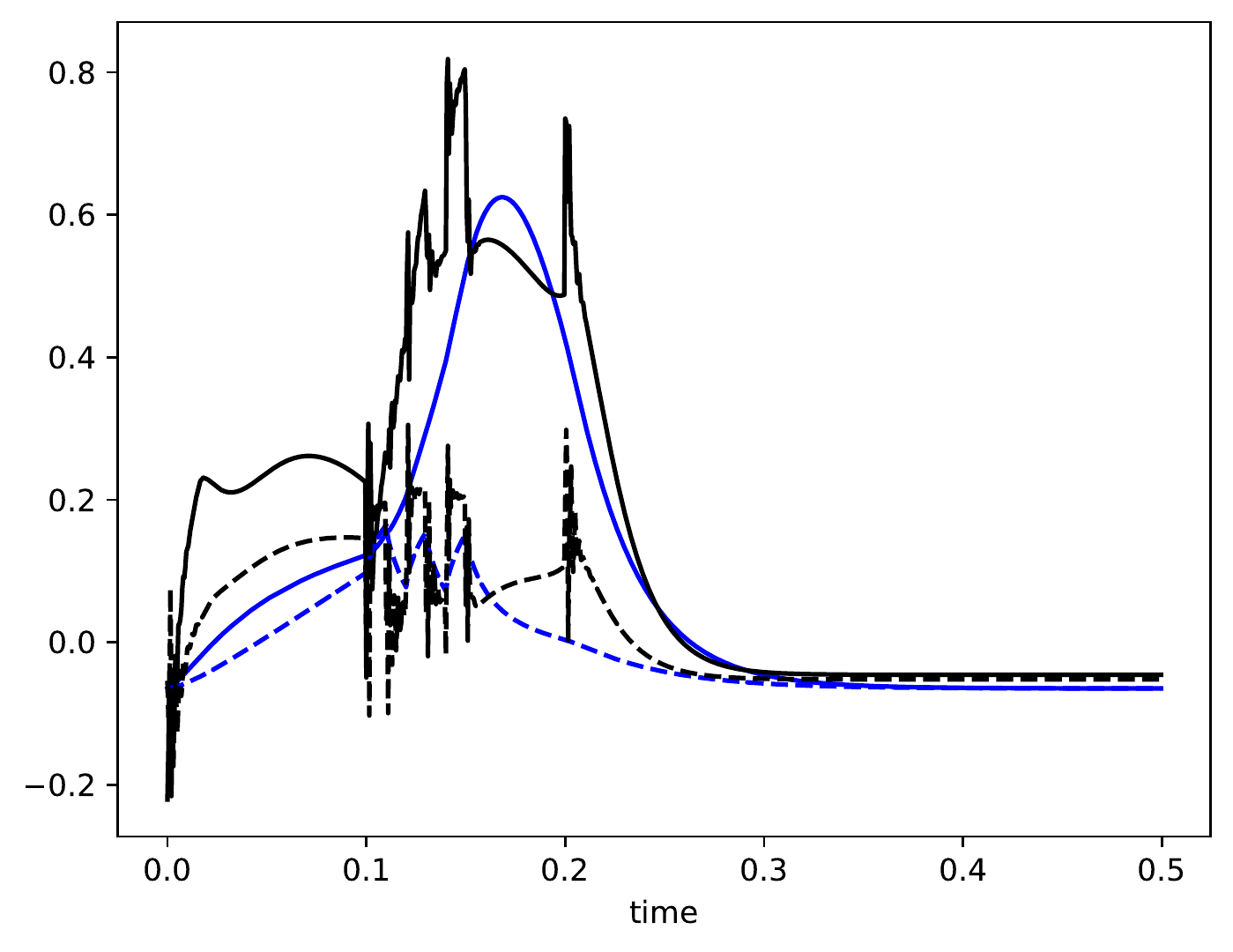} & \includegraphics[width=0.16\textwidth]{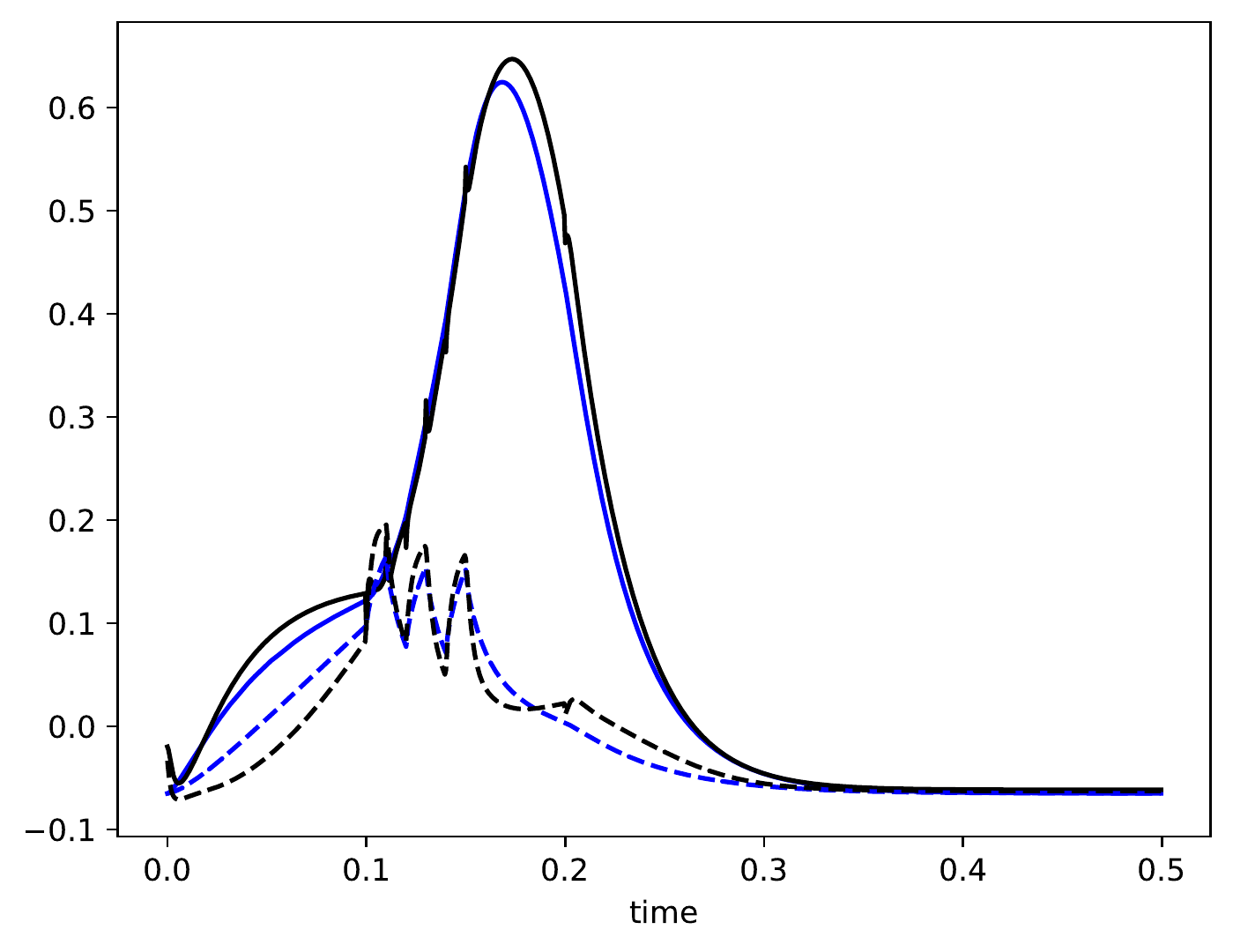} &
		\includegraphics[width=0.16\textwidth]{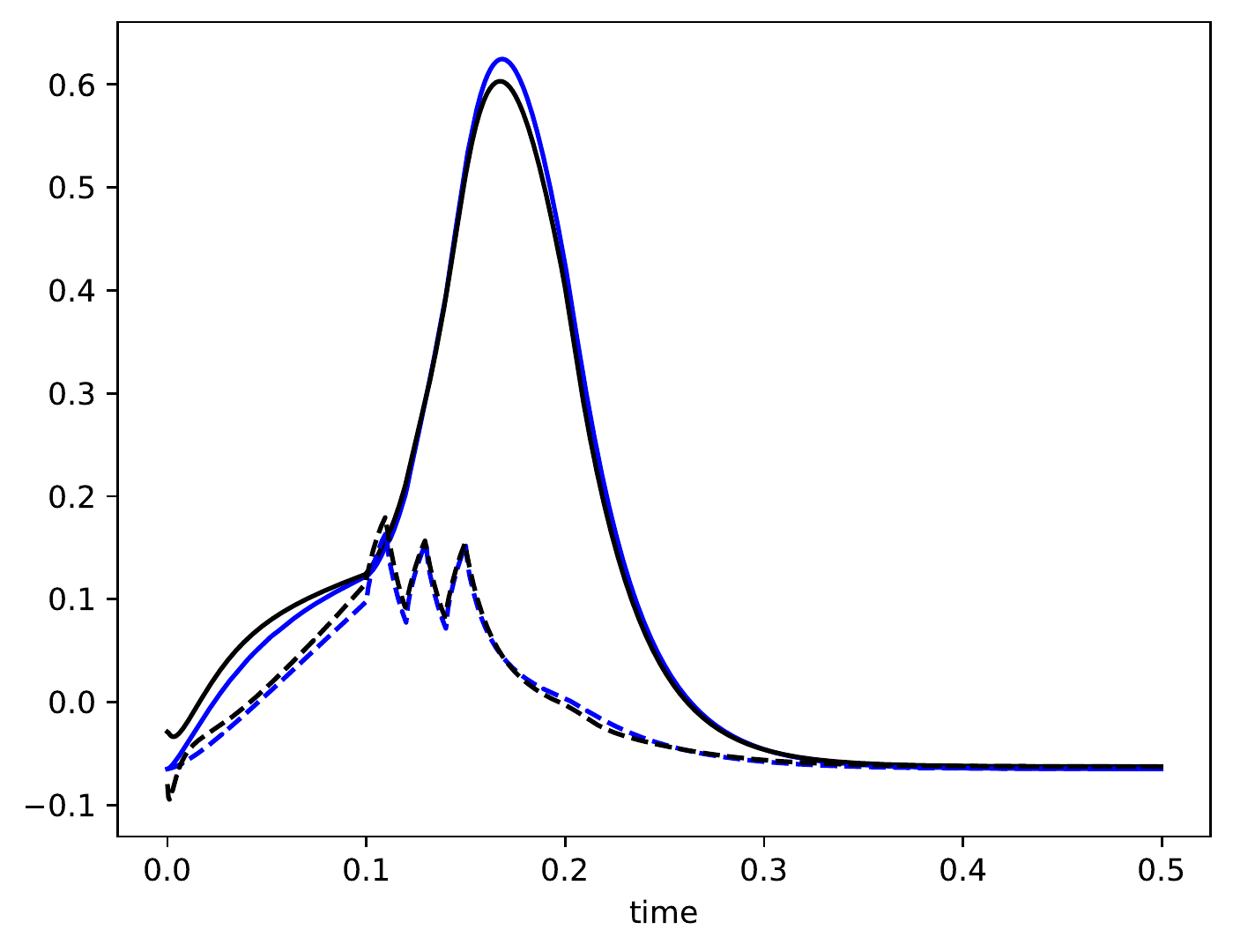} & 
		\includegraphics[width=0.16\textwidth]{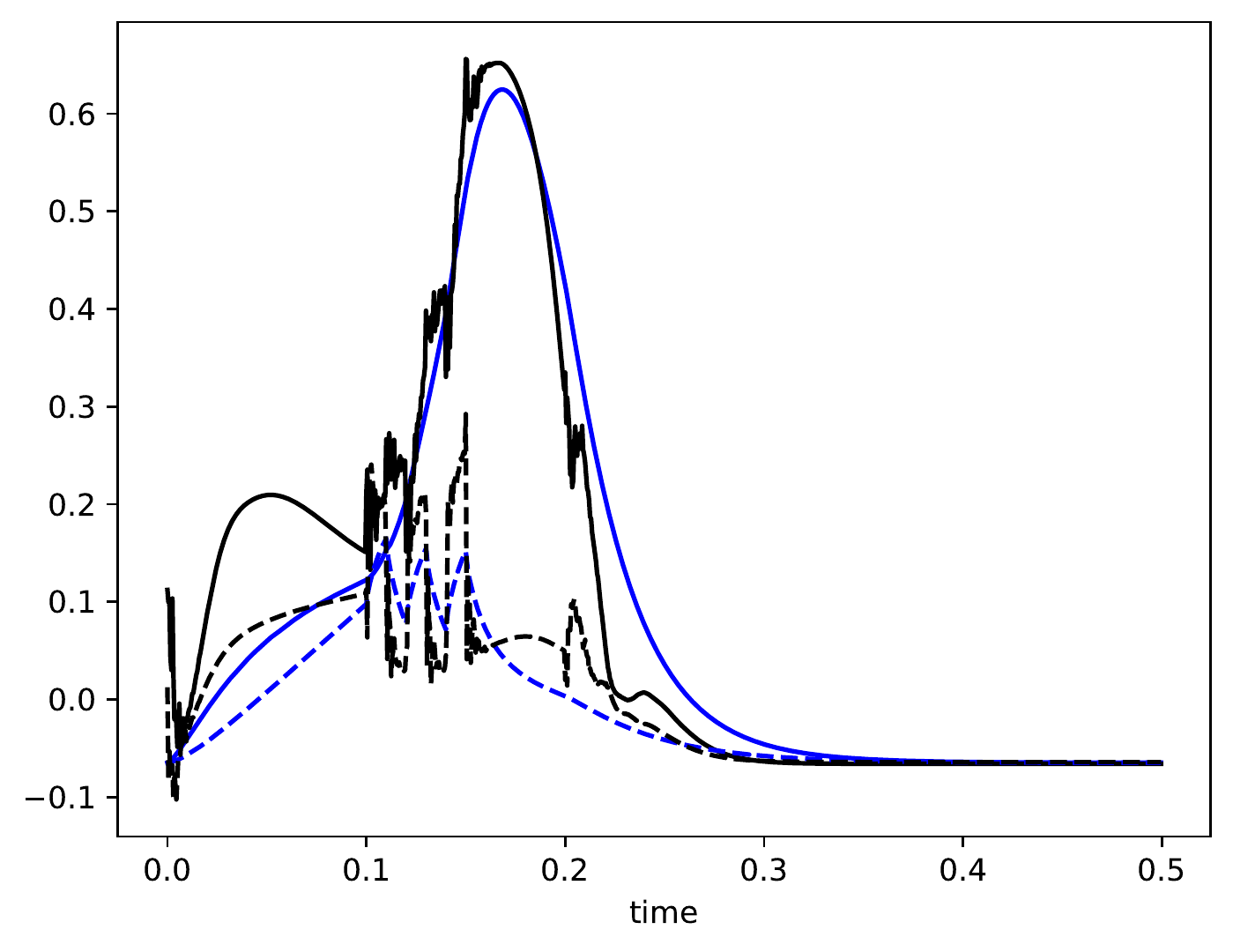} &    \includegraphics[width=0.16\textwidth]{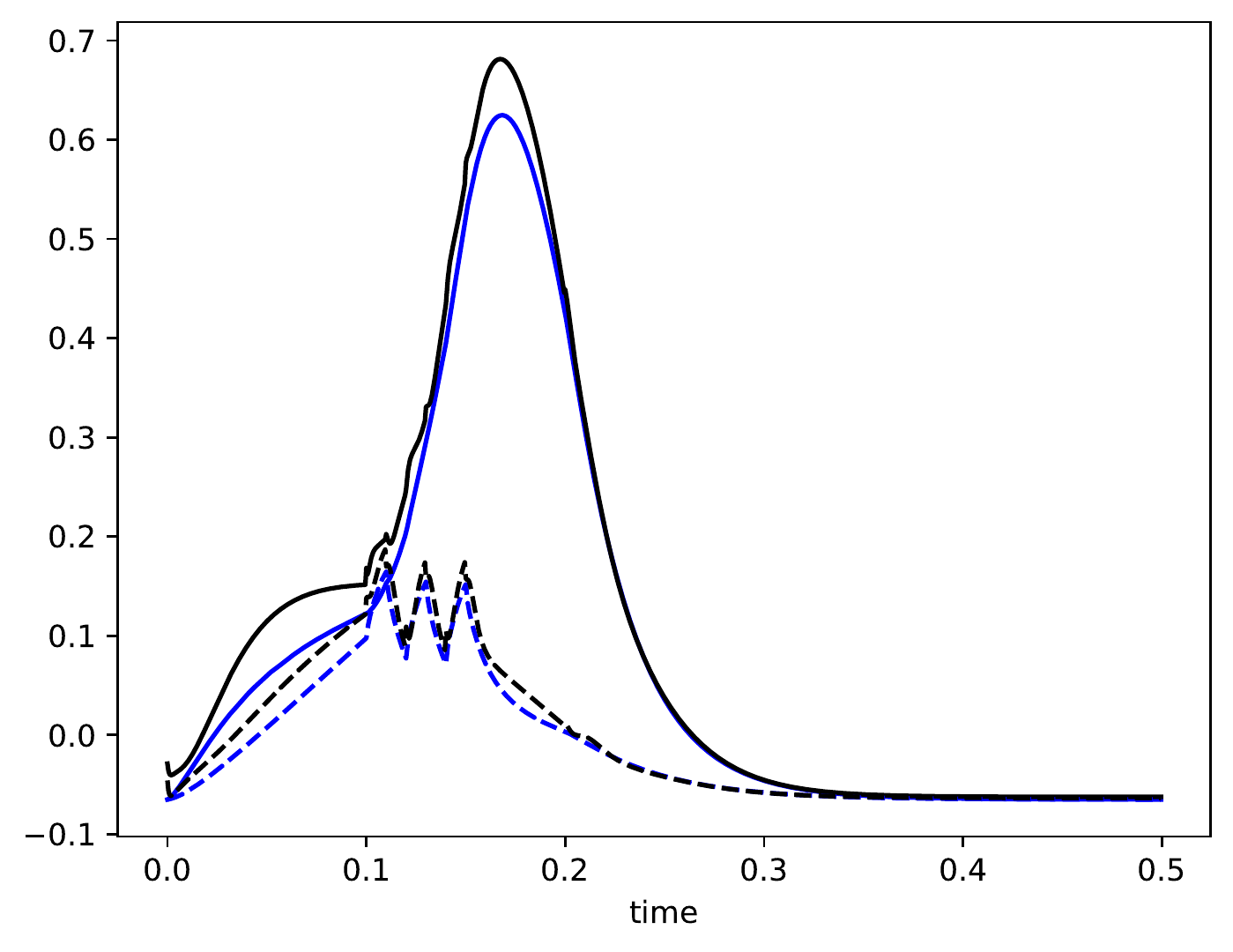} & 
		\includegraphics[width=0.16\textwidth]{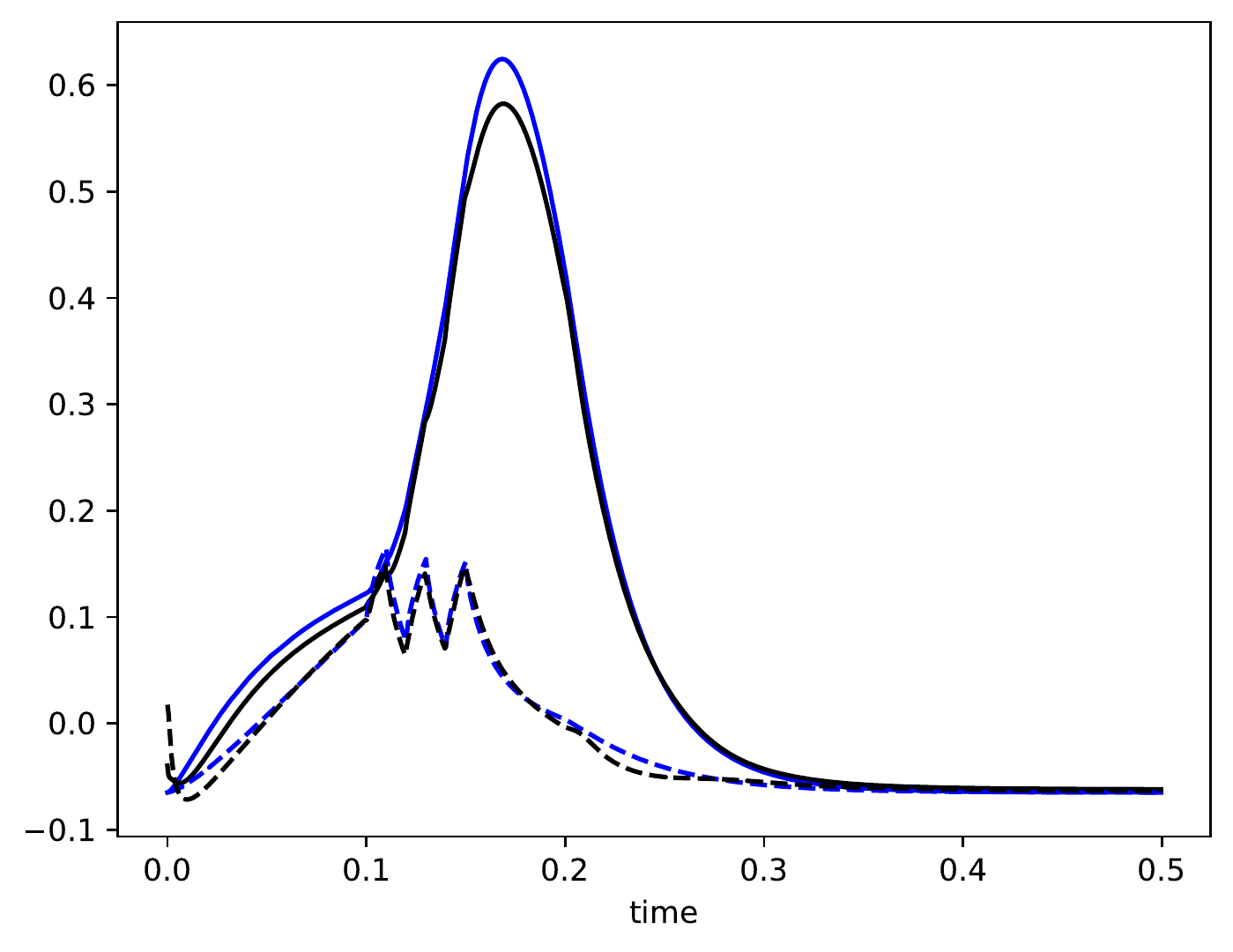}
	\\
		\includegraphics[width=0.16\textwidth]{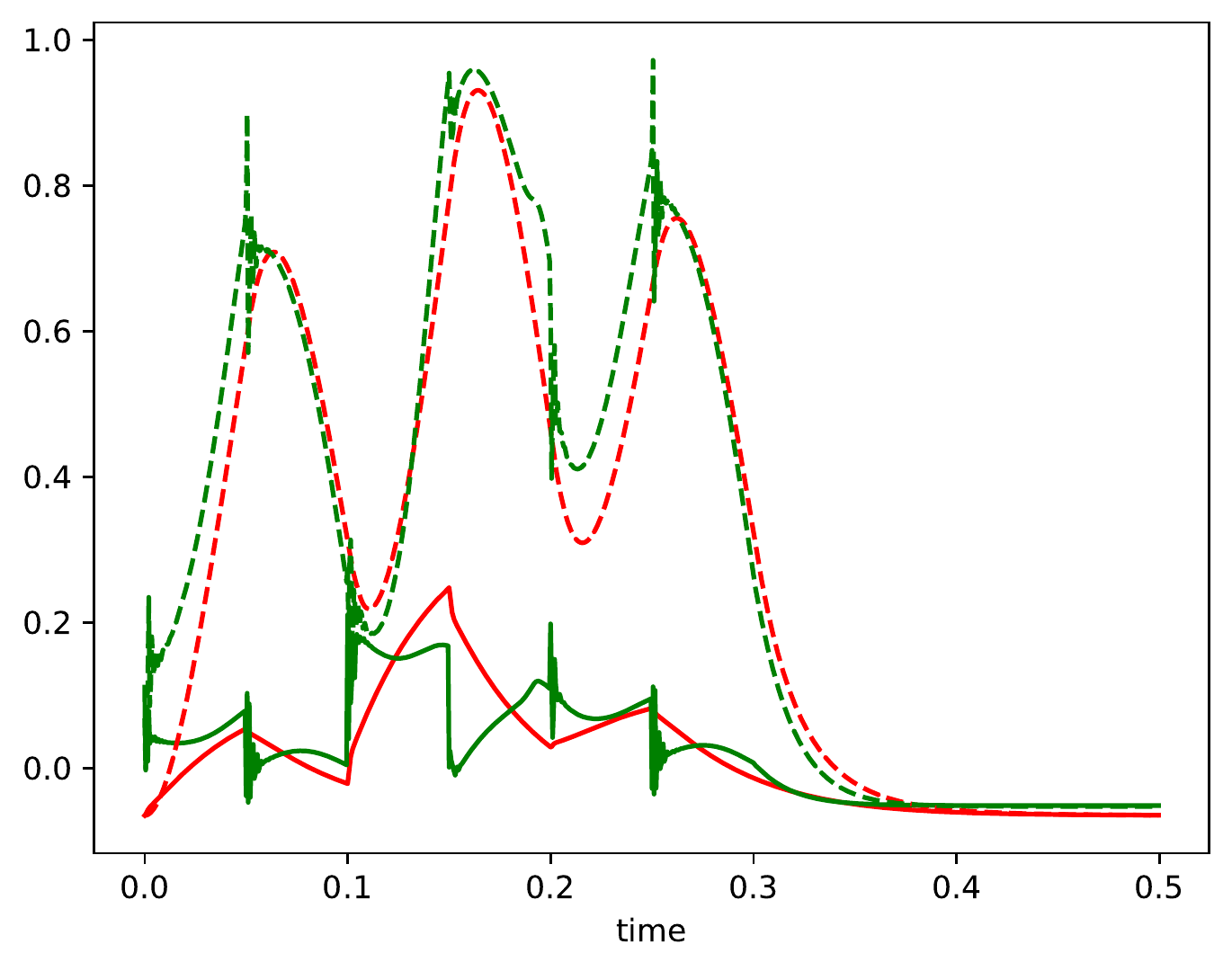} &    \includegraphics[width=0.16\textwidth]{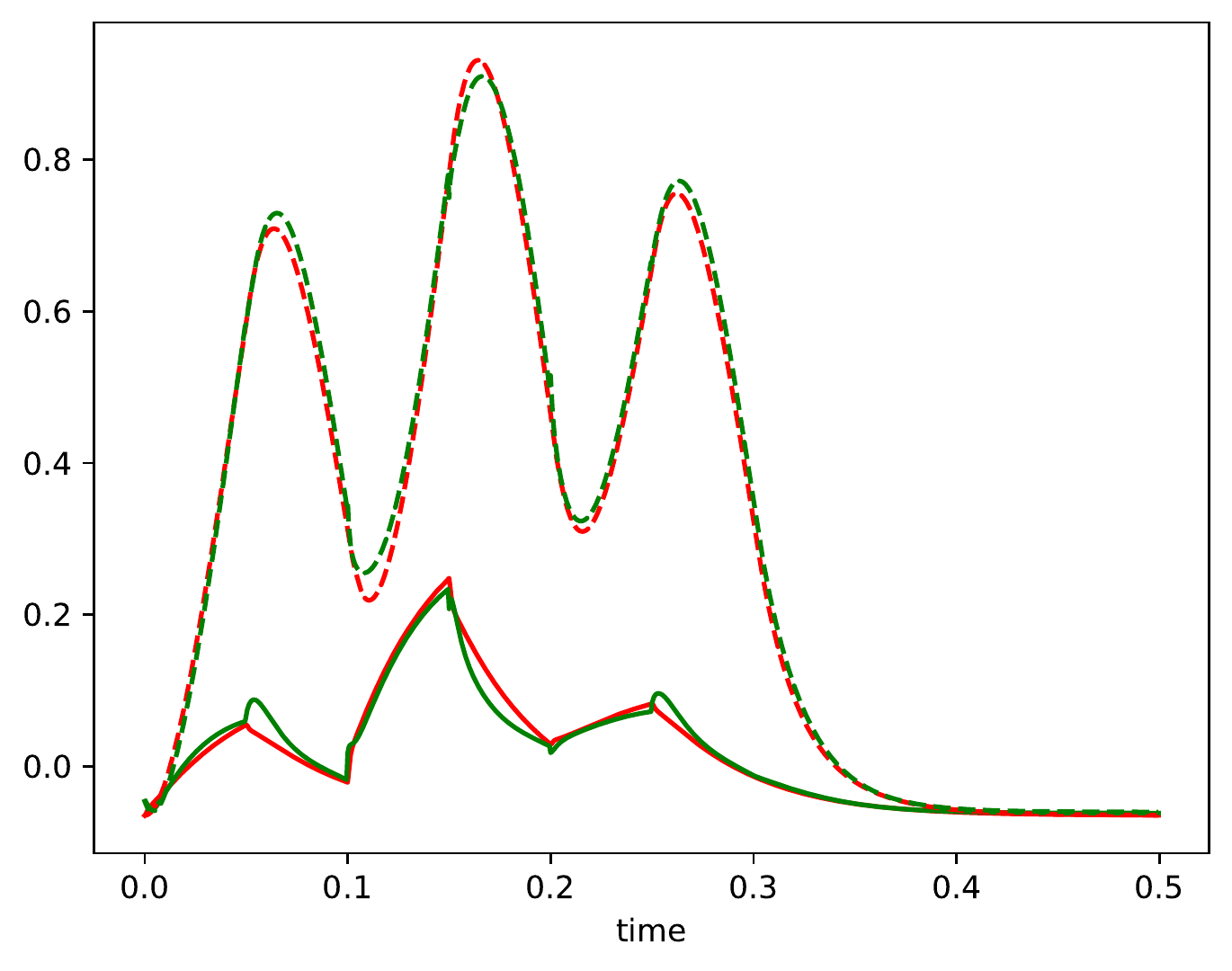} & 
		\includegraphics[width=0.16\textwidth]{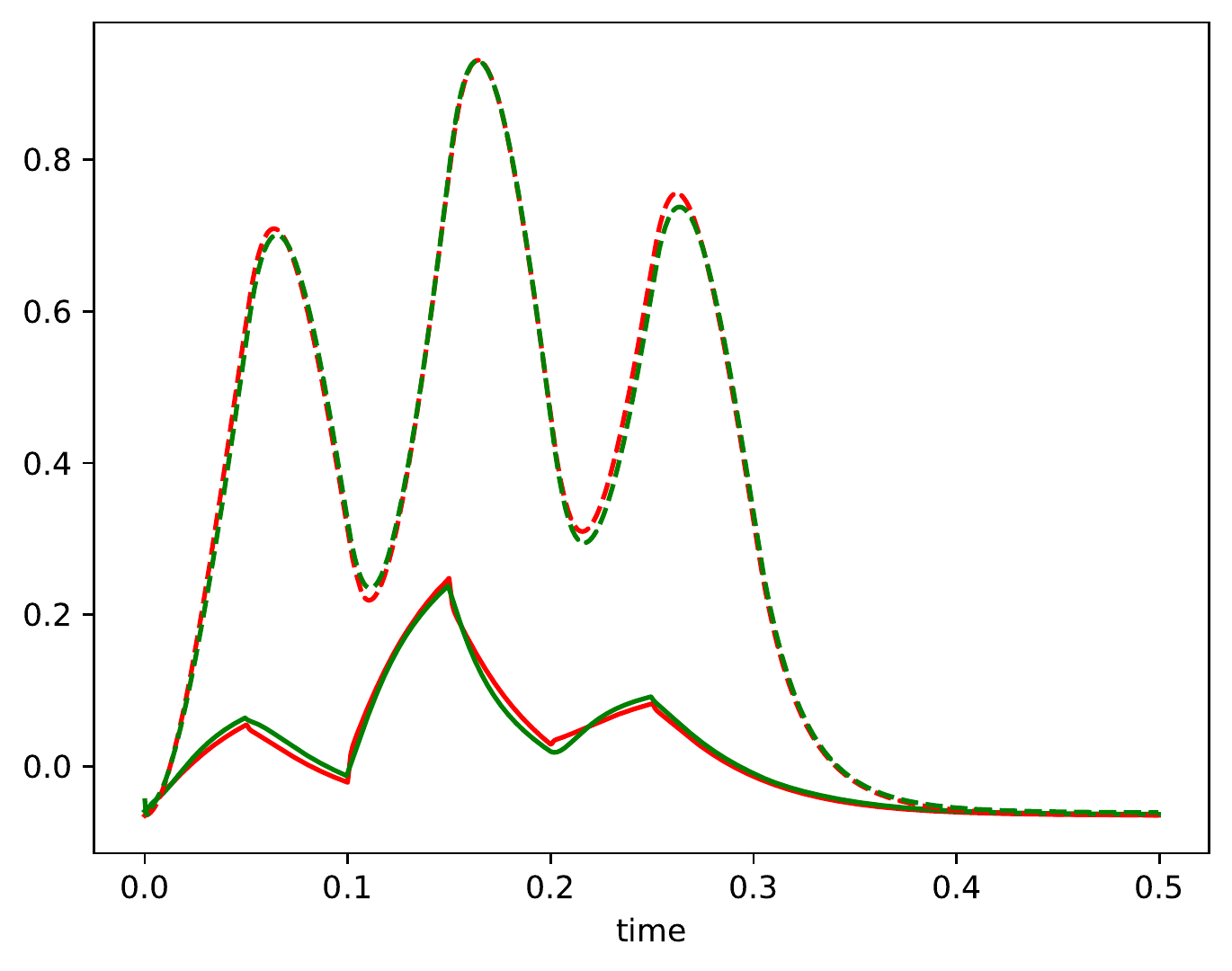} & 
		\includegraphics[width=0.16\textwidth]{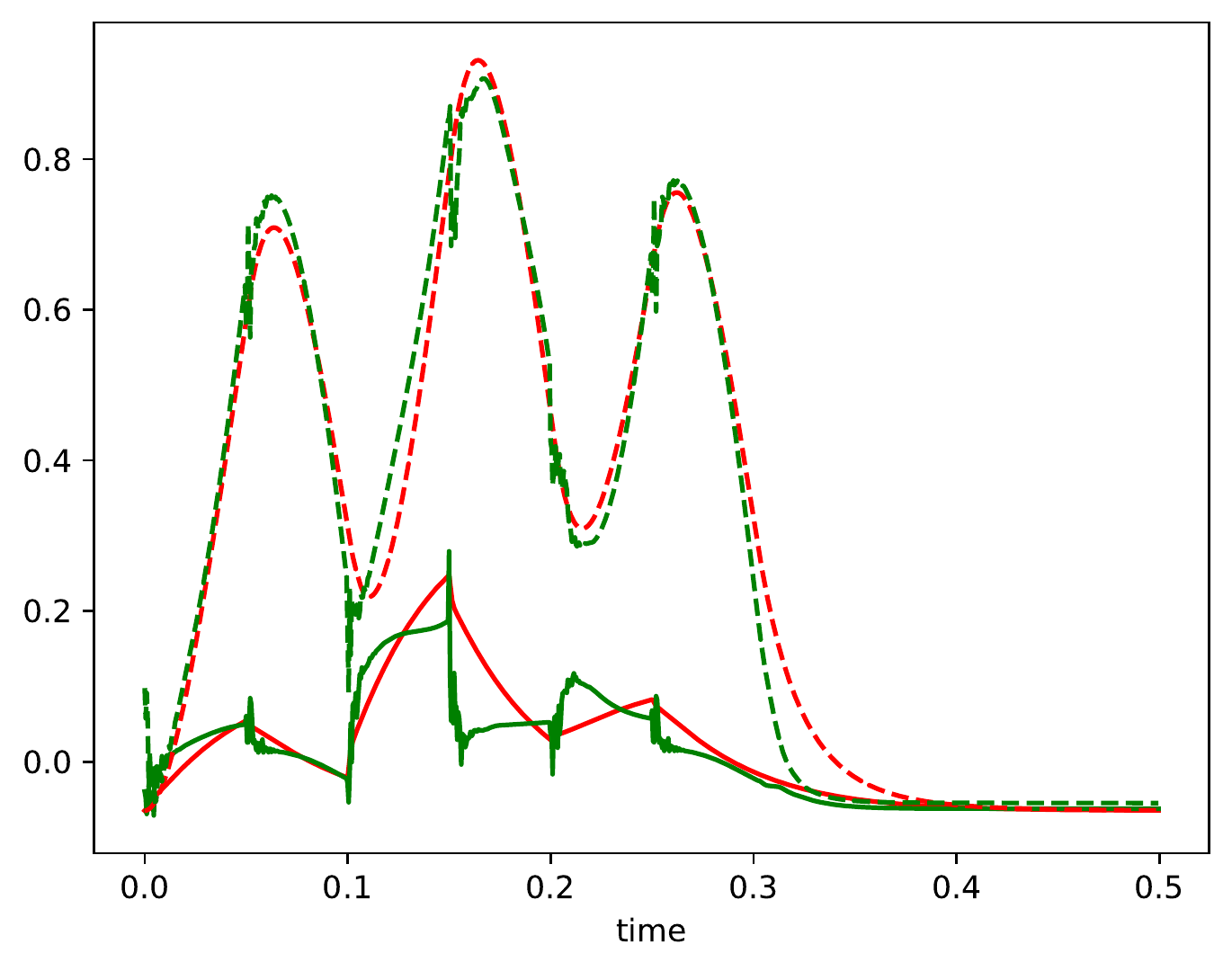} &   \includegraphics[width=0.16\textwidth]{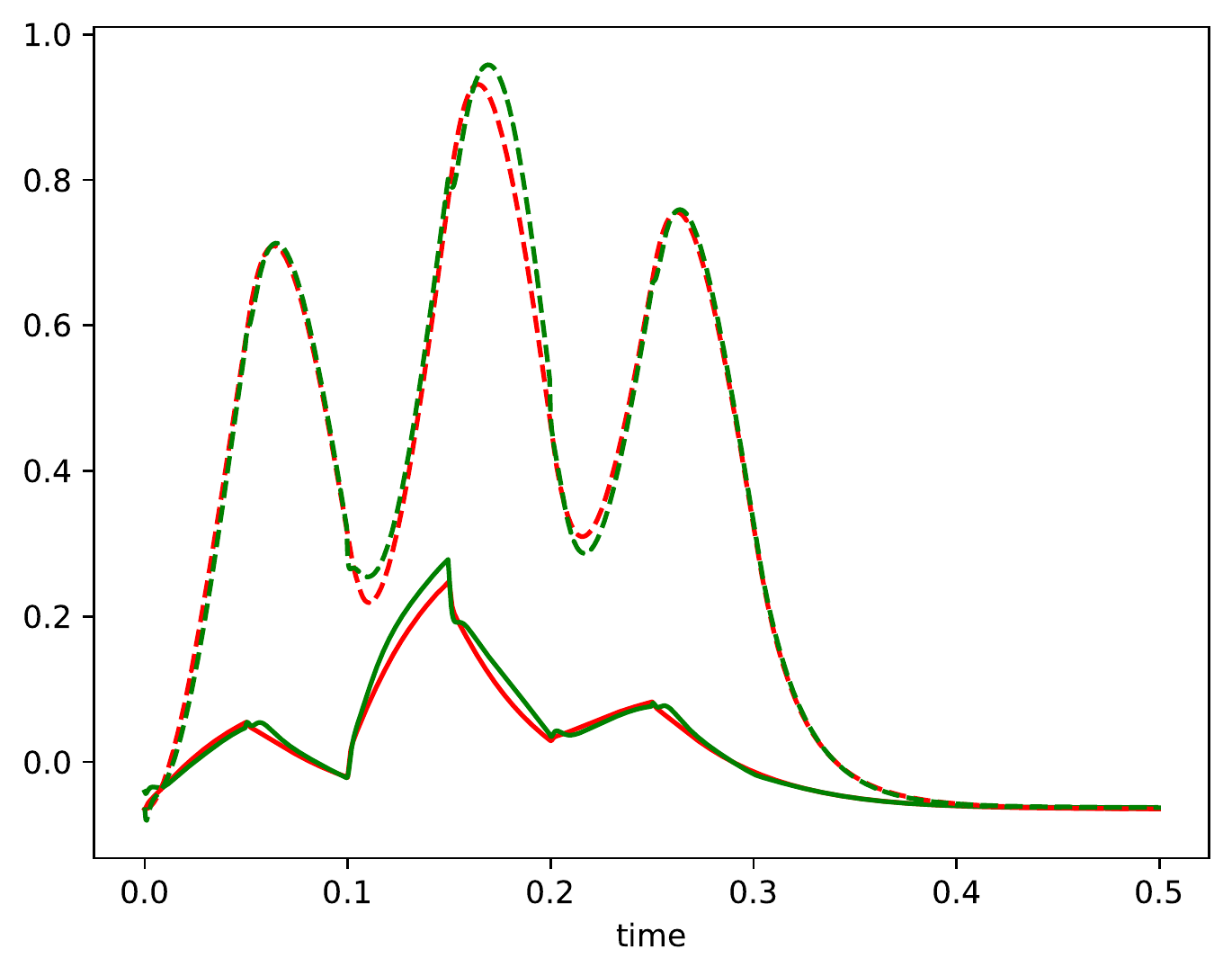} & 
		\includegraphics[width=0.16\textwidth]{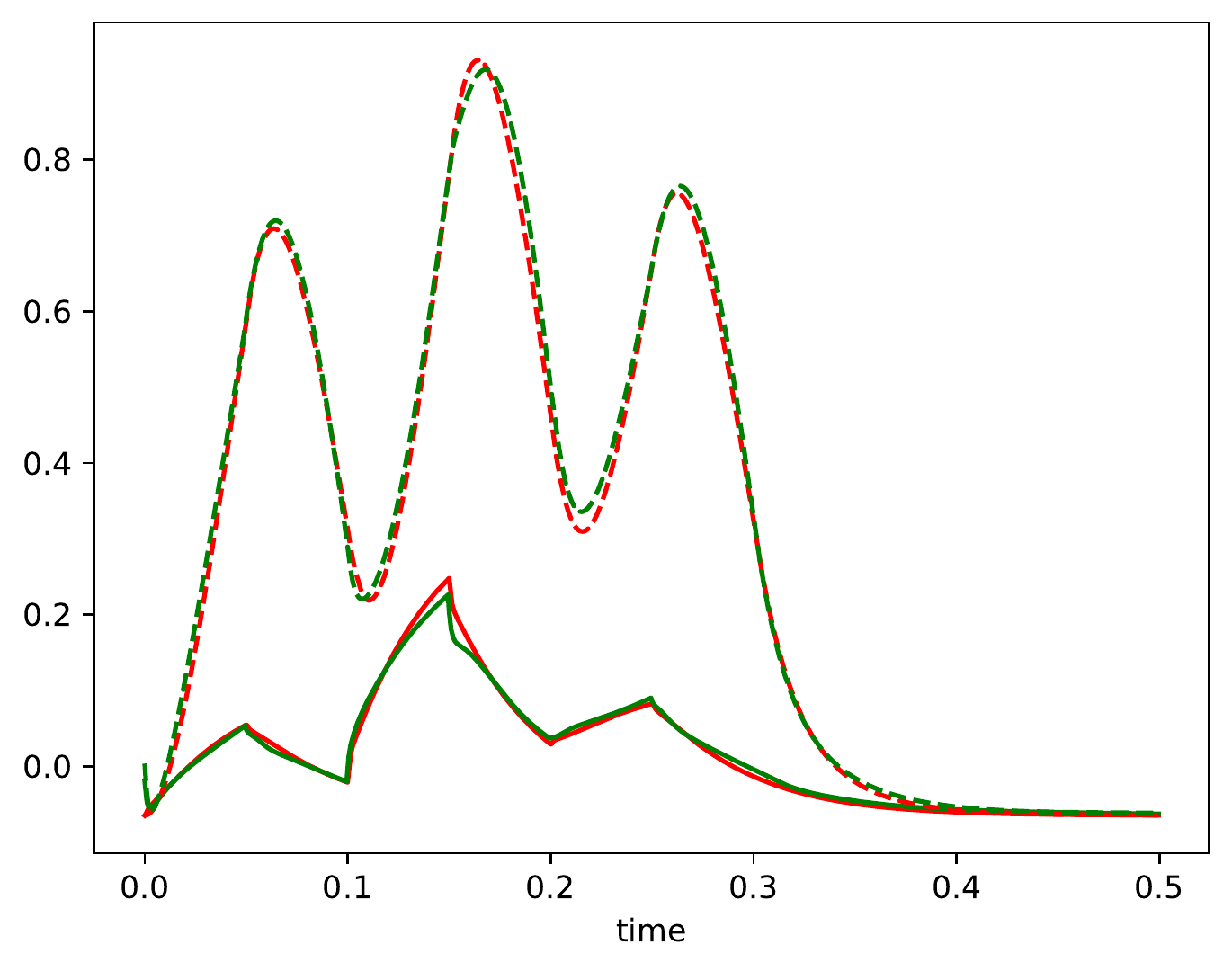}
	\\
		\includegraphics[width=0.16\textwidth]{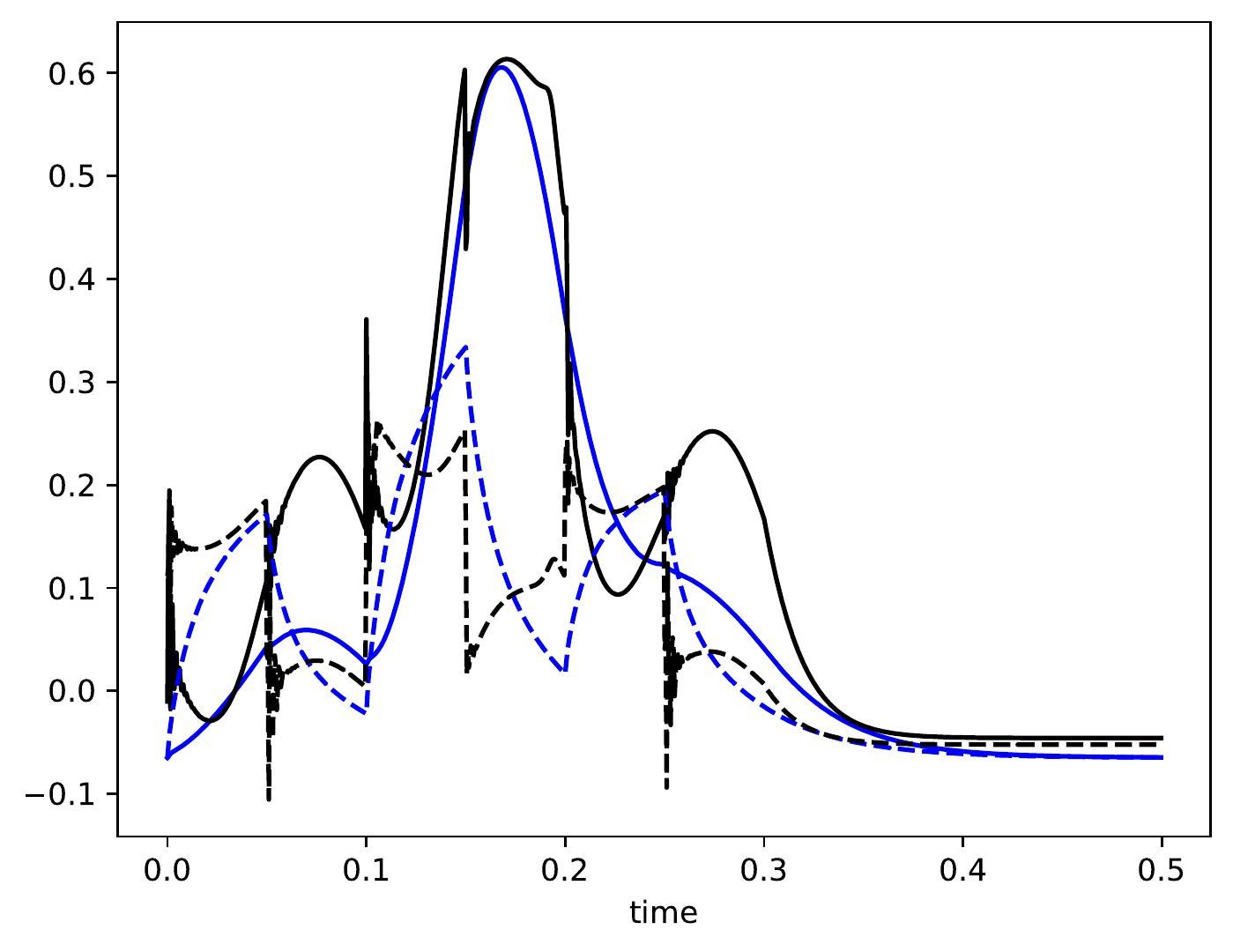} &   \includegraphics[width=0.16\textwidth]{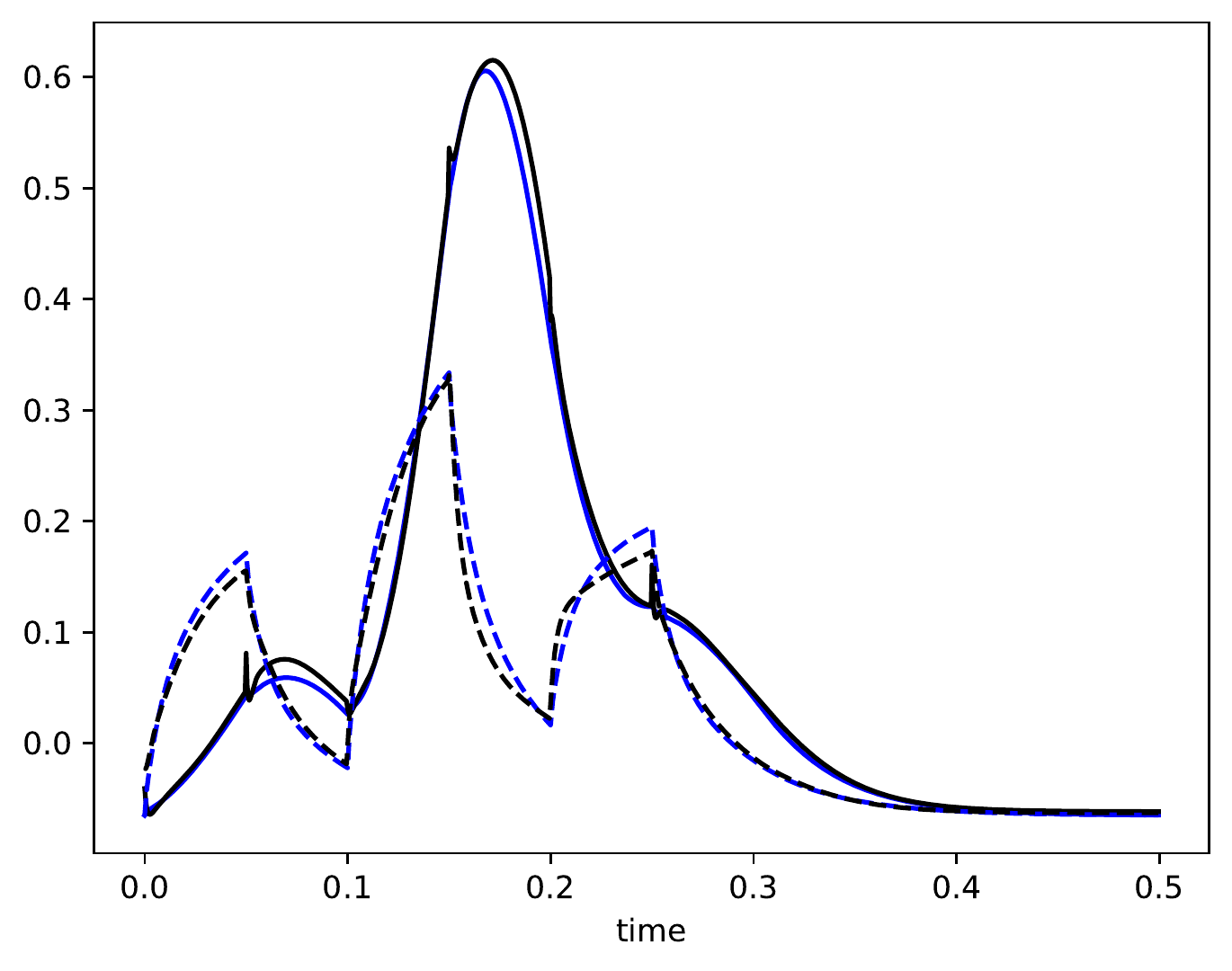} & 
		\includegraphics[width=0.16\textwidth]{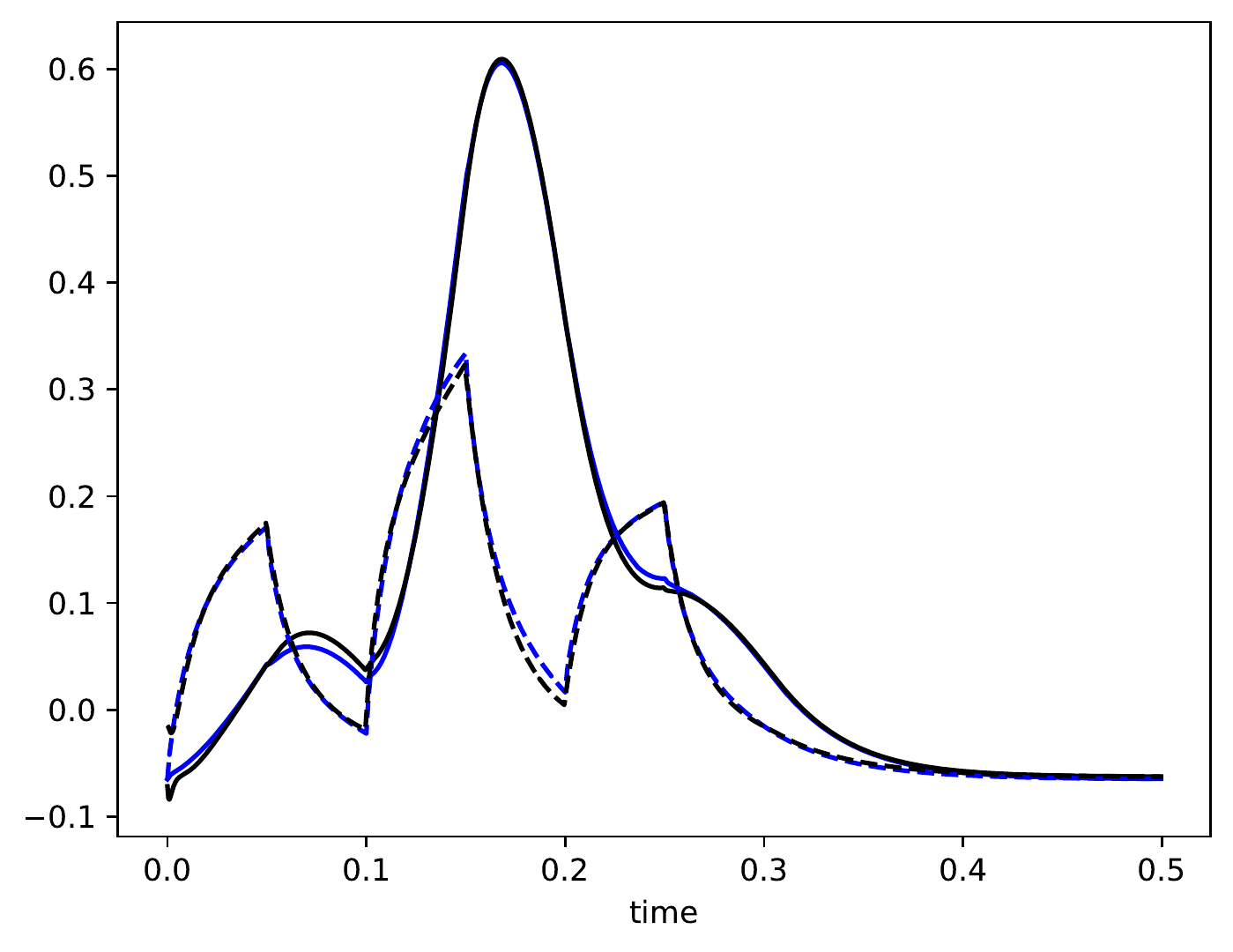} & 
		\includegraphics[width=0.16\textwidth]{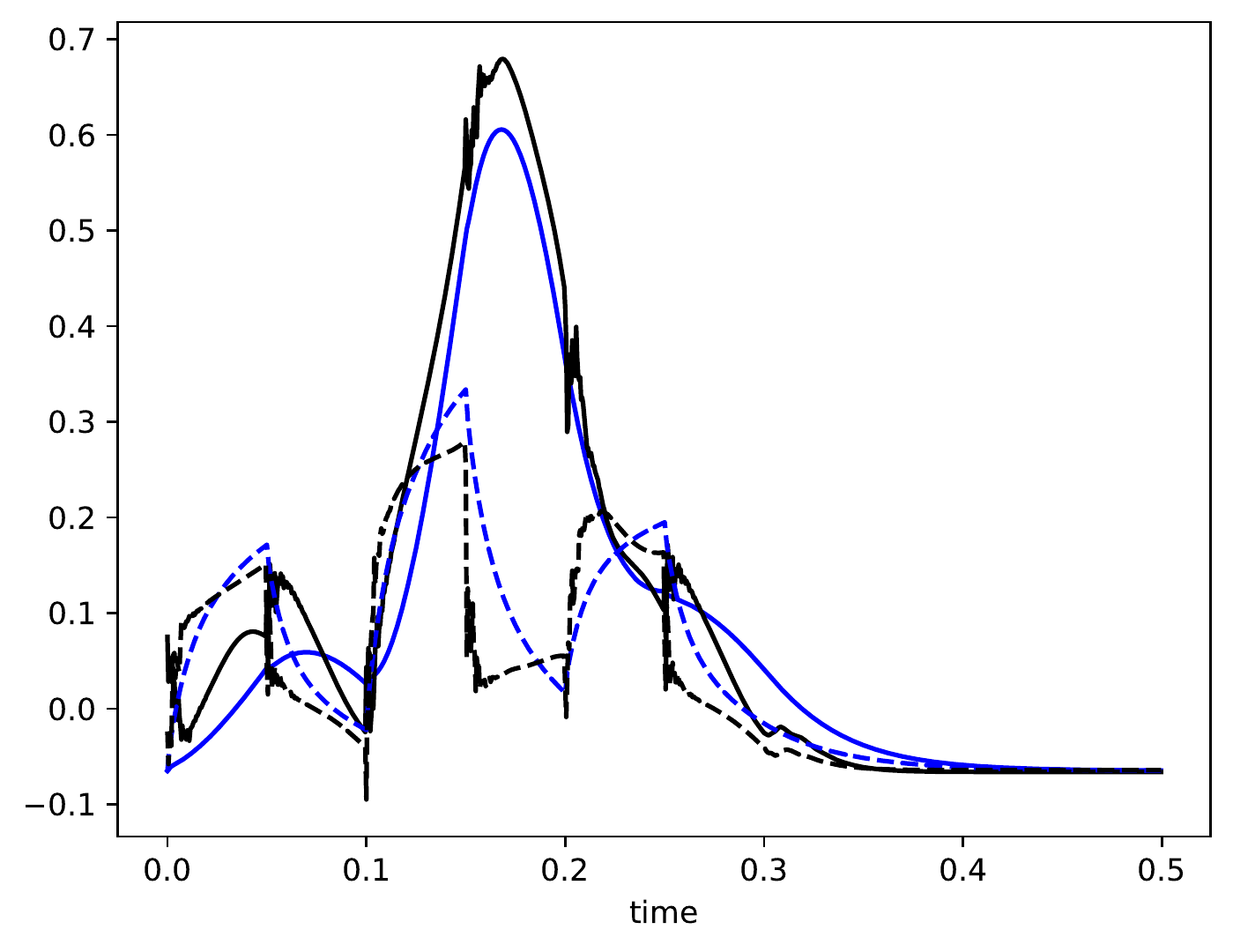} &    \includegraphics[width=0.16\textwidth]{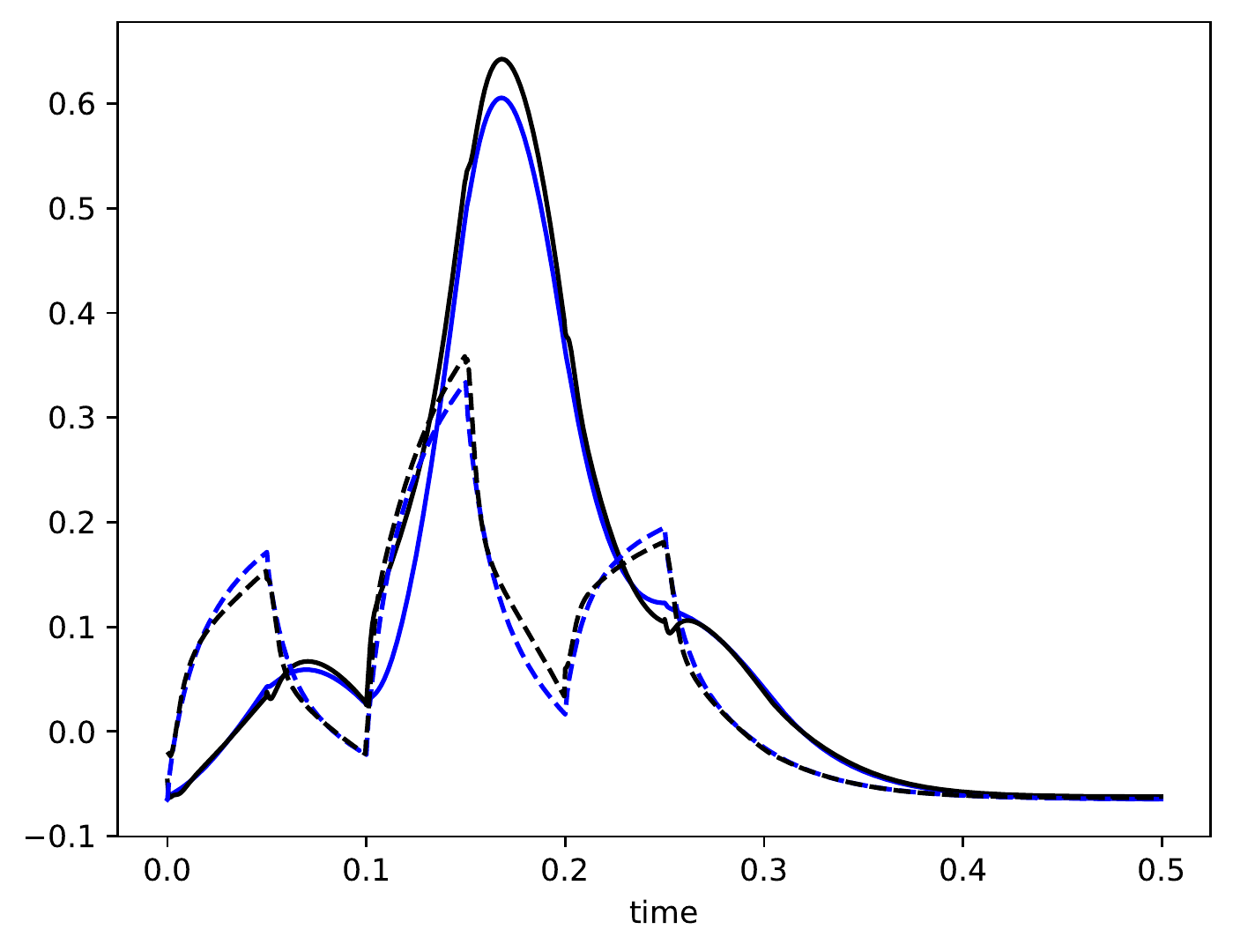} & 
		\includegraphics[width=0.16\textwidth]{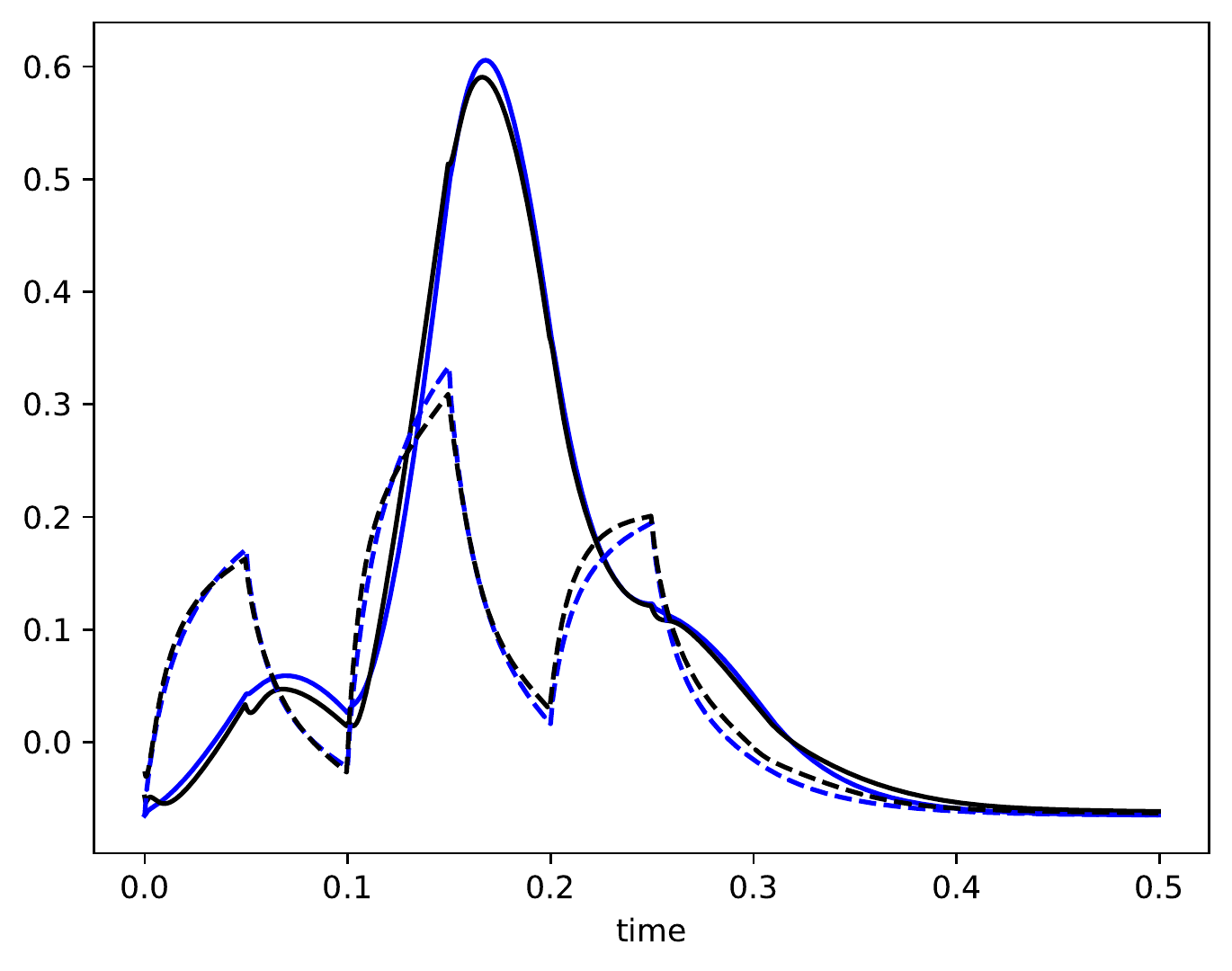}
	\\
  \end{tabular}
   \caption{Real (red \& blue) and predicted (green \& black) sequences on Exp. 1 for DB1 and LUAL ($1^{st}$ and $3^{rd}$ rows) and PVR and VB1 ($2^{nd}$ and $4^{th}$ rows) for two input sequences of the test set.}
  \label{fig:experiment1grid}
\end{figure}

Figure~\ref{fig:exp1loss} shows the evolution of the training and validation losses during the training process. From the plots we can conclude that the simple RNN unit tends to take more time to learn, also not being stable during the end of the training process. Although this is not a good indicator, it is not as alarming as what can be seen on Figure~\ref{fig:experiment1grid}, where it is clear that the simple RNN unit is not able to reproduce the outputs with the expected accuracy, while the LSTM and GRU units perform well. A summary of the results of this experiment is shown in Table~\ref{tab:experiment1}.


Since the simple RNN unit did not perform sufficiently well, not being able to reproduce the output with minimal accuracy, we are left with the LSTMs and GRUs units. Given that the GRU is the less complex unit of the two, we will keep it as the main option, knowing however that if the GRU unit fails to fulfill our expectations, the LSTM unit is also a potential candidate architecture.

\begin{table}
  \caption{The losses of the three architectures, for 16 and 64 hidden units, for the iteration with the smallest validation loss.}
  \label{tab:experiment1}
  \centering
  \begin{tabular}{lllllll}
    \toprule
    & \textbf{RNN-16} & \textbf{LSTM-16} & \textbf{GRU-16} & \textbf{RNN-64} & \textbf{LSTM-64} & \textbf{GRU-64} \\ \midrule
    Training    & 4.1694e-03         & 1.0566e-04      & 4.8521e-05     & 8.9530e-04     & 7.0319e-05      & 5.1814e-05     \\ \midrule
    Validation  & 6.9184e-03         & 1.4021e-03      & 1.2415e-03     & 3.0161e-03     & 1.1483e-03      & 1.2251e-03     \\ \midrule
    Test        & 8.0882e-03         & 3.7738e-04      & 1.1487e-04     & 3.0835e-03     & 5.3662e-04      & 1.7551e-04     \\
    \bottomrule
  \end{tabular}
\end{table}

\subsection{Experiment 2} \label{sec:experiment2}

From the previous example, the GRU, due to its accuracy and relative simplicity, emerged as the prime candidate unit for our modelling purposes. However, to determine how small the models can be without compromising the accuracy, we need to determine the required number of units for the recurrent layer.
The focus of this second experiment is therefore to test different sizes of the recurrent layer and determine the smallest size that is still able to generate a model with sufficient accuracy.

We test both LSTM and GRU units using the dataset with the coarser time step of \(0.5~\si{\milli\second}\). The LSTM does not produce noticeable improvement over the GRU and therefore we only report here the results obtained with the GRU unit for
6 different sizes of the recurrent layer: \(2\), \(4\), \(8\), \(16\), \(32\), \(64\).

\begin{figure}
    \centering
    \includegraphics[scale=0.375]{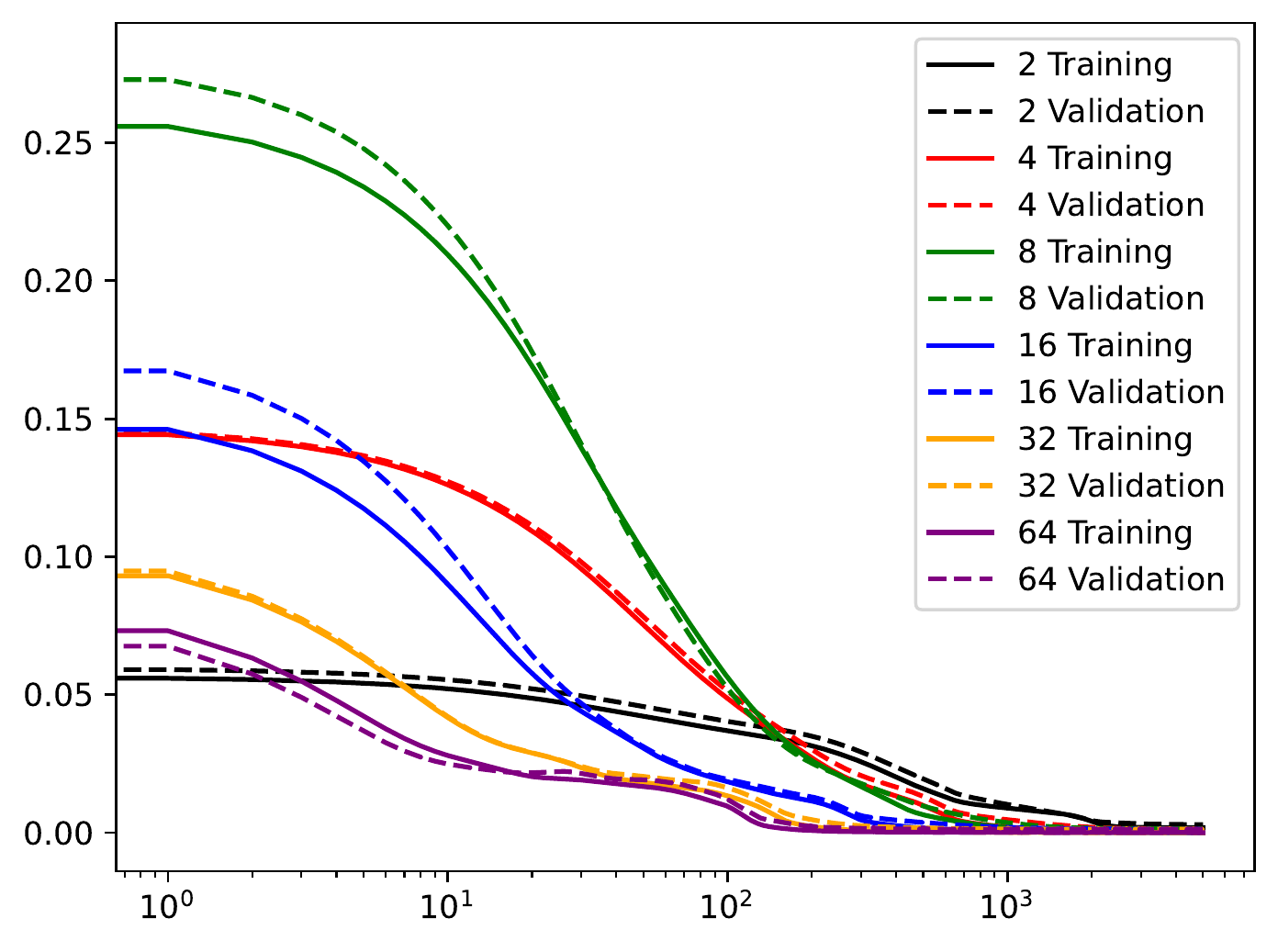}
    \caption{Training and validation loss for 6 different hidden sizes of a GRU-based recurrent layer.}
    \label{fig:exp2loss}
\end{figure}

\begin{figure}
 \setlength\tabcolsep{2pt}
  \centering
  \begin{tabular}{ccccccc}
    \textbf{2} & \textbf{4} & \textbf{8} & \textbf{16} & \textbf{32} & \textbf{64} \\
        \includegraphics[width=0.16\textwidth]{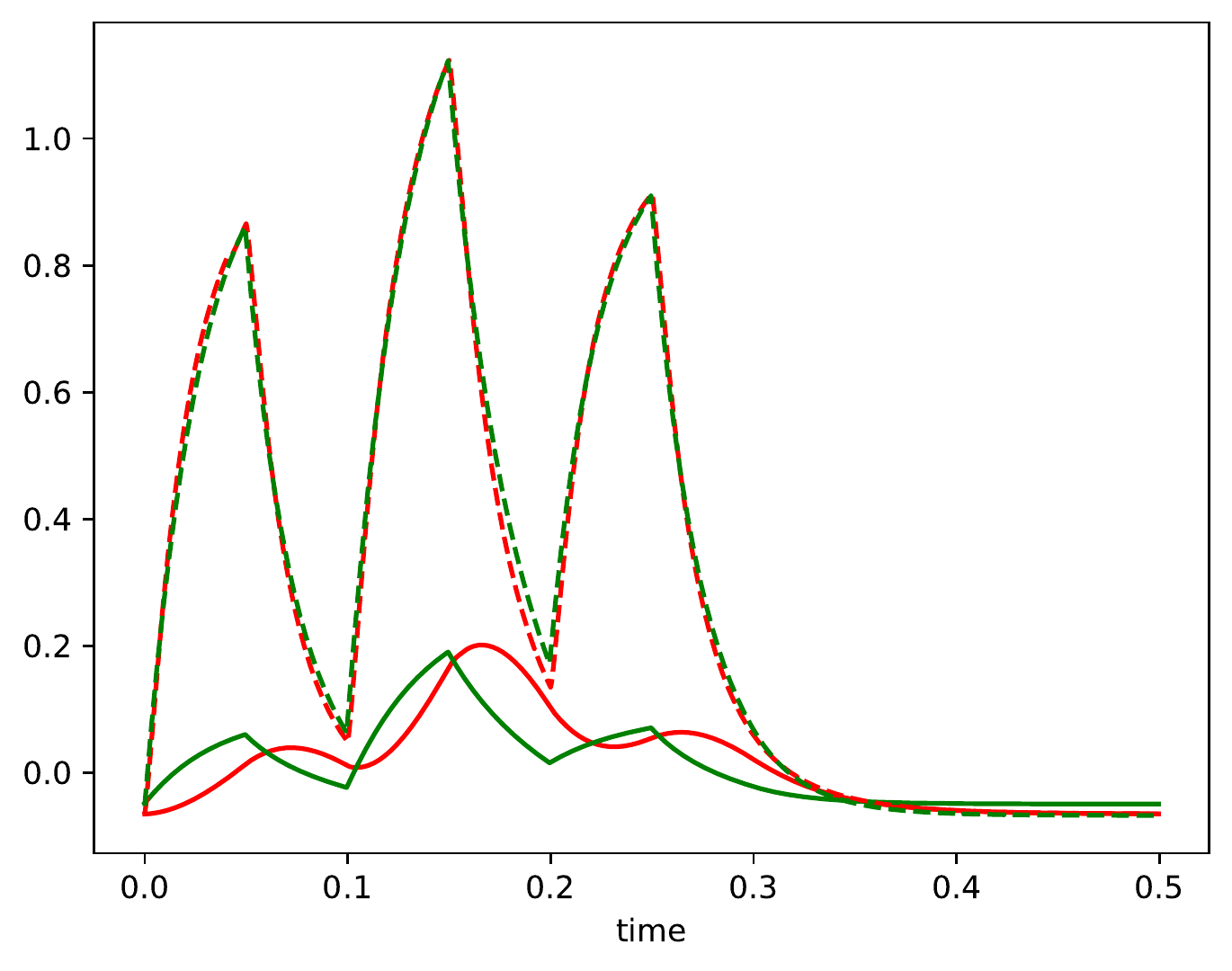} &
		\includegraphics[width=0.16\textwidth]{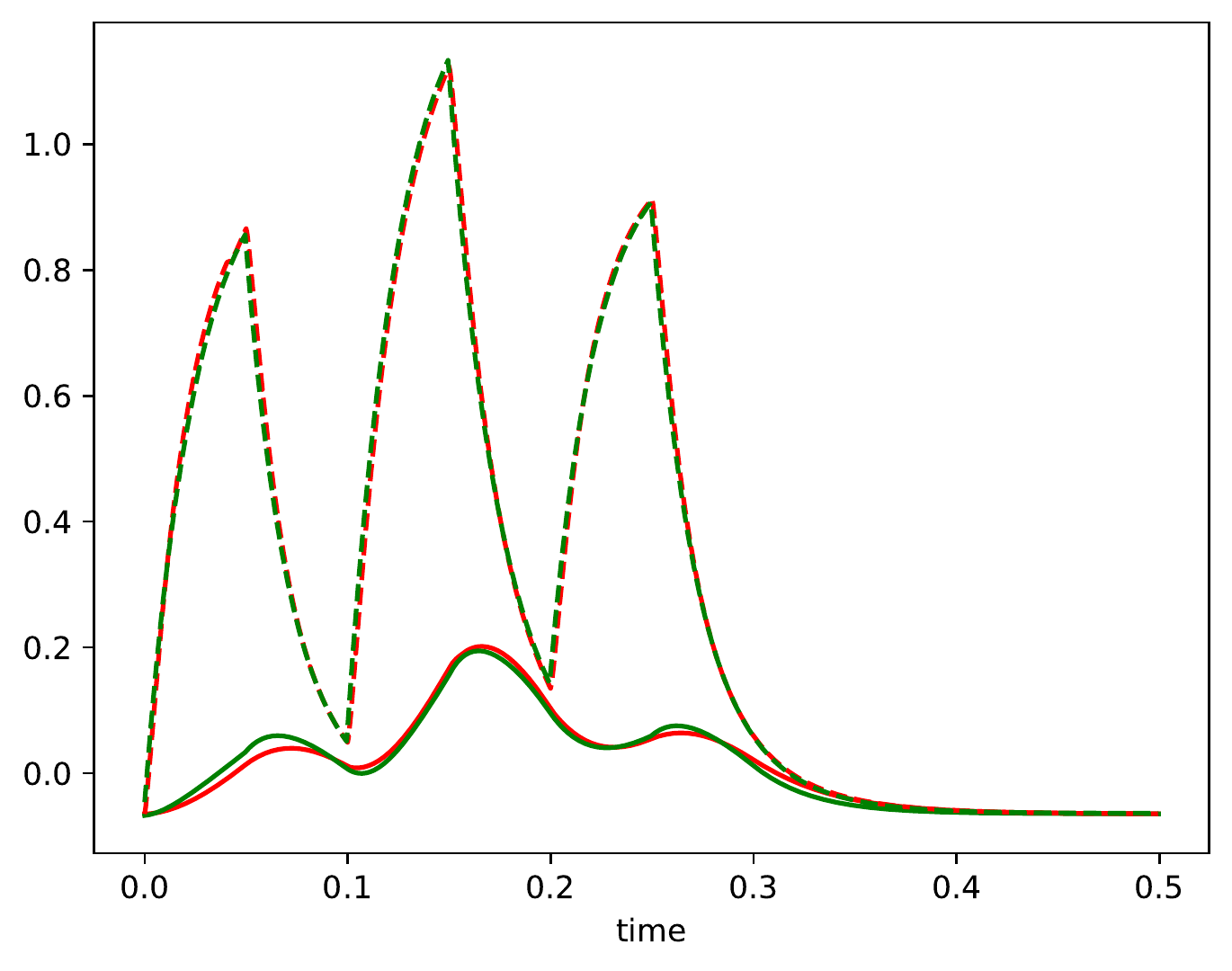} &  \includegraphics[width=0.16\textwidth]{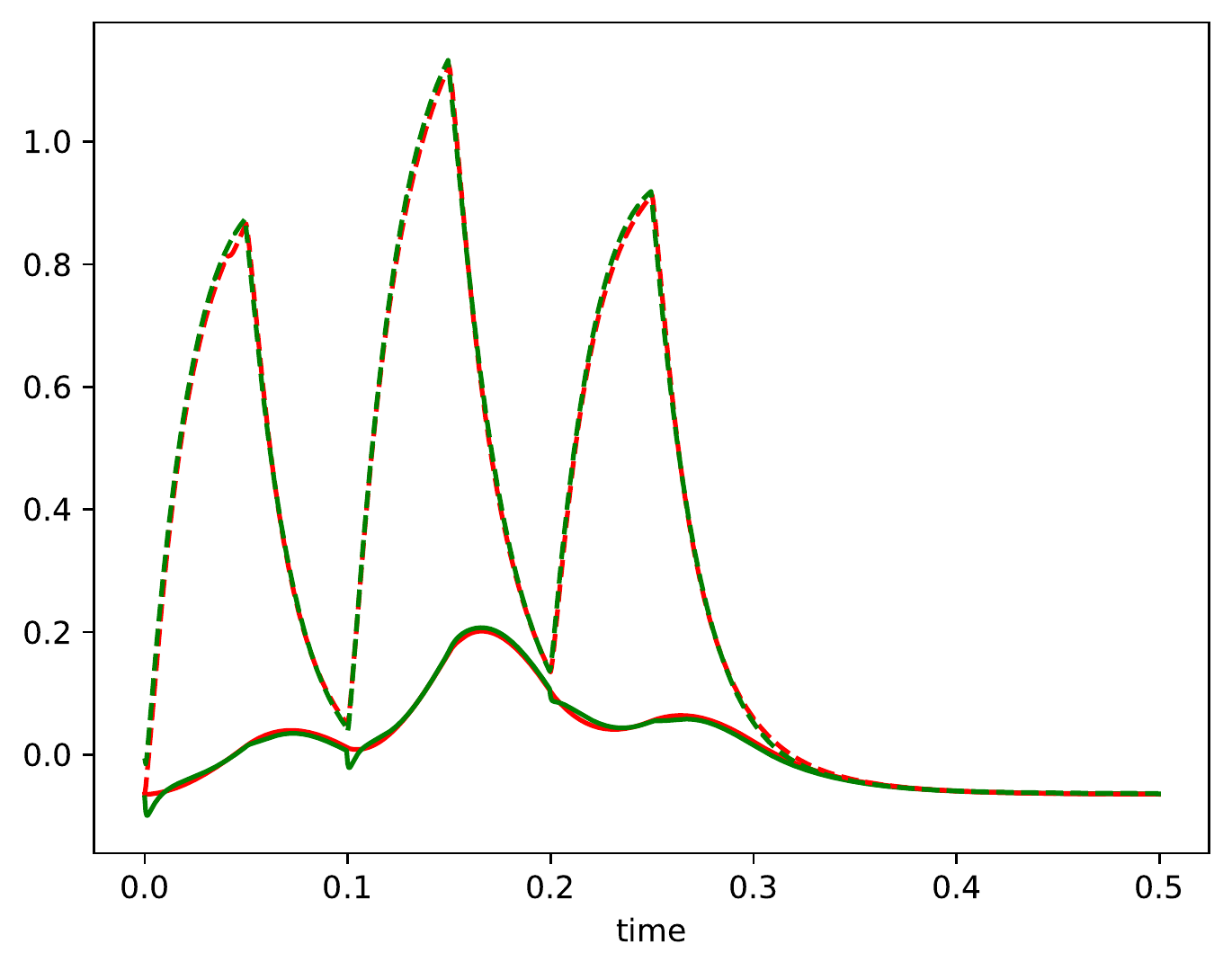} & 
		\includegraphics[width=0.16\textwidth]{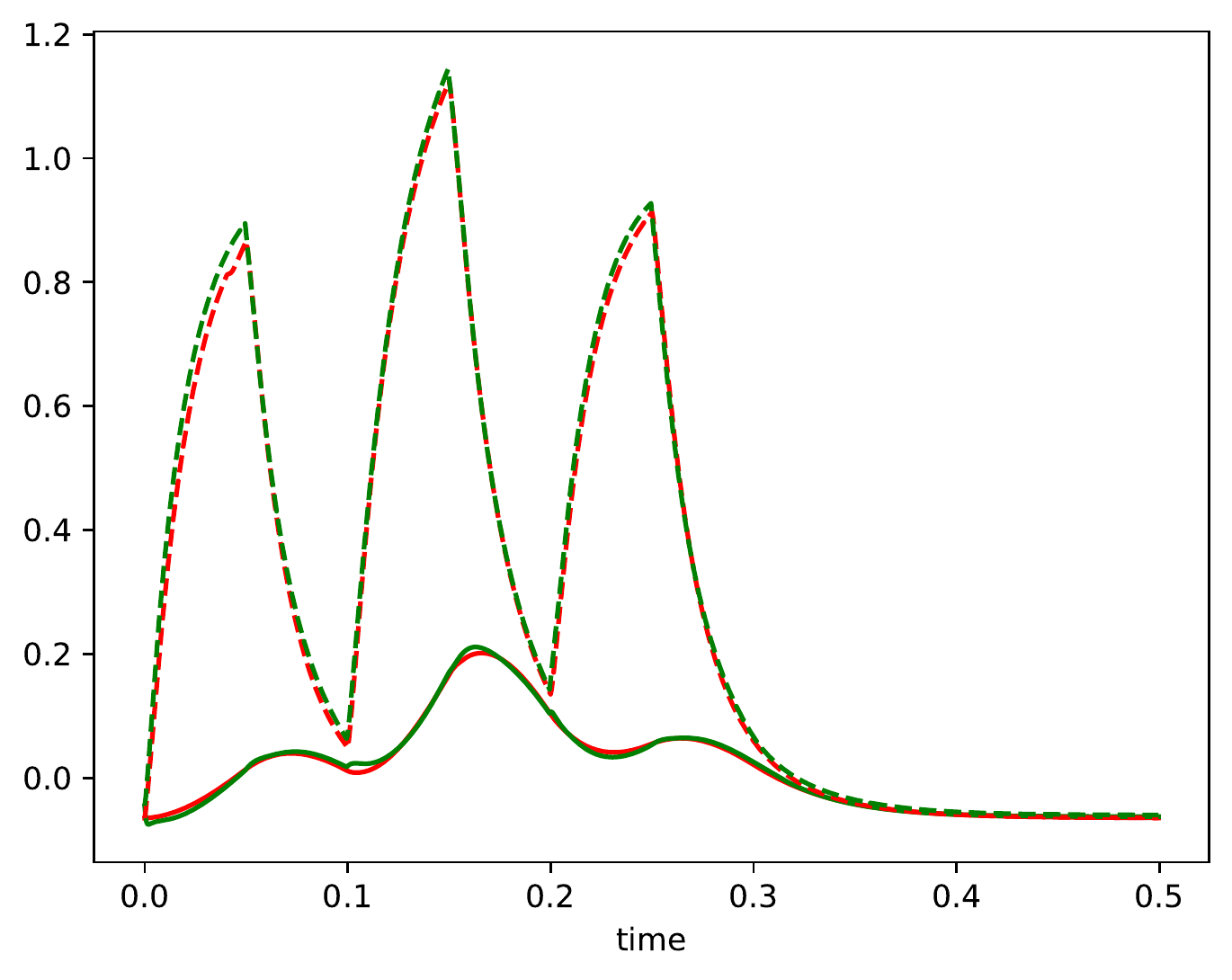} & 
		\includegraphics[width=0.16\textwidth]{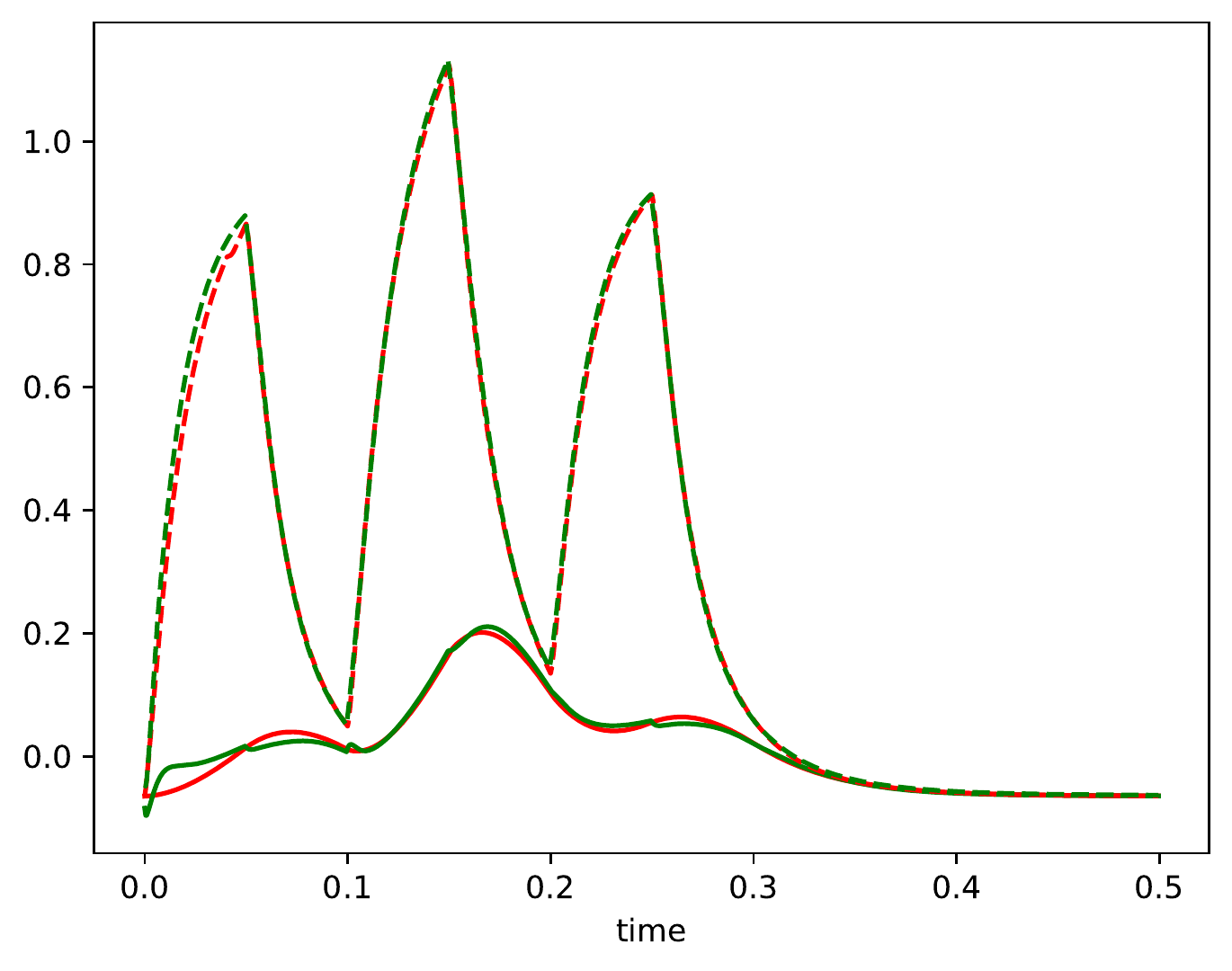} &  \includegraphics[width=0.16\textwidth]{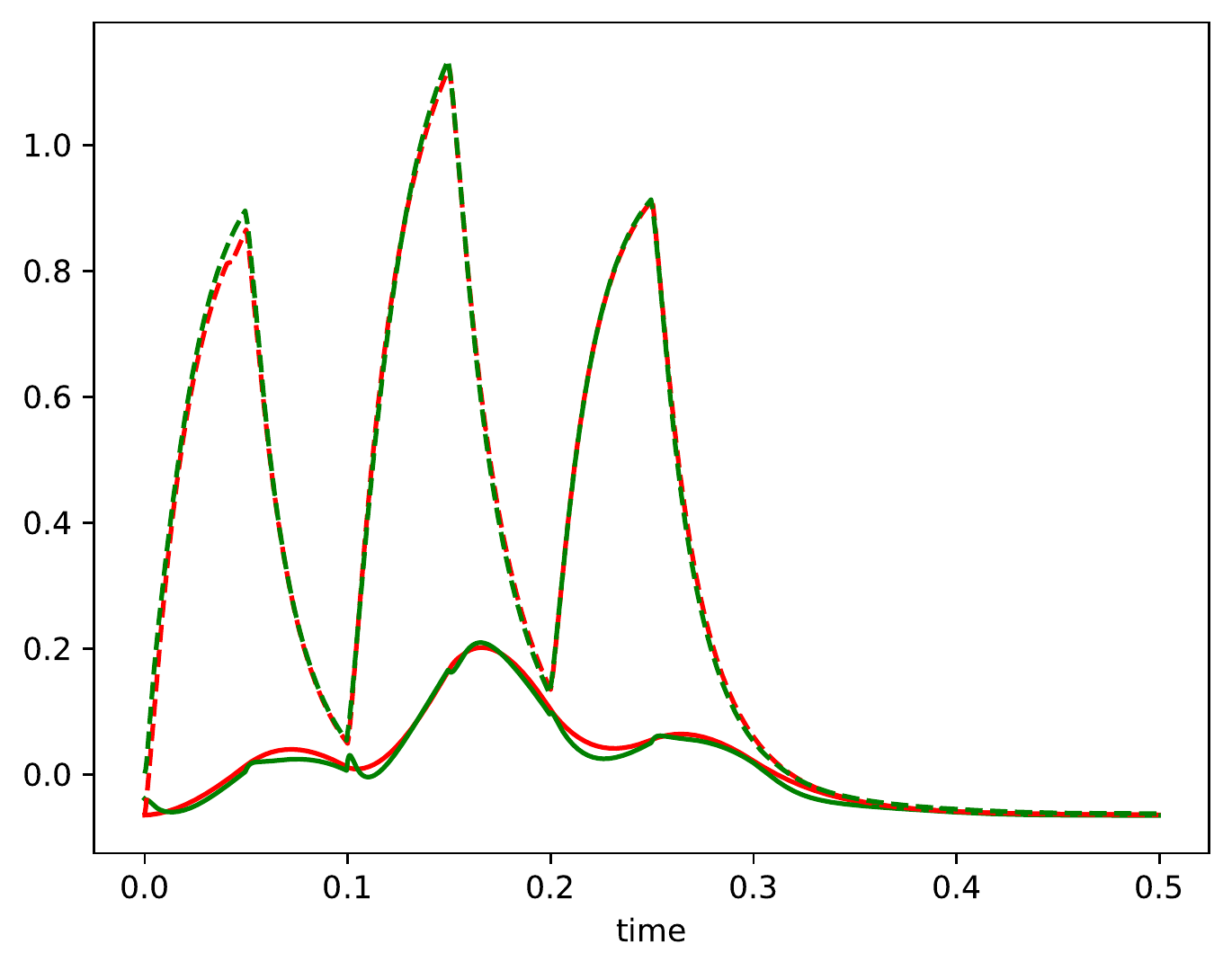}
	\\
	    \includegraphics[width=0.16\textwidth]{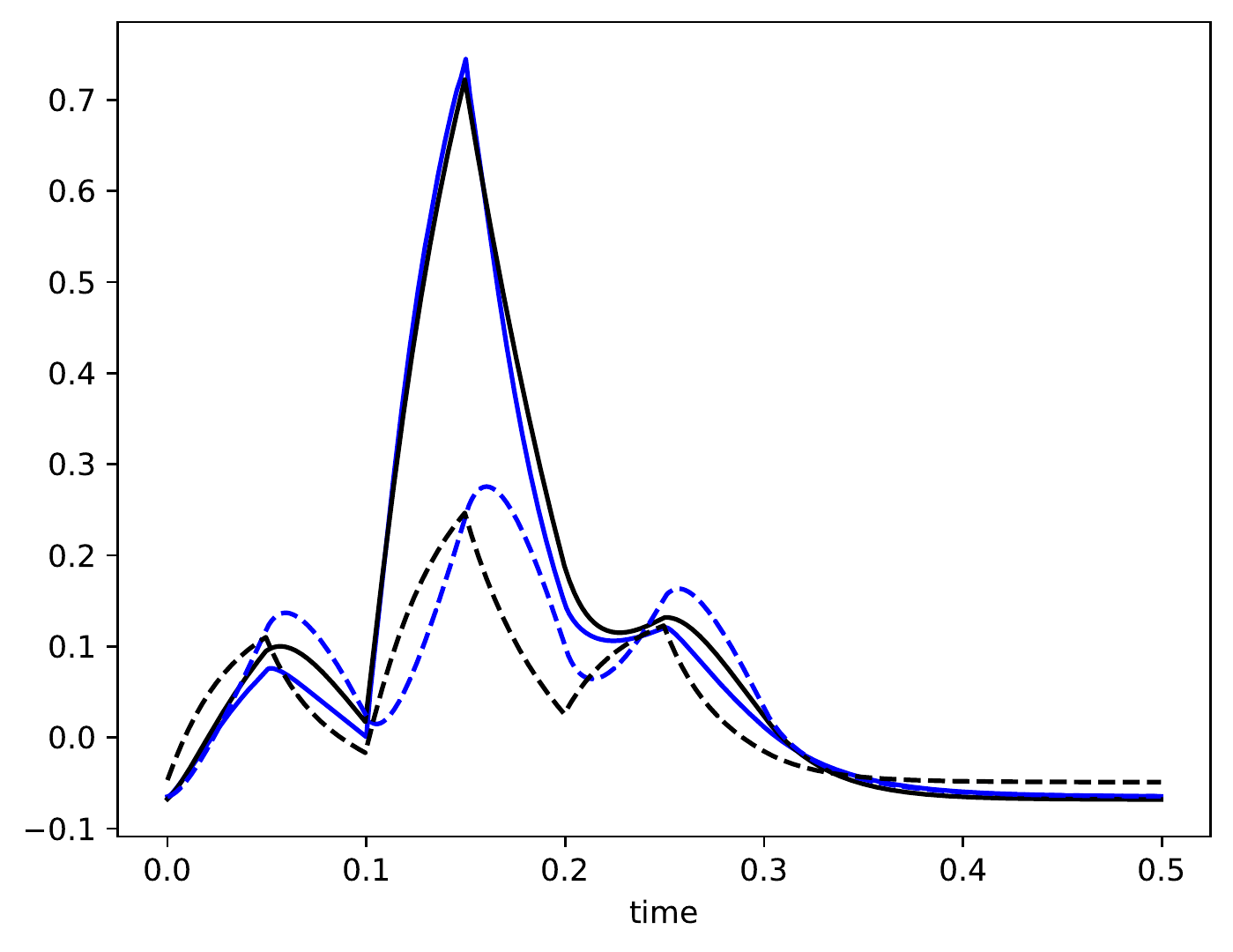} &
		\includegraphics[width=0.16\textwidth]{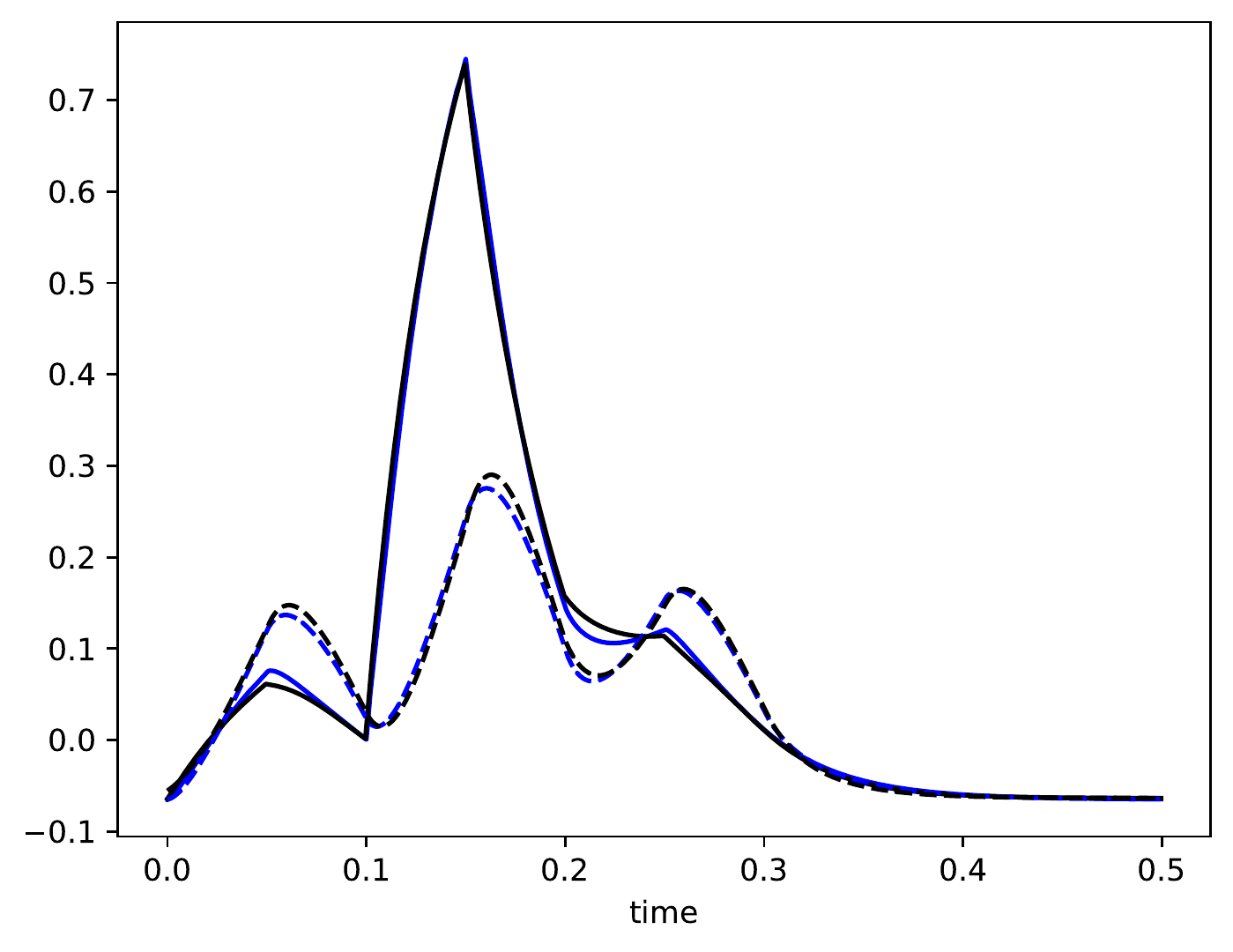} &  \includegraphics[width=0.16\textwidth]{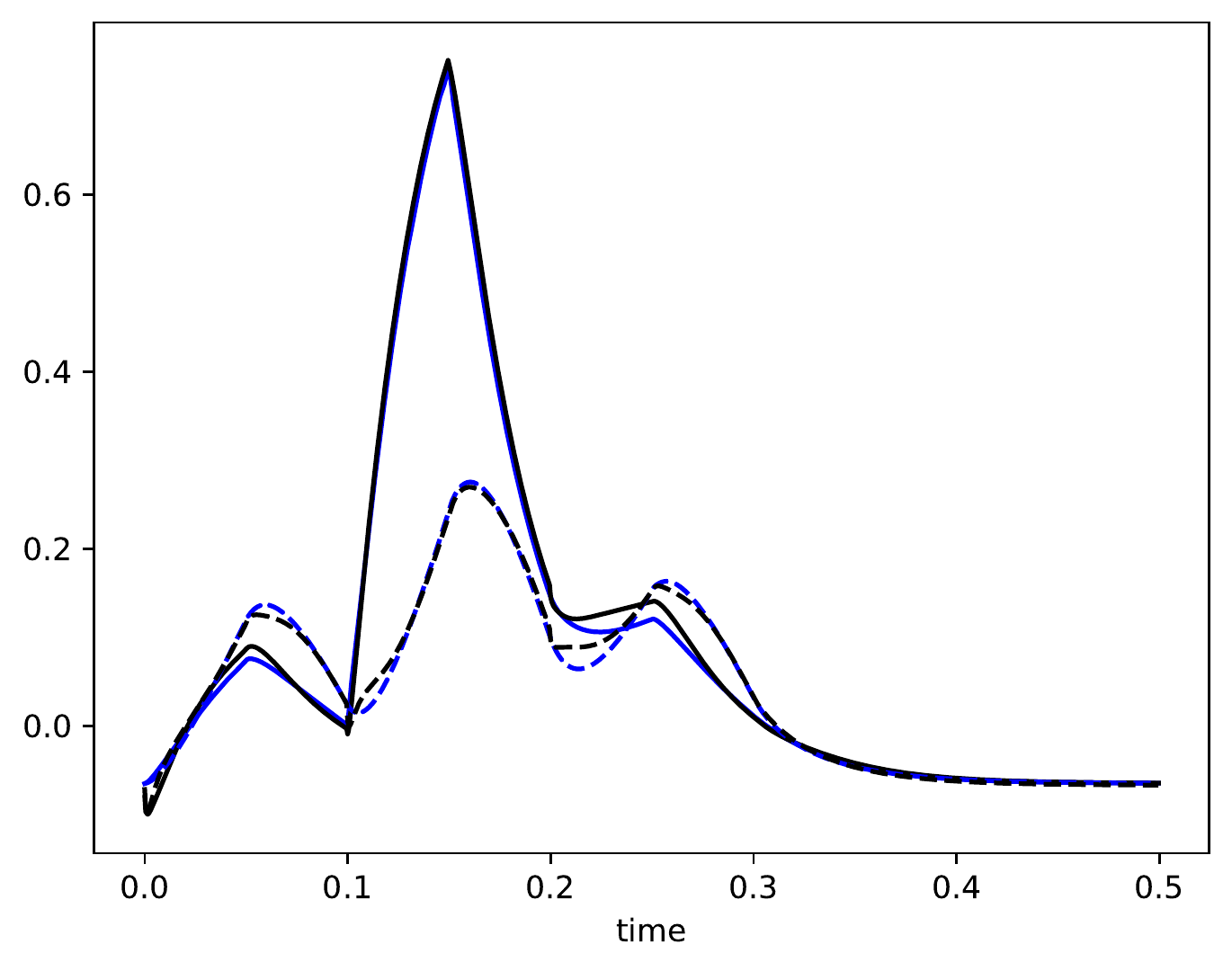} & 
		\includegraphics[width=0.16\textwidth]{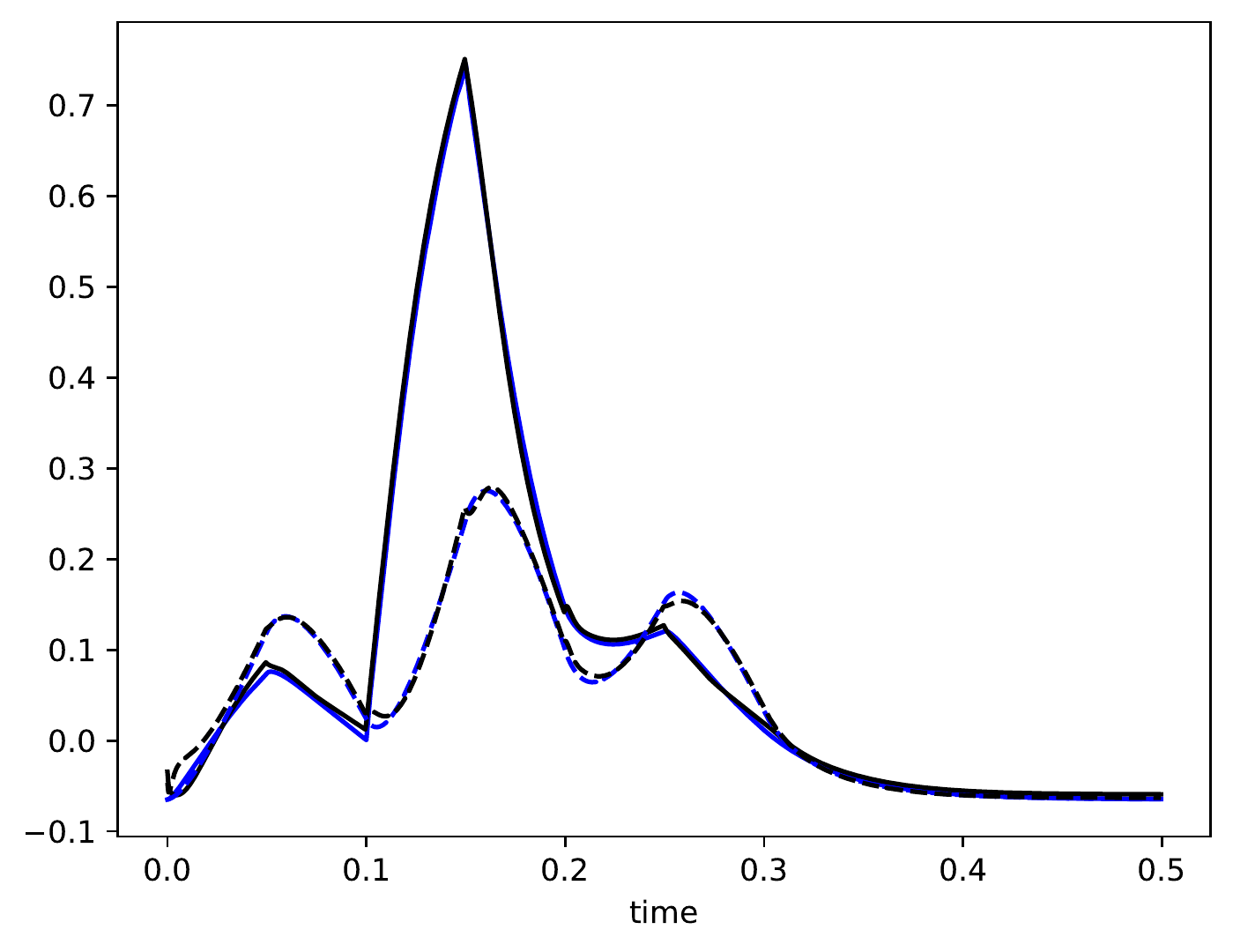} & 
		\includegraphics[width=0.16\textwidth]{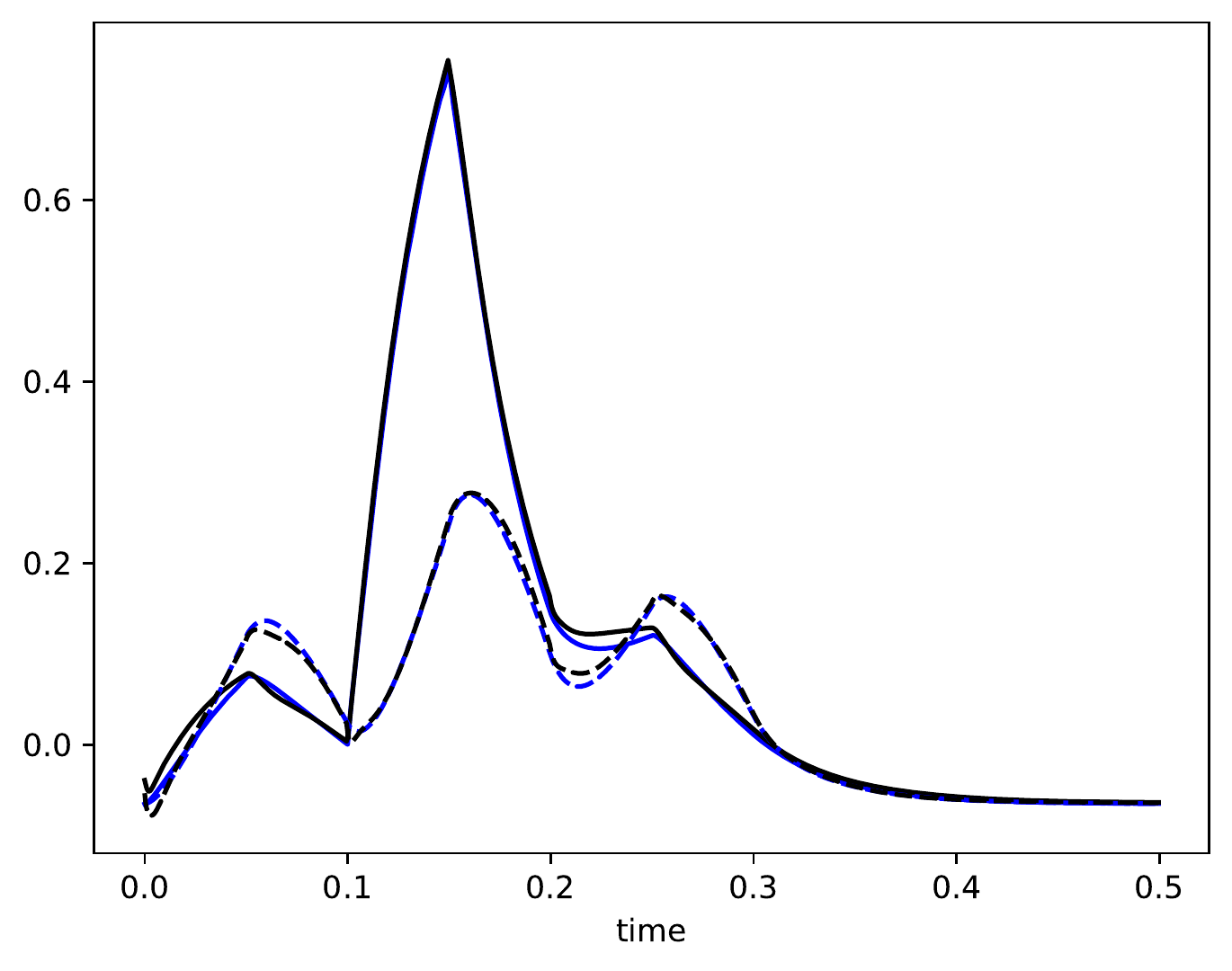} &  \includegraphics[width=0.16\textwidth]{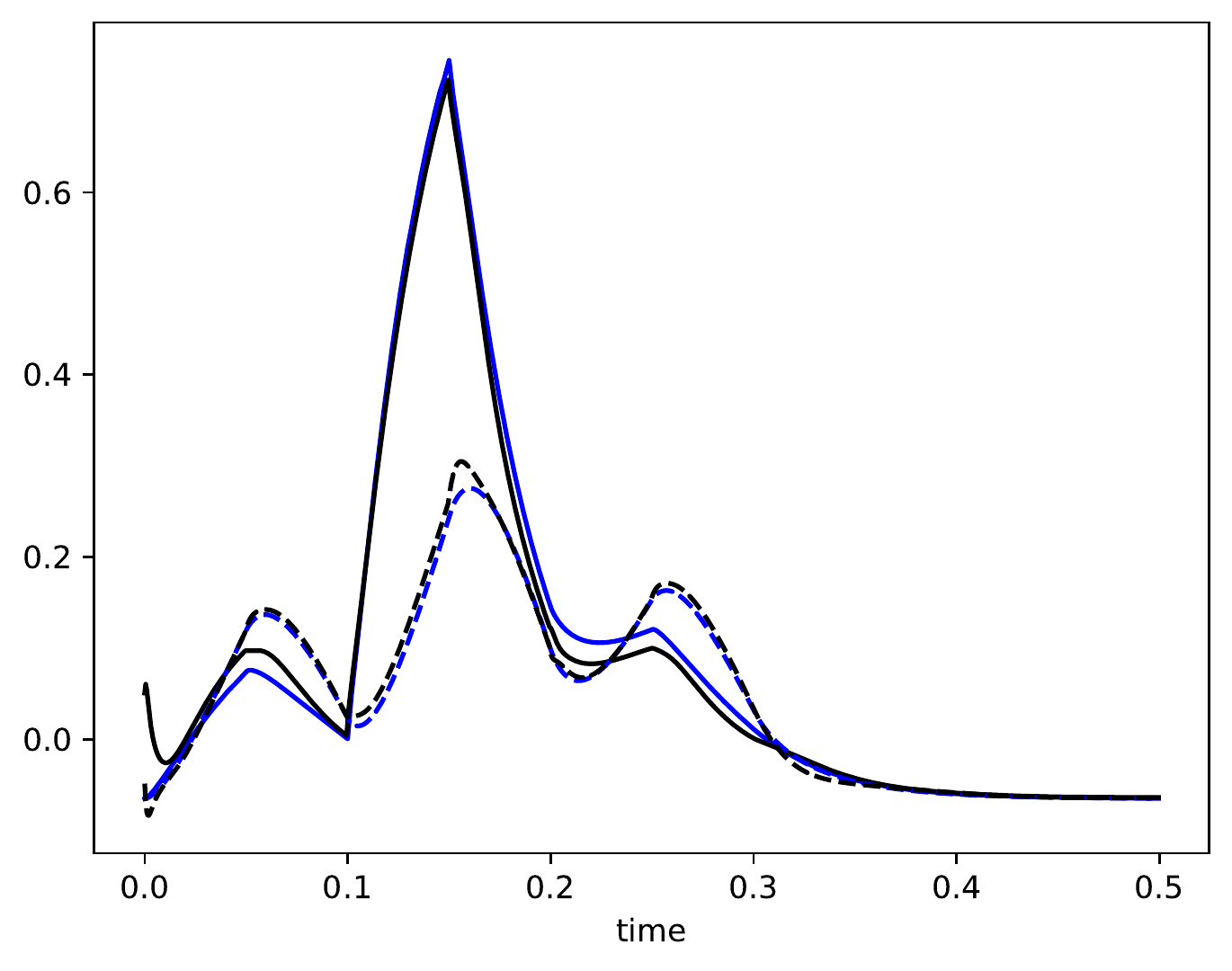}
	\\
	    \includegraphics[width=0.16\textwidth]{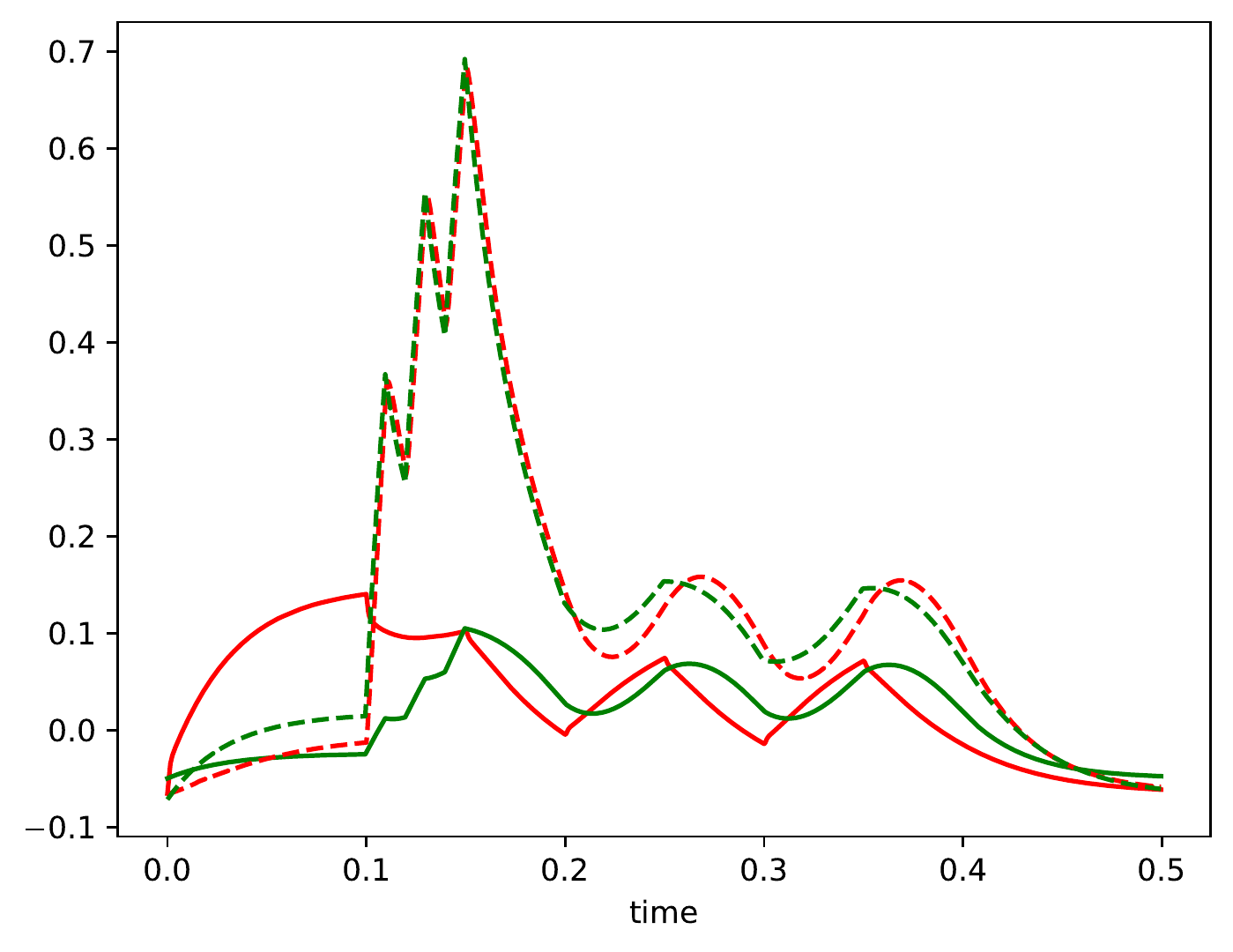} &
		\includegraphics[width=0.16\textwidth]{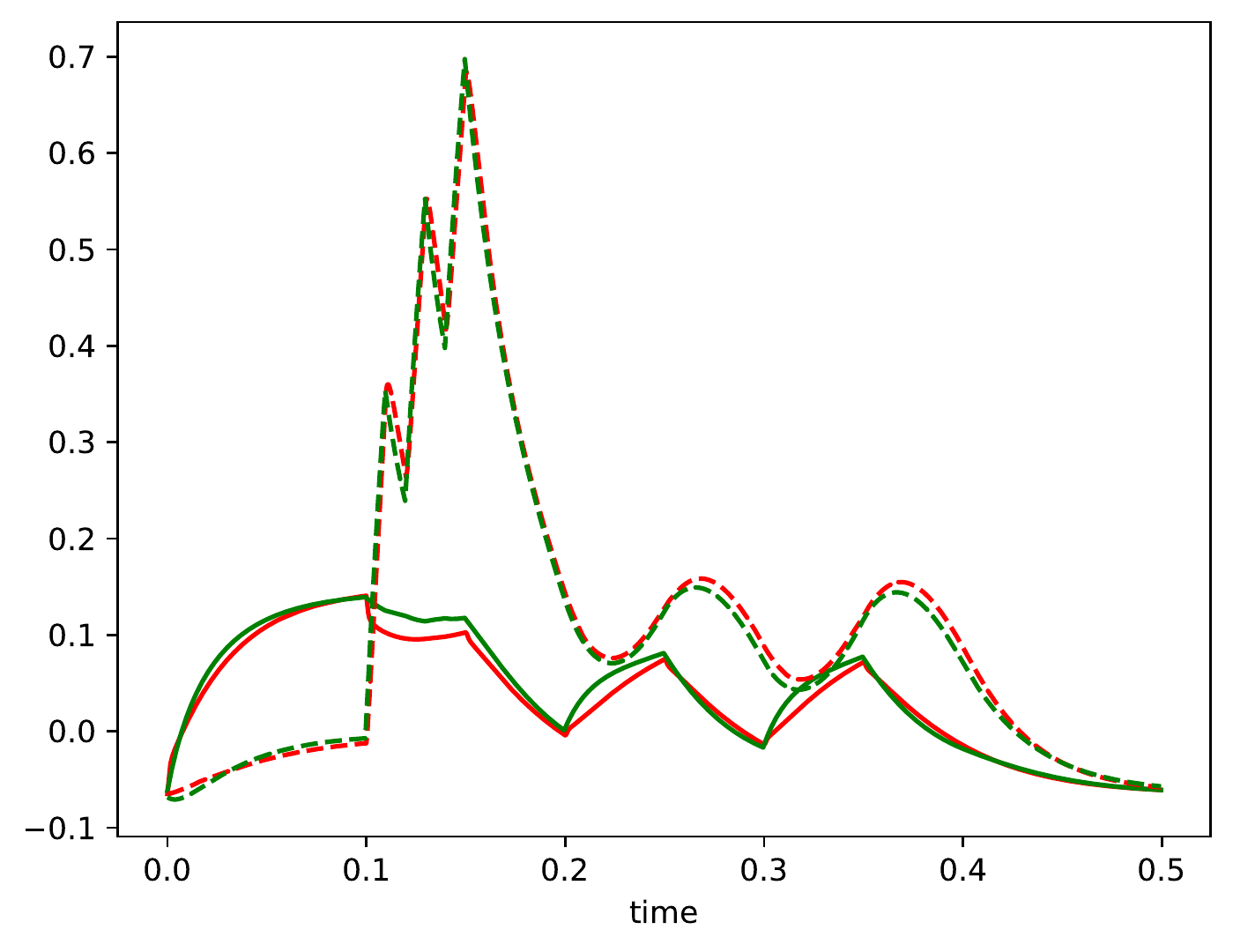} &  \includegraphics[width=0.16\textwidth]{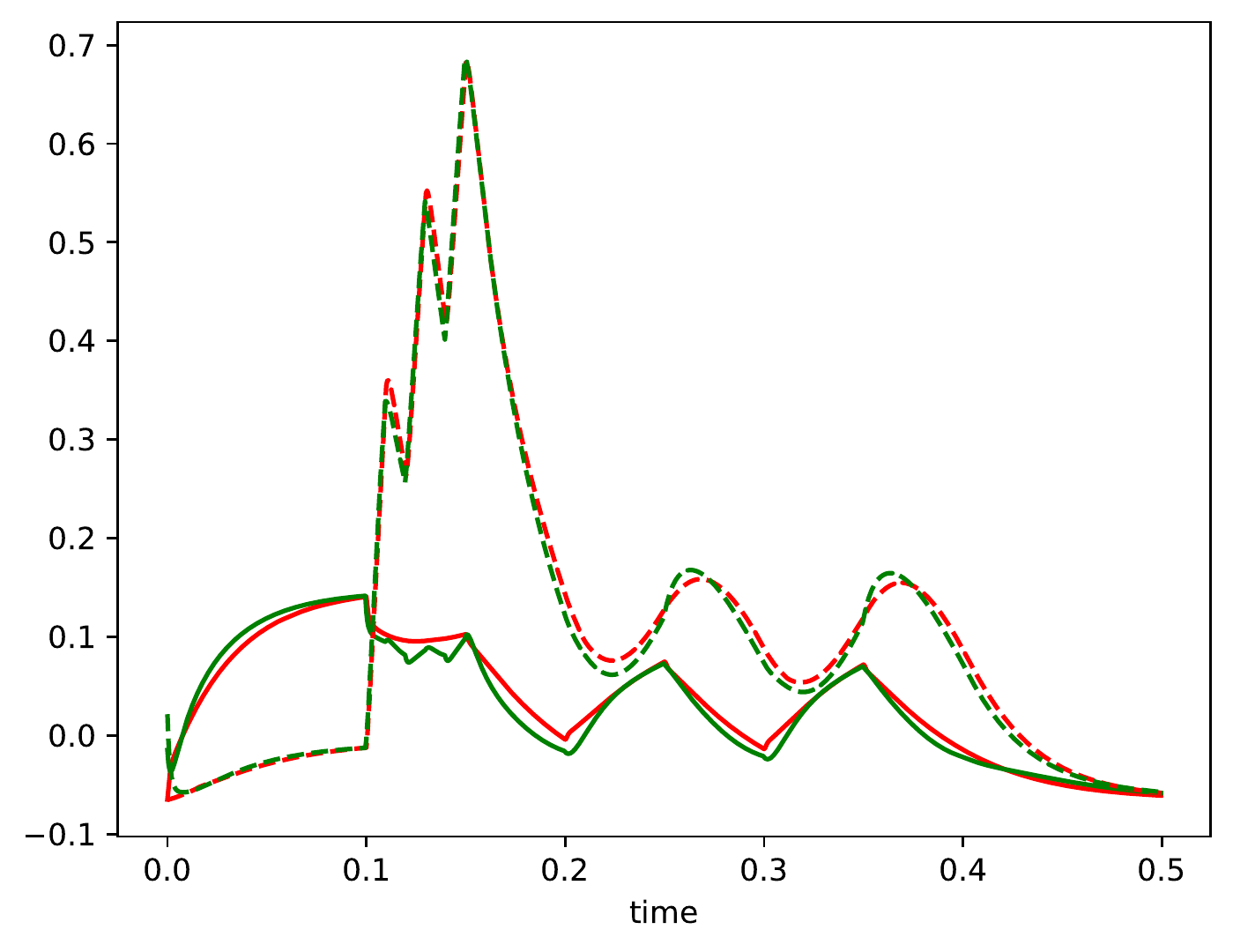} & 
		\includegraphics[width=0.16\textwidth]{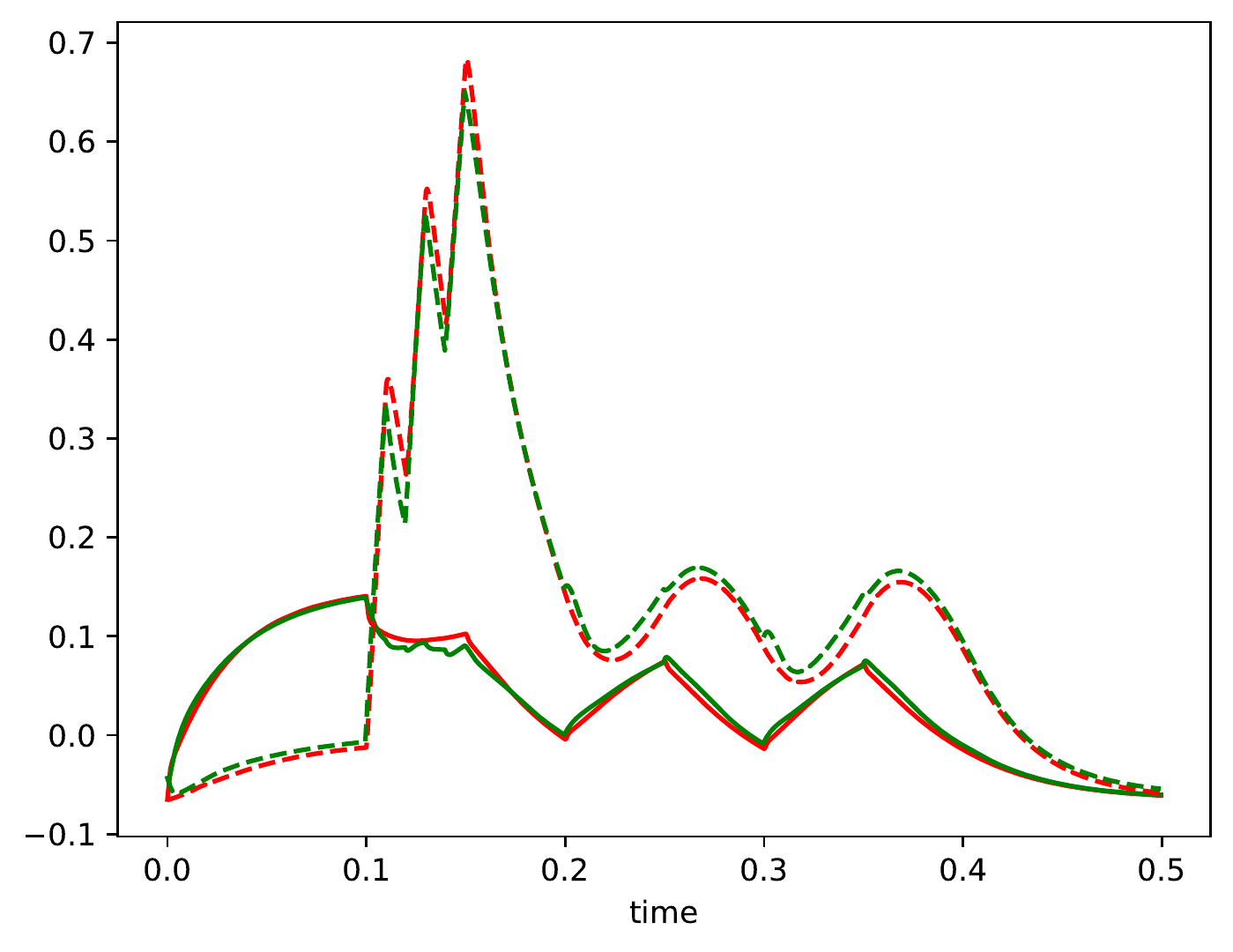} & 
		\includegraphics[width=0.16\textwidth]{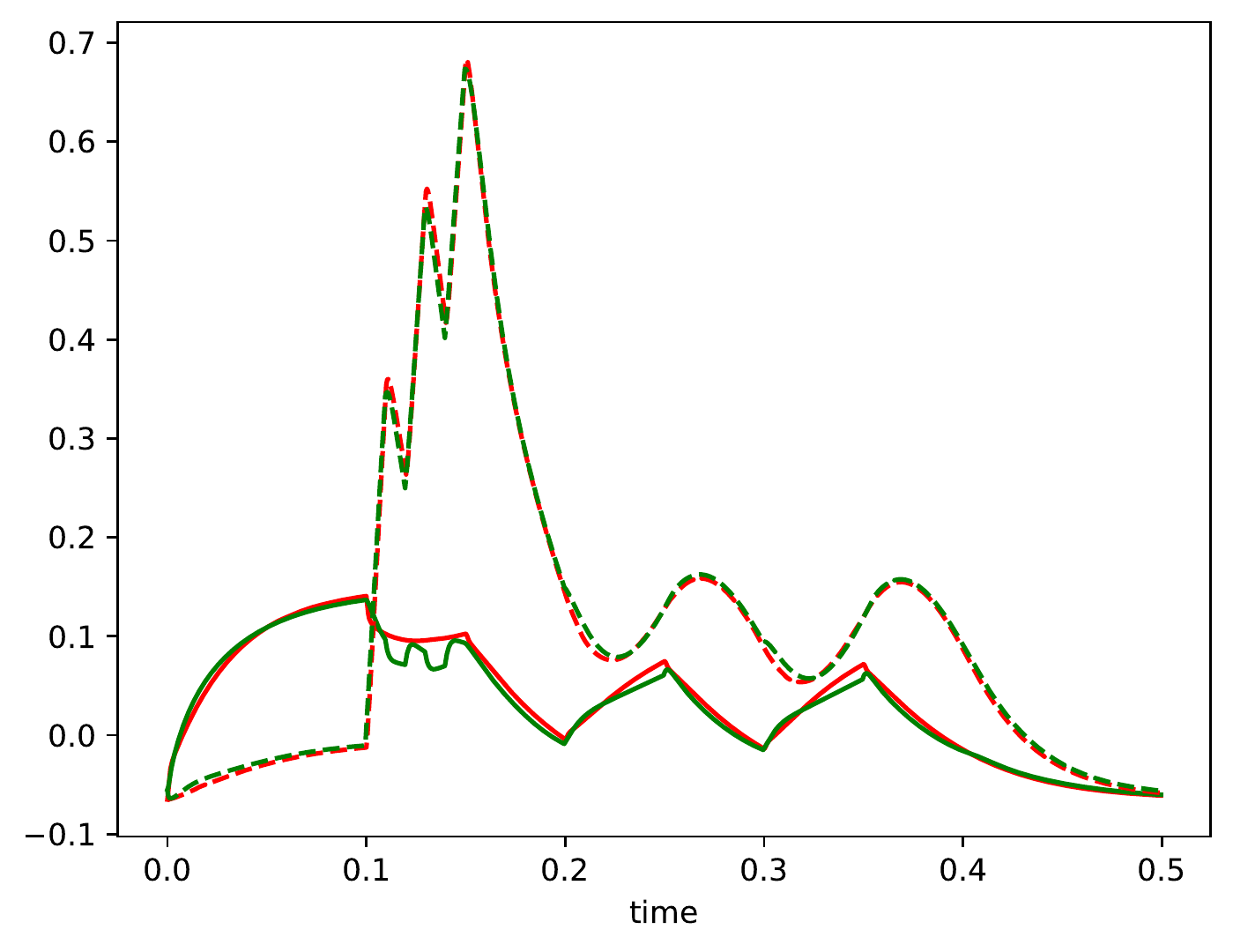} &  \includegraphics[width=0.16\textwidth]{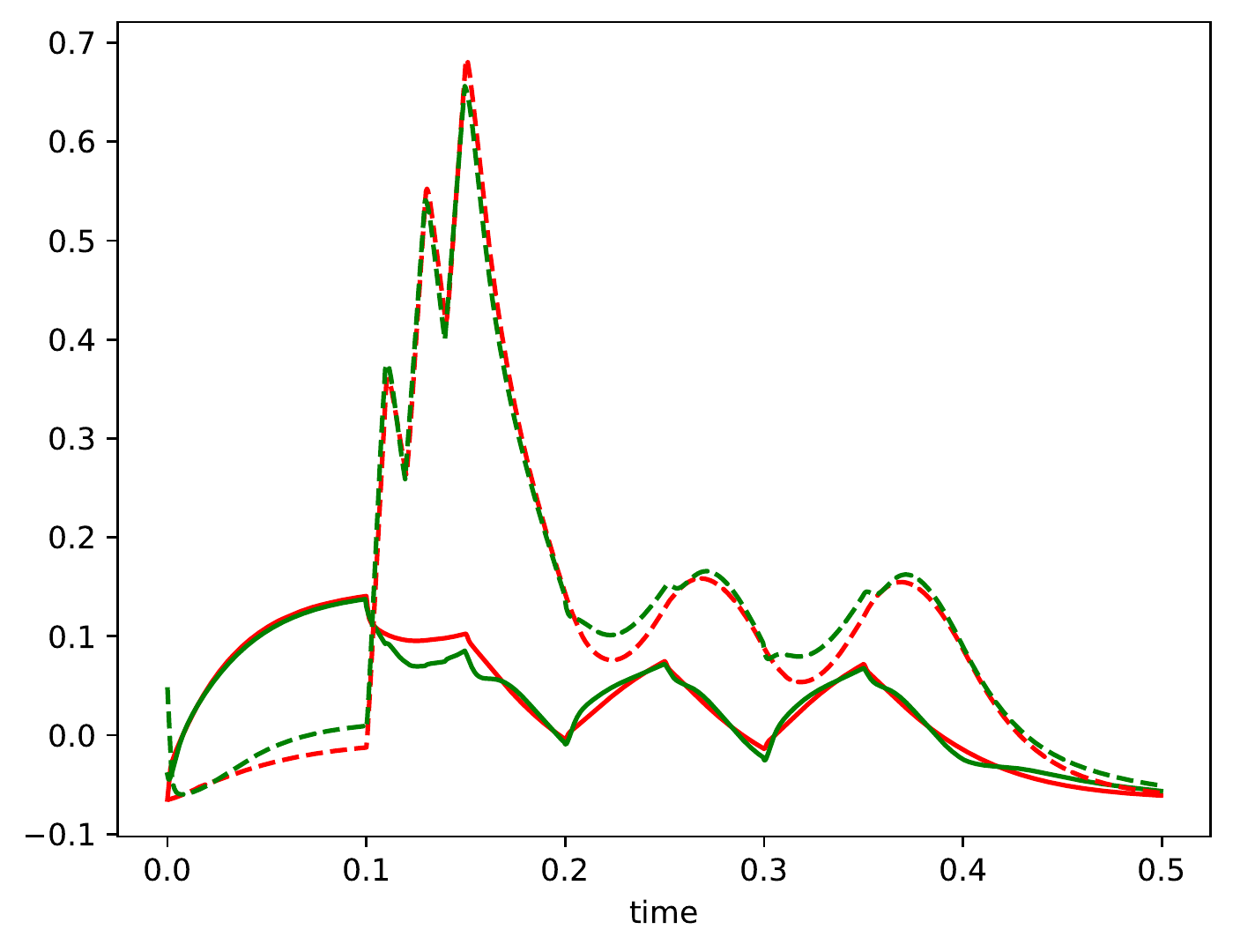}
	\\
	    \includegraphics[width=0.16\textwidth]{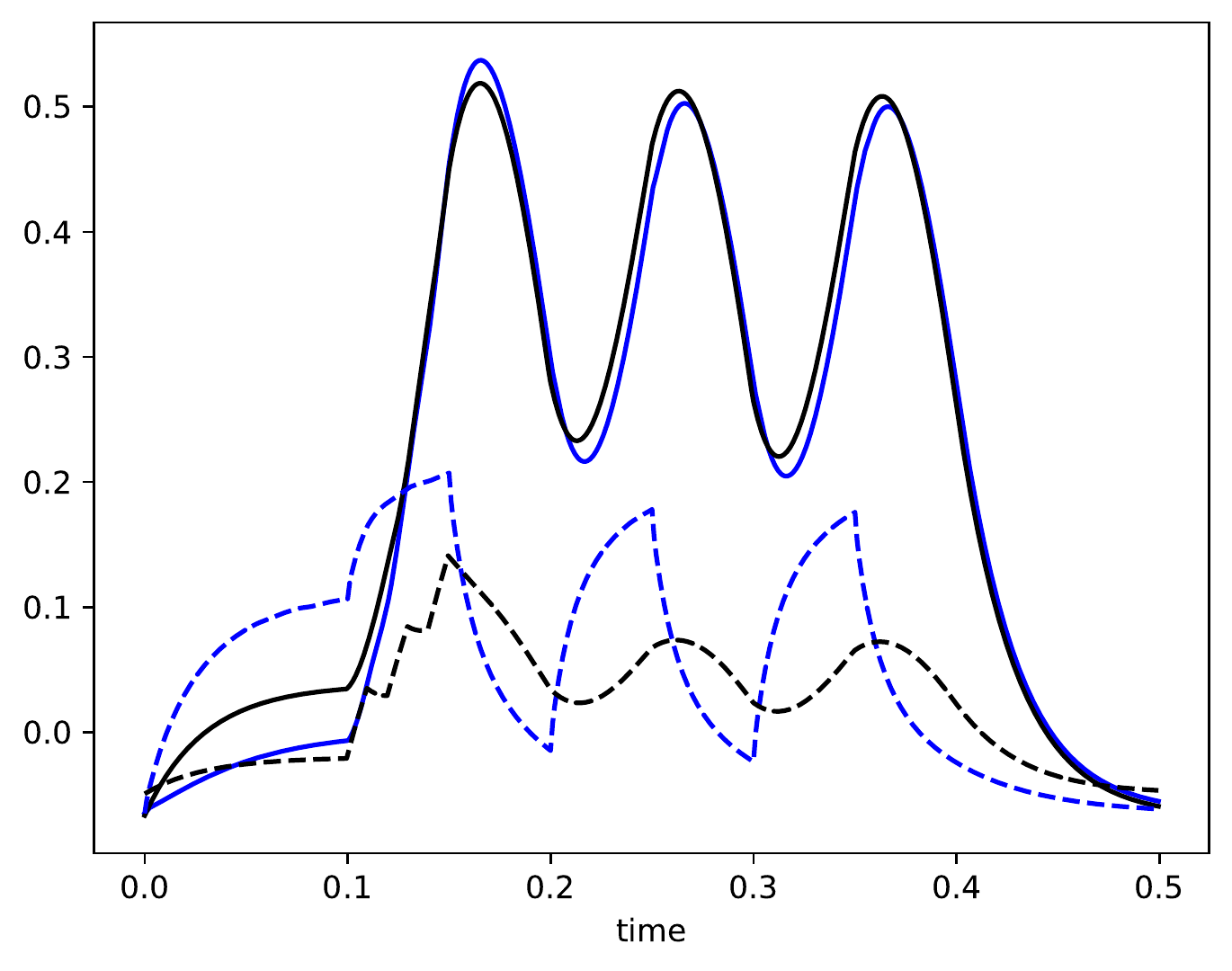} &
		\includegraphics[width=0.16\textwidth]{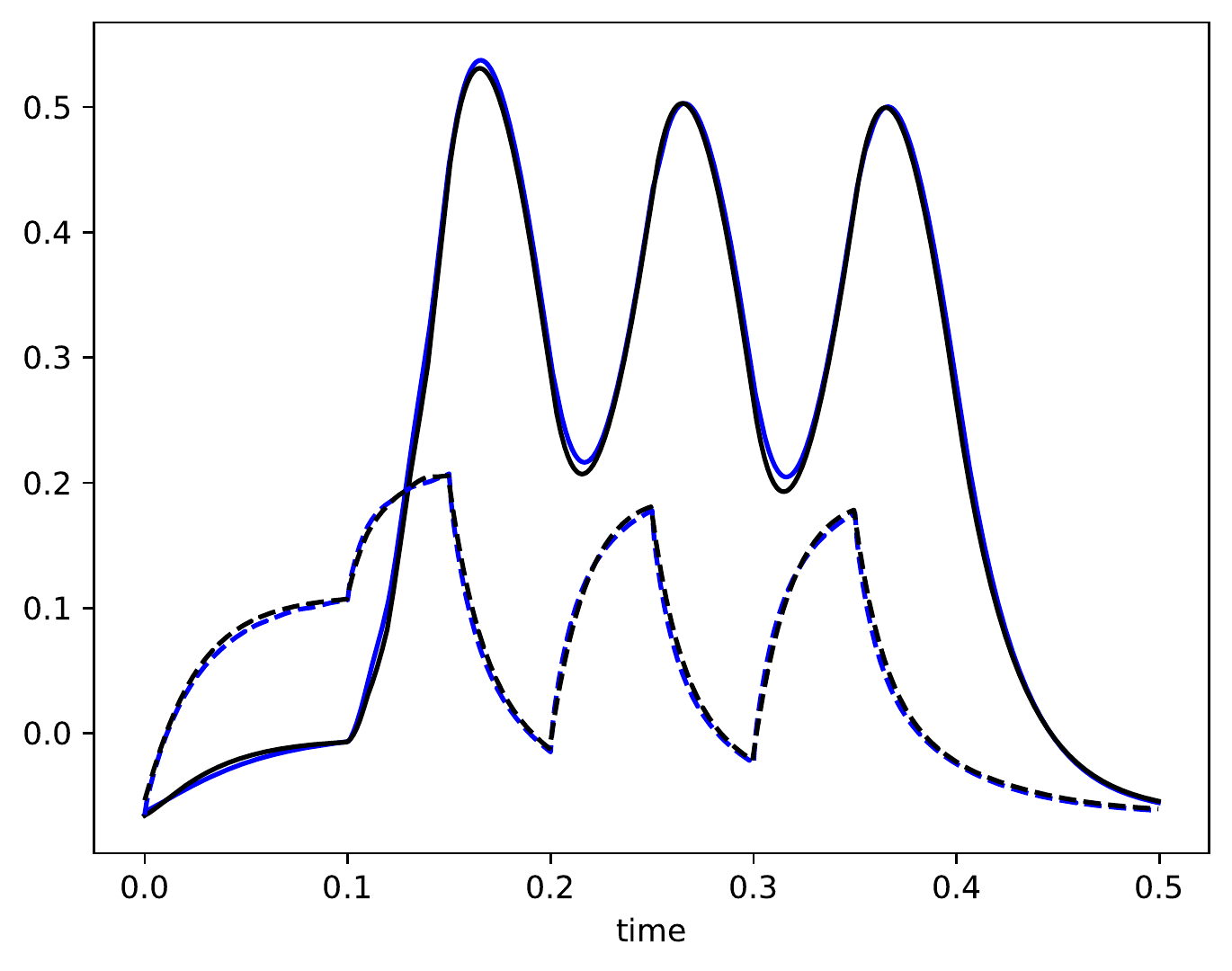} &  \includegraphics[width=0.16\textwidth]{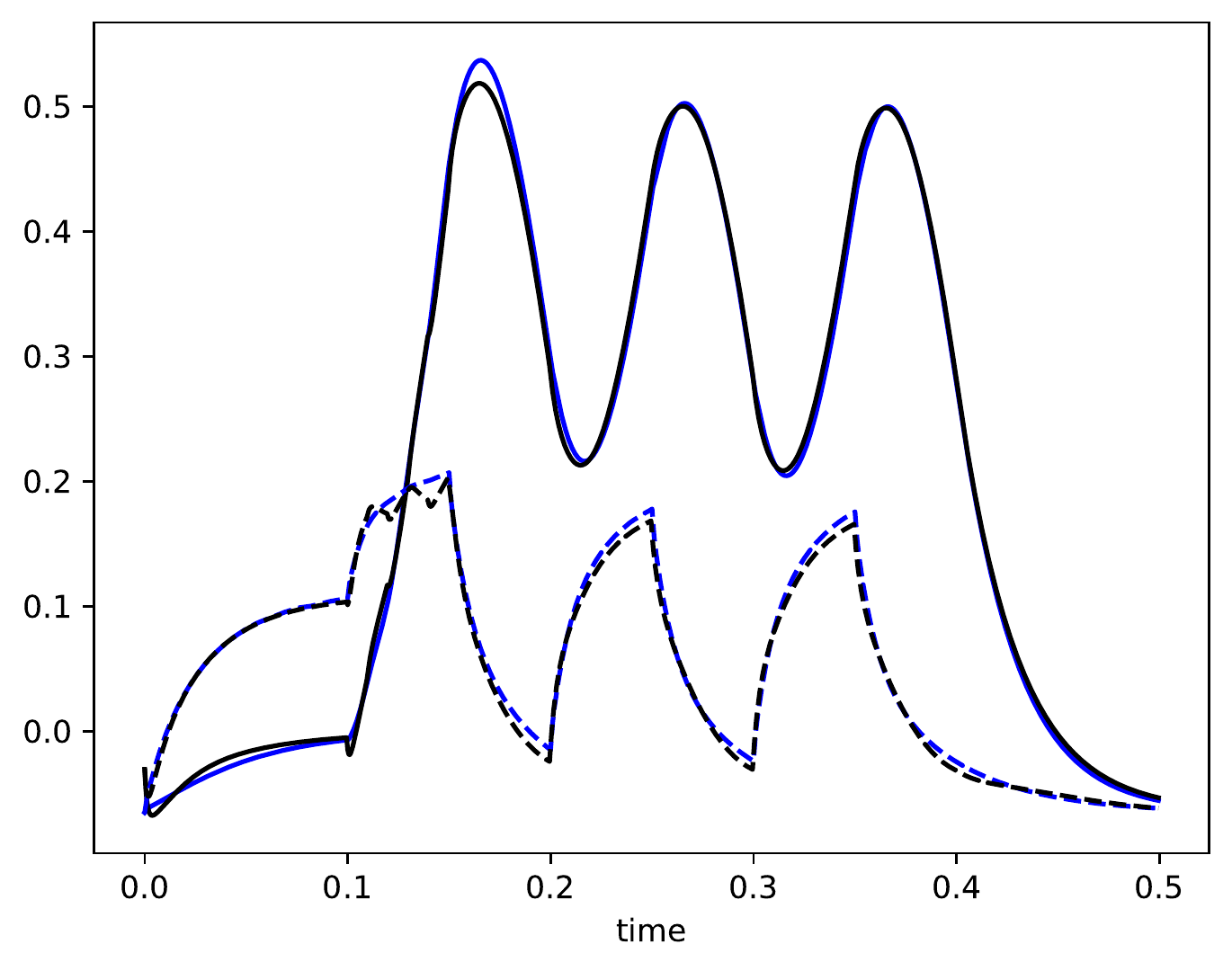} & 
		\includegraphics[width=0.16\textwidth]{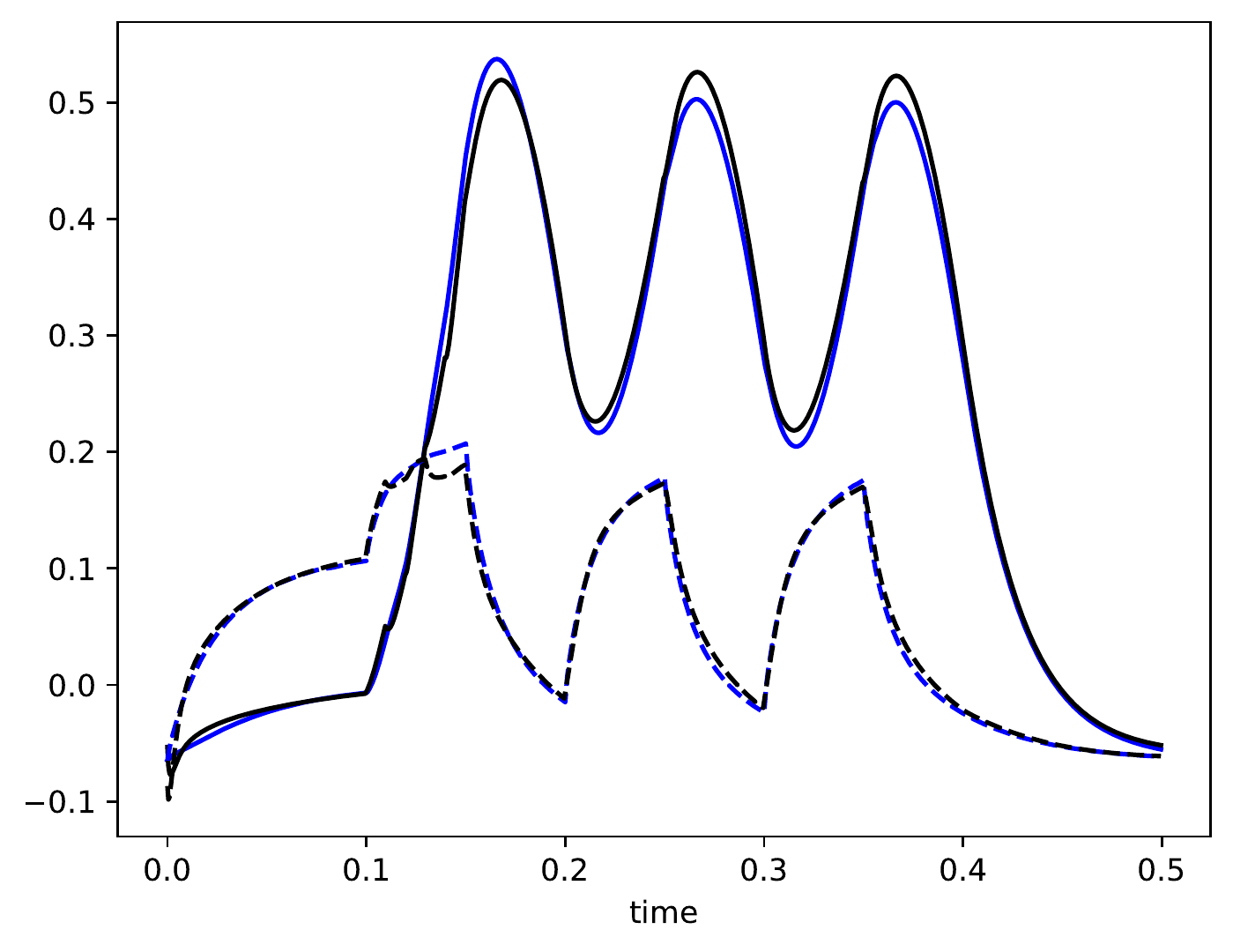} & 
		\includegraphics[width=0.16\textwidth]{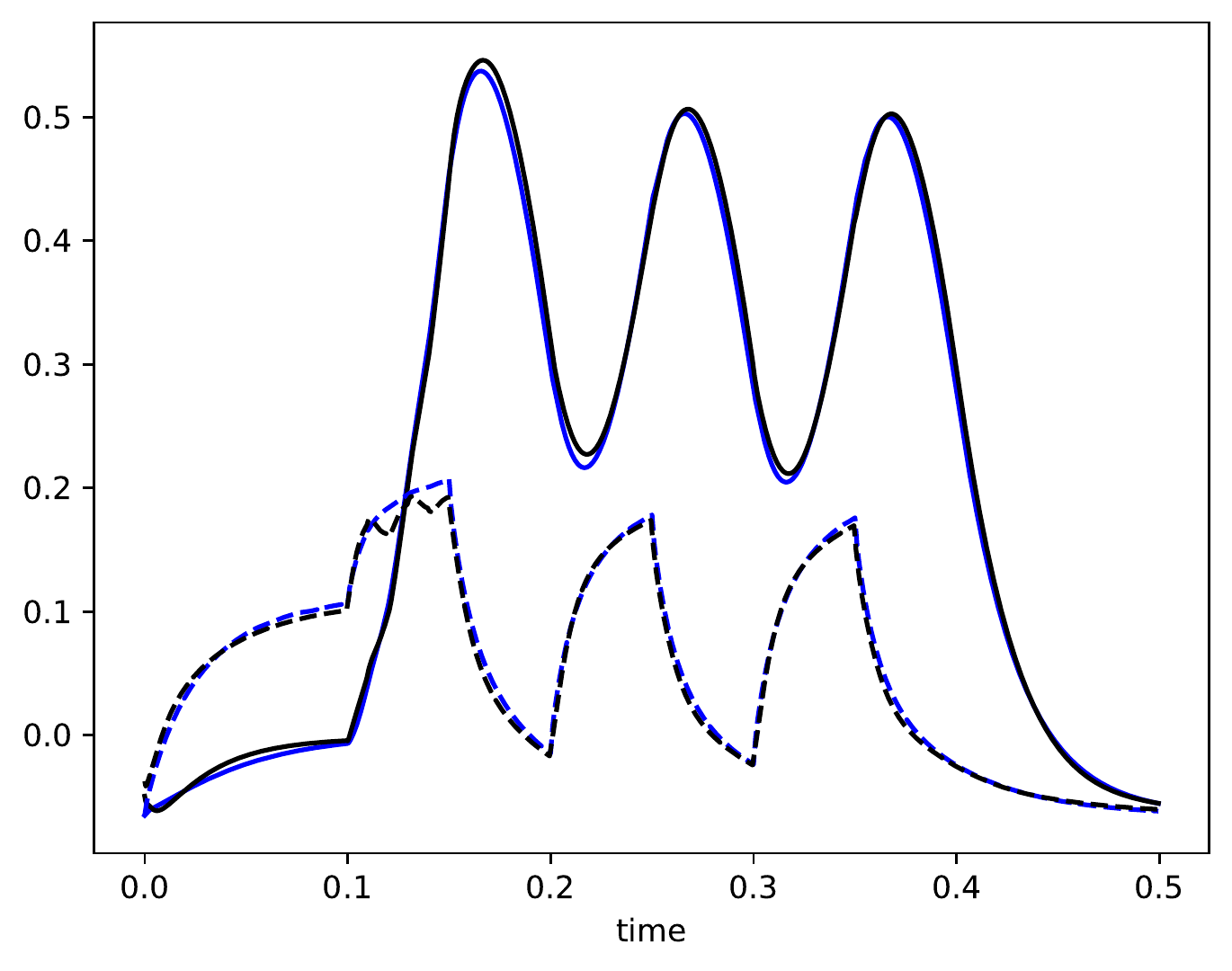} &  \includegraphics[width=0.16\textwidth]{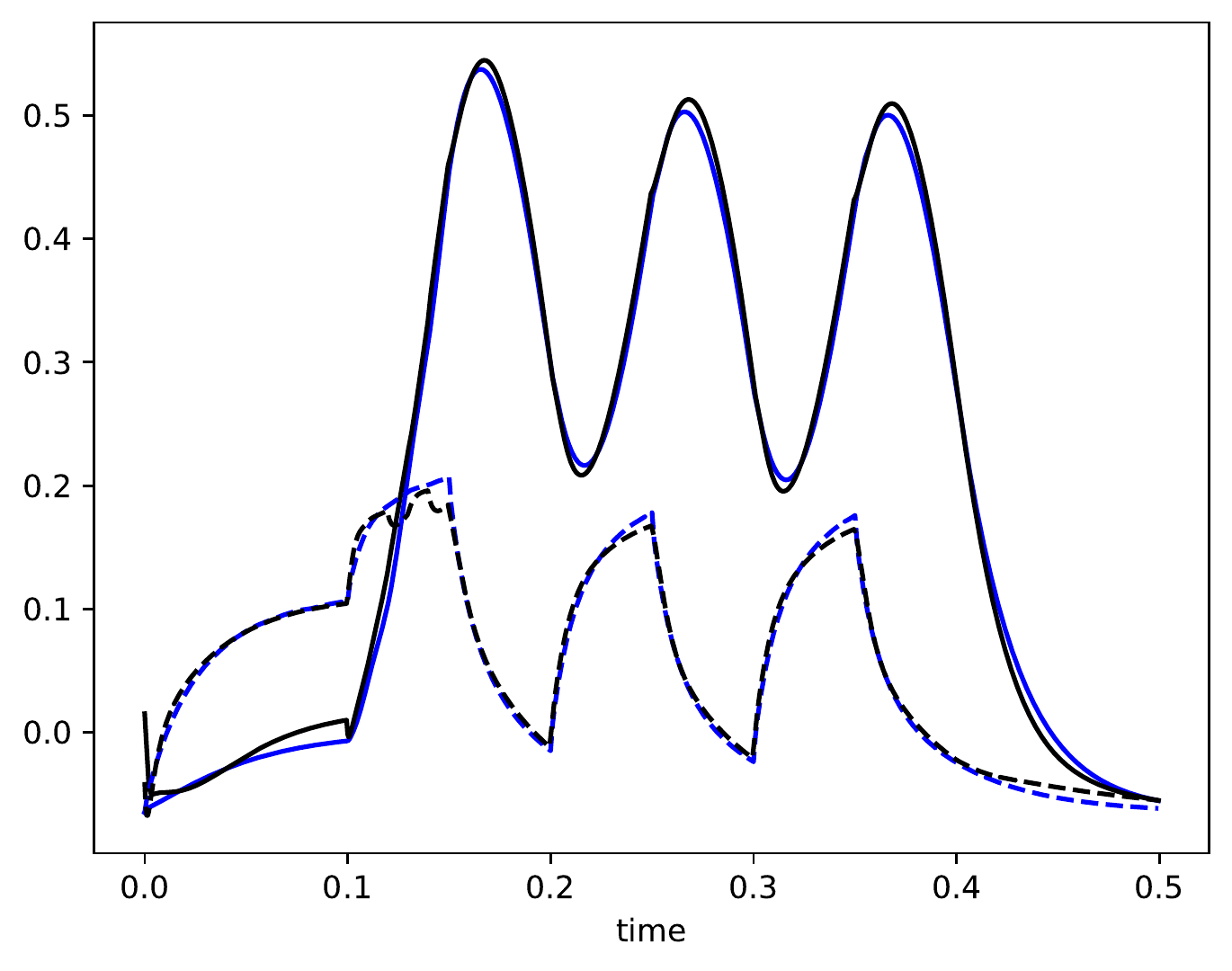}
	\\
  \end{tabular}
   \caption{Real (red \& blue) and predicted (green \& black) sequences on Exp. 2 for DB1 and LUAL ($1^{st}$ and $3^{rd}$ rows) and PVR and VB1 ($2^{nd}$ and $4^{th}$ rows) for two input sequences of the test set.}
  \label{fig:experiment2grid}
\end{figure}

Figure~\ref{fig:exp2loss} illustrates the evolution of the training and validation losses during the learning process, where one can see that for hidden sizes under 32 the model takes a little more time to converge, but still reaches a stable and low loss. 
However, from the data in Figure~\ref{fig:experiment2grid} and Table~\ref{tab:experiment2} we can see that the model with 2 units is not accurate enough but that a hidden size of 4 units can already reproduce the outputs with high accuracy.


\begin{figure}
    \centering
    \includegraphics[scale=0.375]{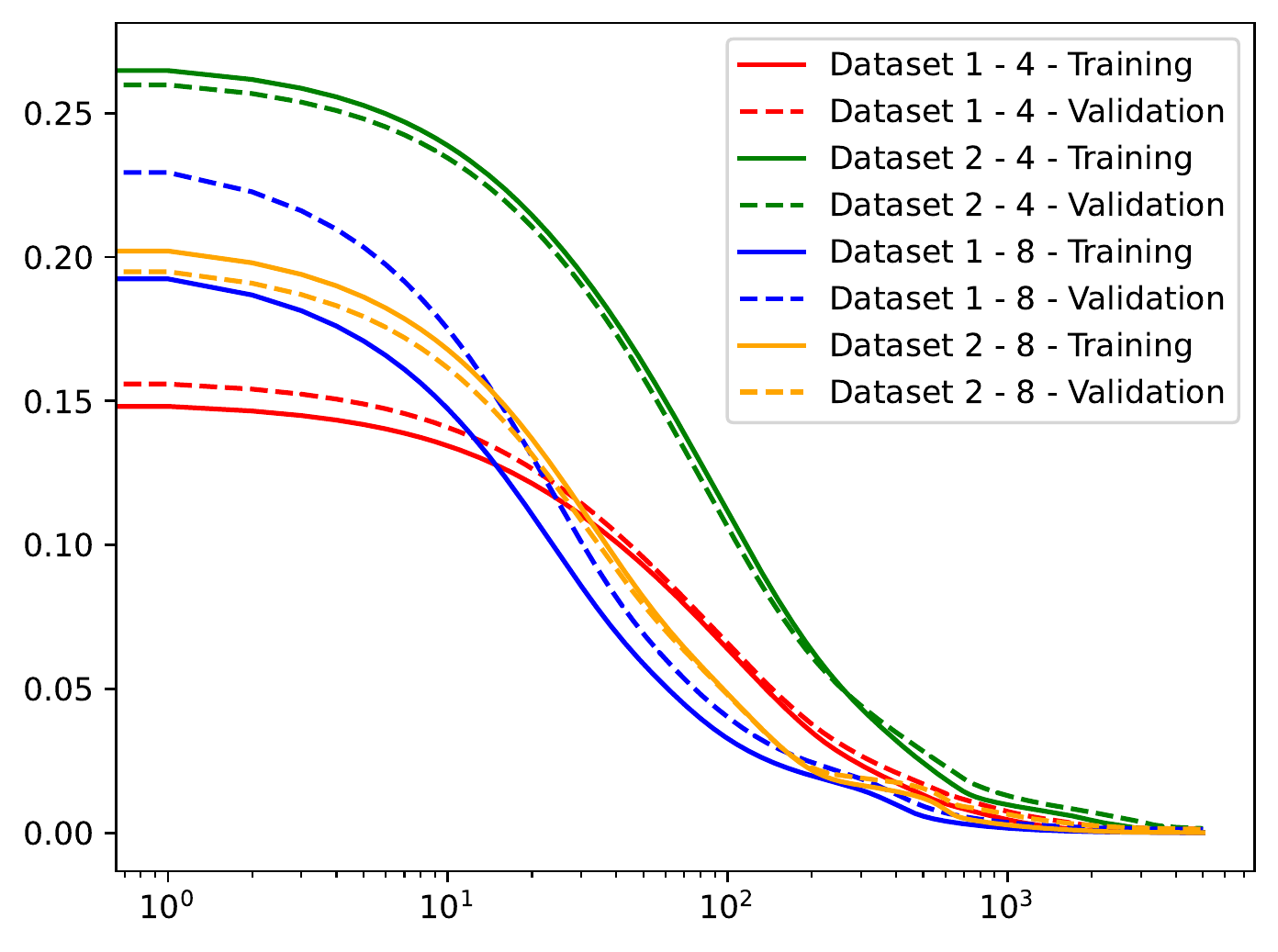}
    \caption{Training and validation loss for the two datasets with two different hidden sizes with a recurrent layer composed of GRUs.}
    \label{fig:exp3loss}
\end{figure}

\begin{table}
  \caption{The losses obtained for the GRU model, for different sizes of the recurrent layer, for the iteration with the smallest validation loss.}
  \label{tab:experiment2}
  \centering
  \begin{tabular}{lllllll}
    \toprule
& \textbf{2 Units} & \textbf{4 Units} & \textbf{8 Units} & \textbf{16 Units} & \textbf{32 Units} & \textbf{64 Units}   \\ \midrule
Training   & 1.8245e-03 & 1.1258e-04 & 7.2544e-05 & 6.4518e-05 & 3.6889e-05 & 9.6526e-05 \\ \midrule
Validation & 2.9376e-03 & 1.2652e-03 & 1.3360e-03 & 1.2396e-03 & 1.2415e-03 & 1.2074e-03 \\ \midrule
Test       & 3.6064e-03 & 1.3699e-04 & 1.4633e-04 & 1.3248e-04 & 1.1083e-04 & 2.0769e-04 \\
    \bottomrule
  \end{tabular}
\end{table}
%



\subsection{Experiment 3} \label{sec:experiment3}

In this experiment we explore the models' behaviour for data sampled with different time steps as this leads to longer sequences. 
From a methodology standpoint this is important since, even though time-wise the dynamics do not change, the temporal dependencies that the model has to learn are farther back in the sequence, which increases the difficulty of the learning process.




We run the model for the two datasets, one with the coarser (\(0.5~\si{\milli\second}\)) time step and one with a finer (\(0.1~\si{\milli\second}\)) time step. The experiment is done only for the GRU and for each dataset we test two different sizes of the recurrent layer, \(4\) and \(8\) units.


\begin{figure}
 \setlength\tabcolsep{2pt}
  \centering
  \begin{tabular}{cccccc}
    \textbf{Dataset1 - 4} & \textbf{Dataset2 - 4} & \textbf{Dataset1 - 8} & \textbf{Dataset2 - 8}  \\
		\includegraphics[width=0.16\textwidth]{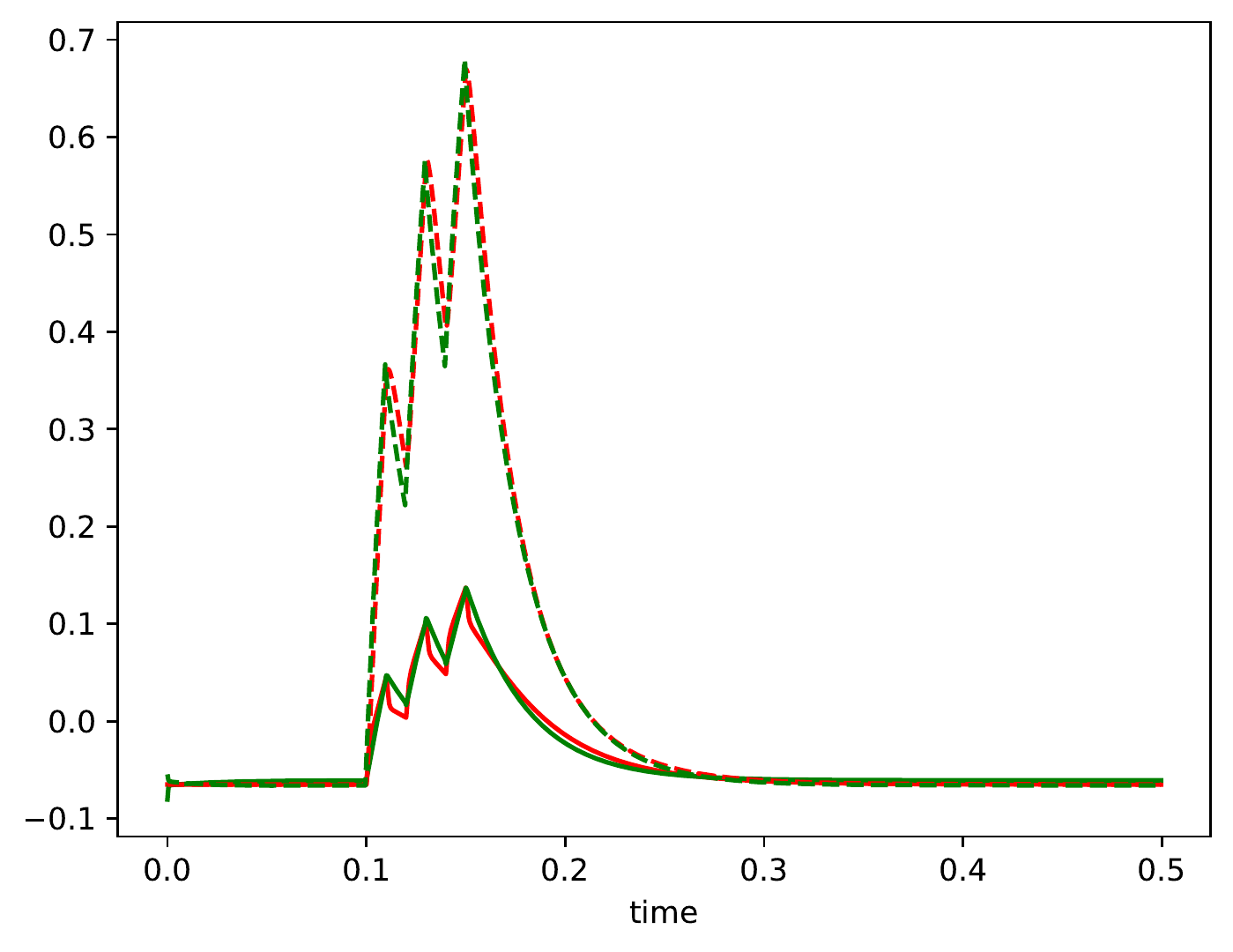} & 
		\includegraphics[width=0.16\textwidth]{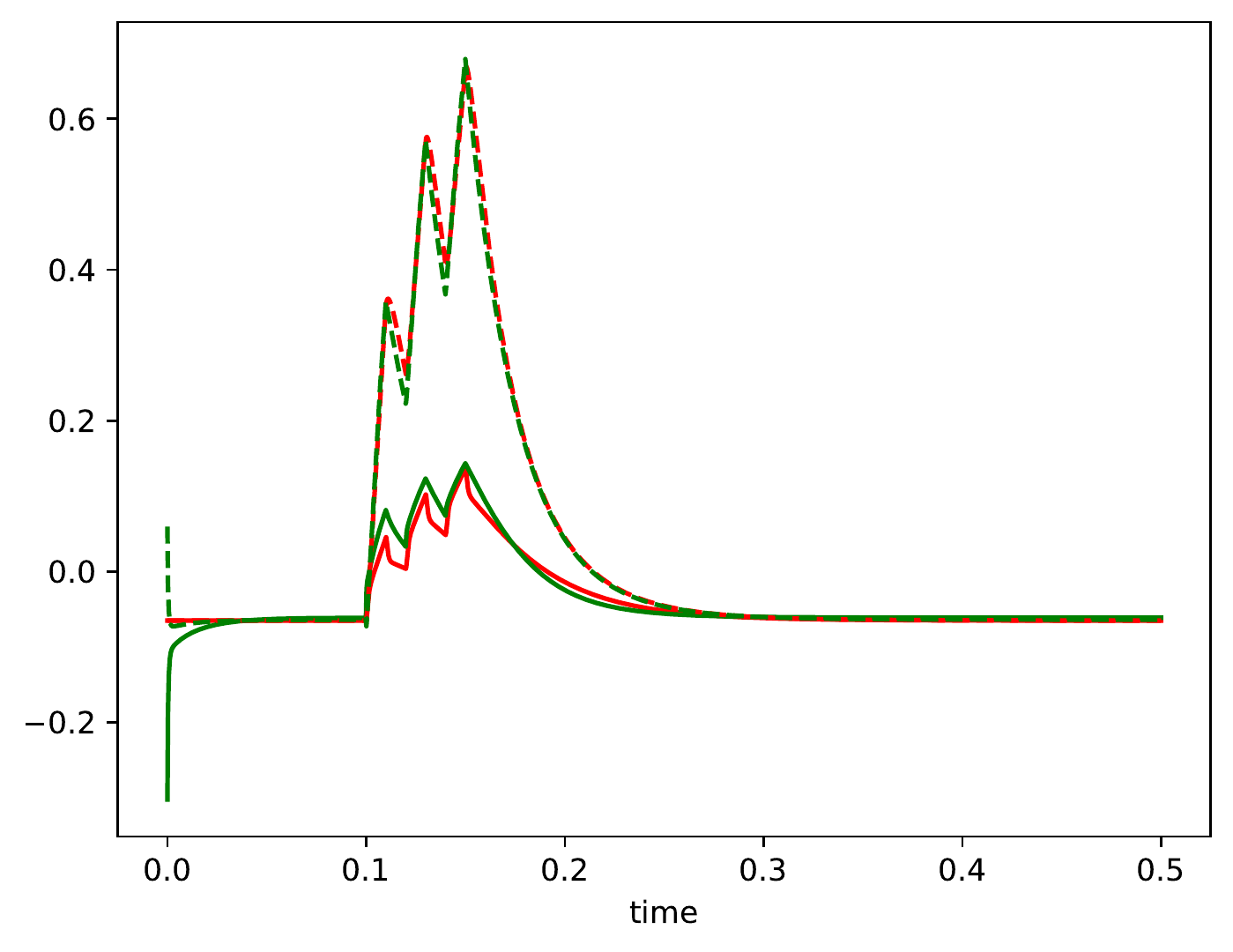} & 
		\includegraphics[width=0.16\textwidth]{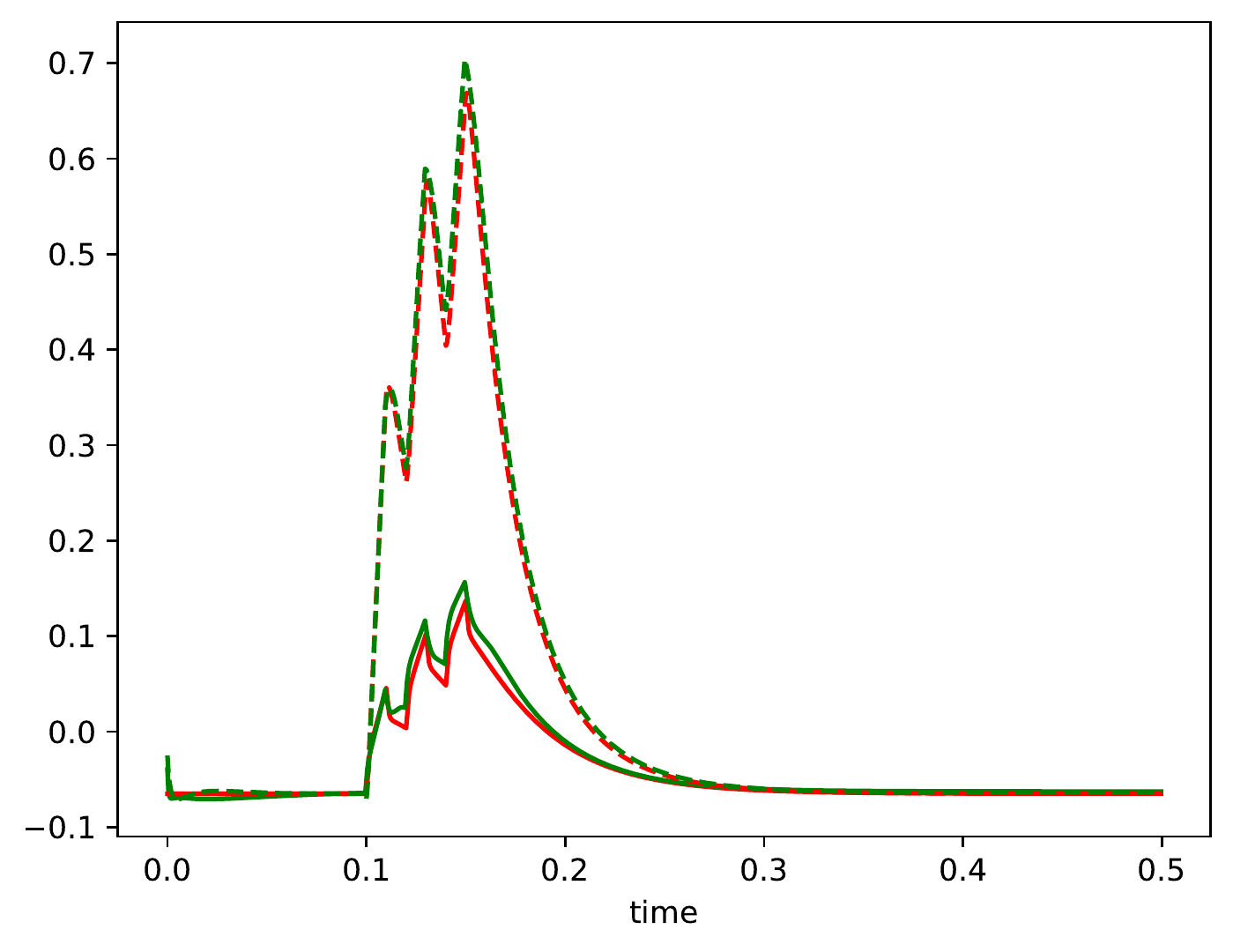} & 
		\includegraphics[width=0.16\textwidth]{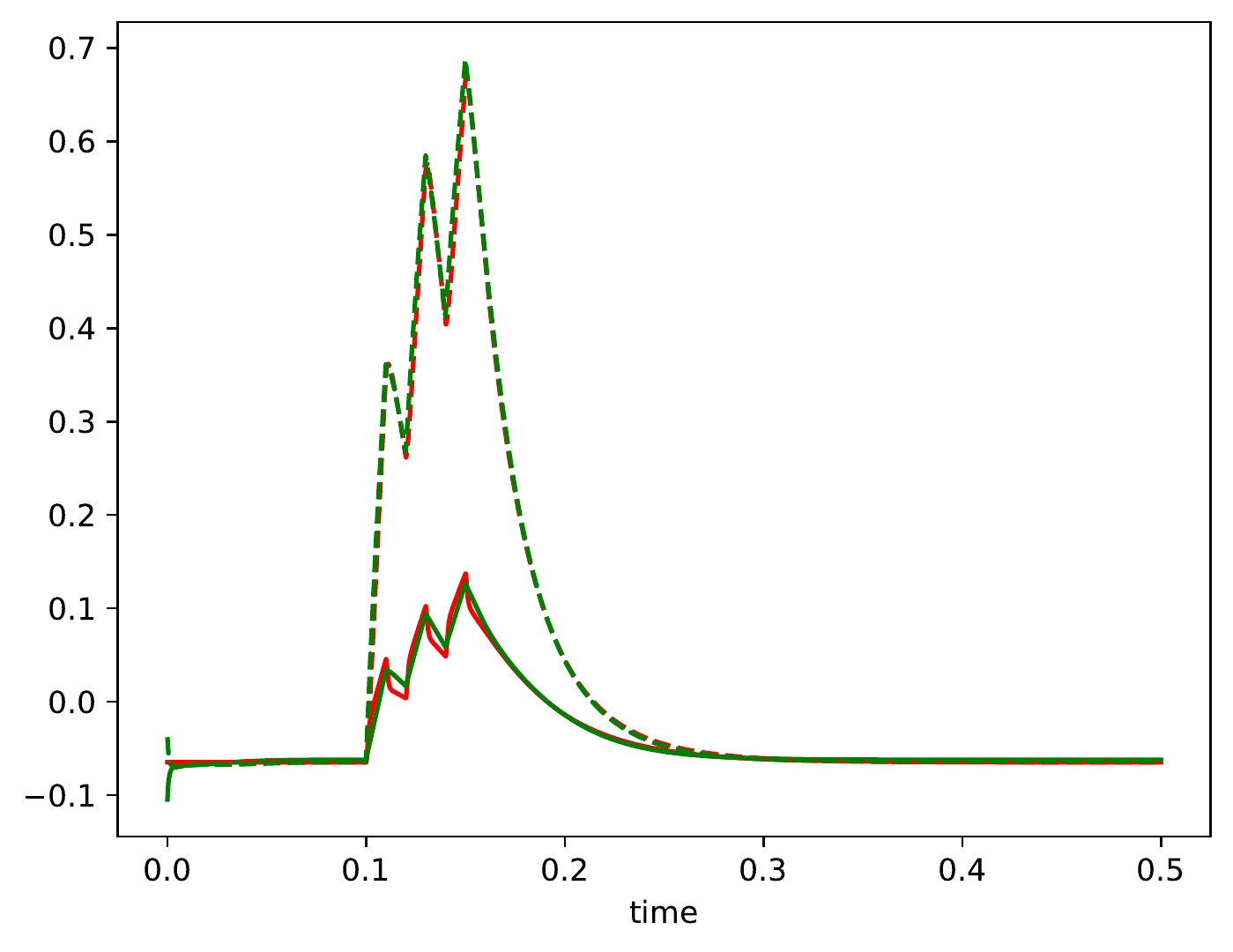} & 
	\\
		\includegraphics[width=0.16\textwidth]{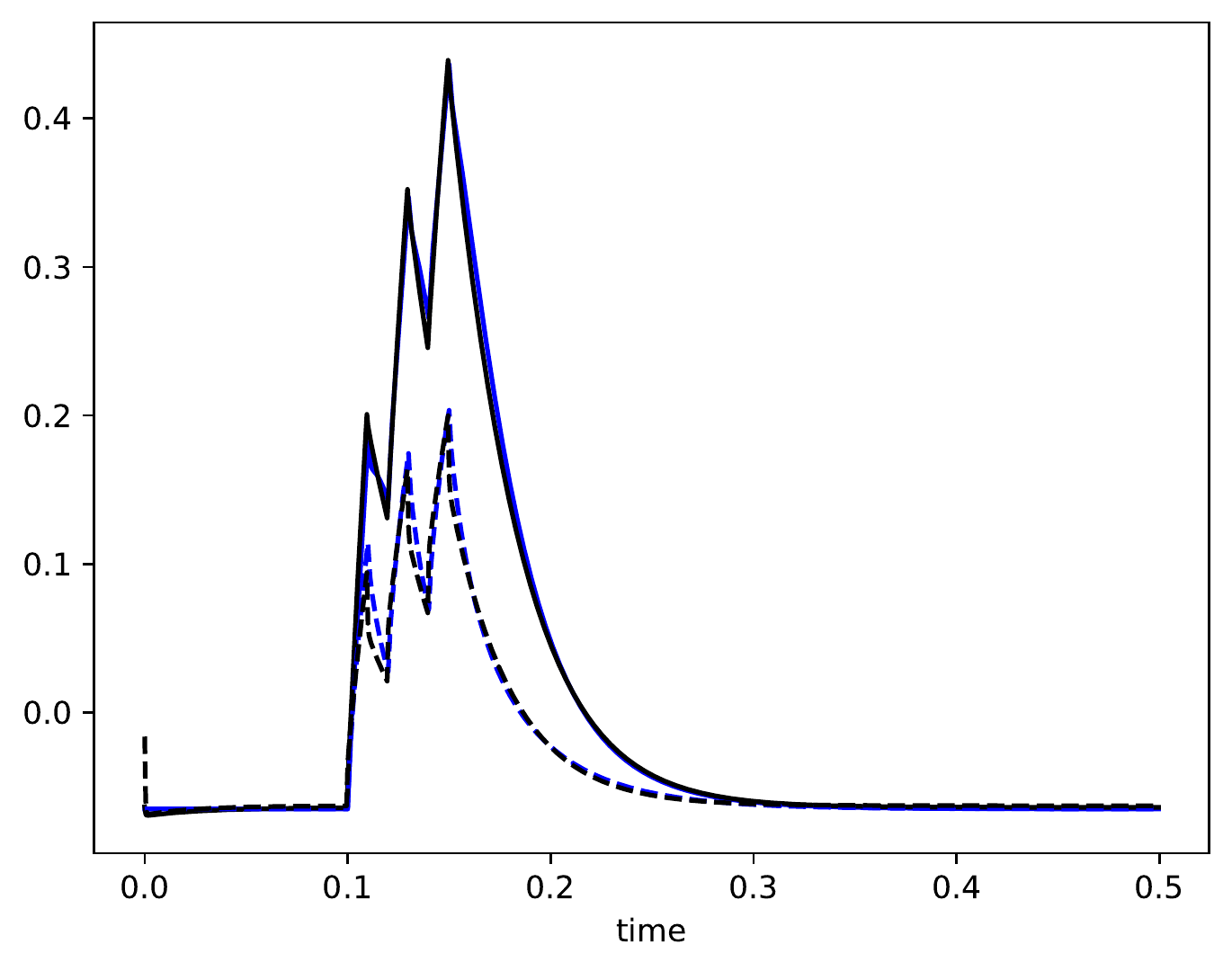} & 
		\includegraphics[width=0.16\textwidth]{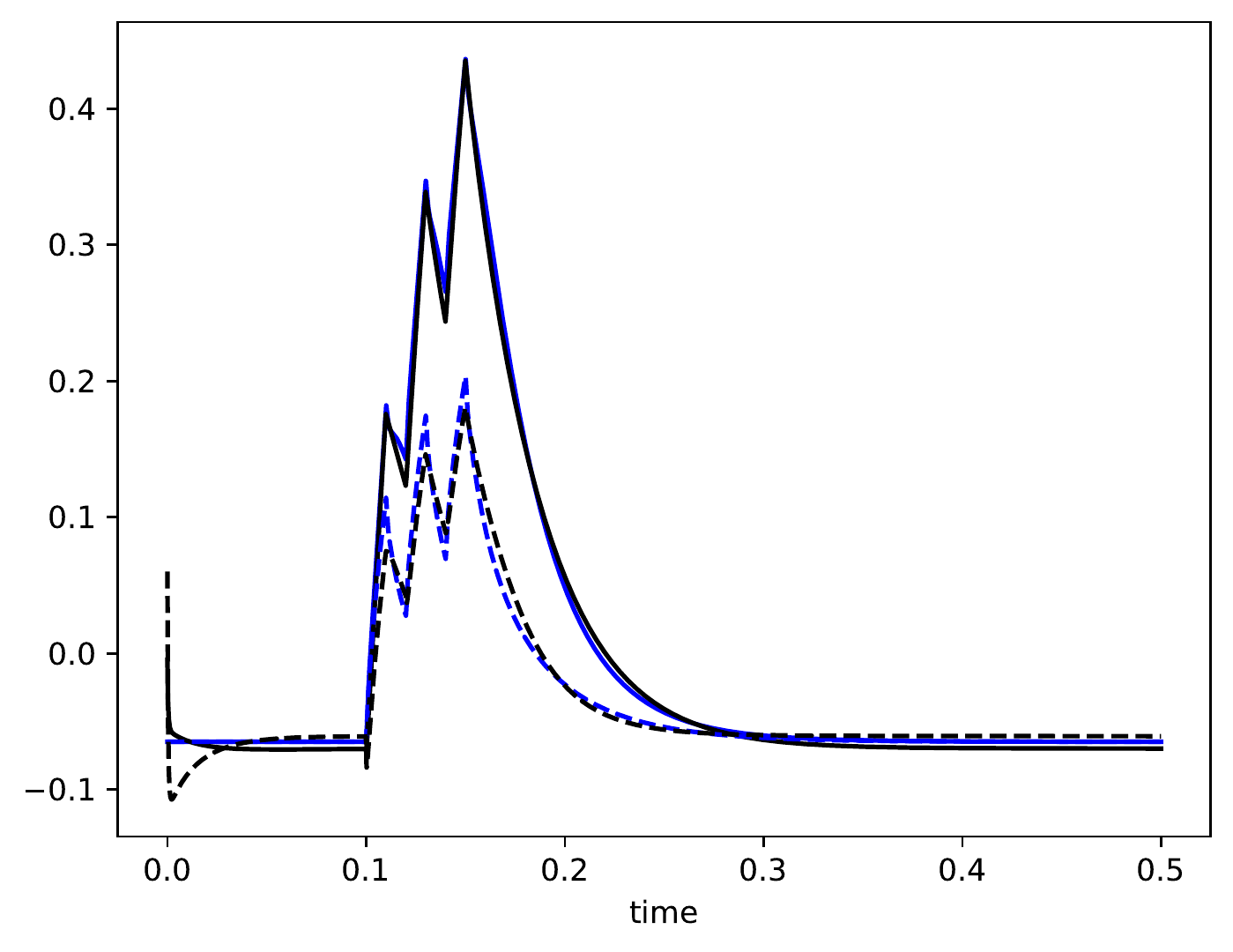} & 
		\includegraphics[width=0.16\textwidth]{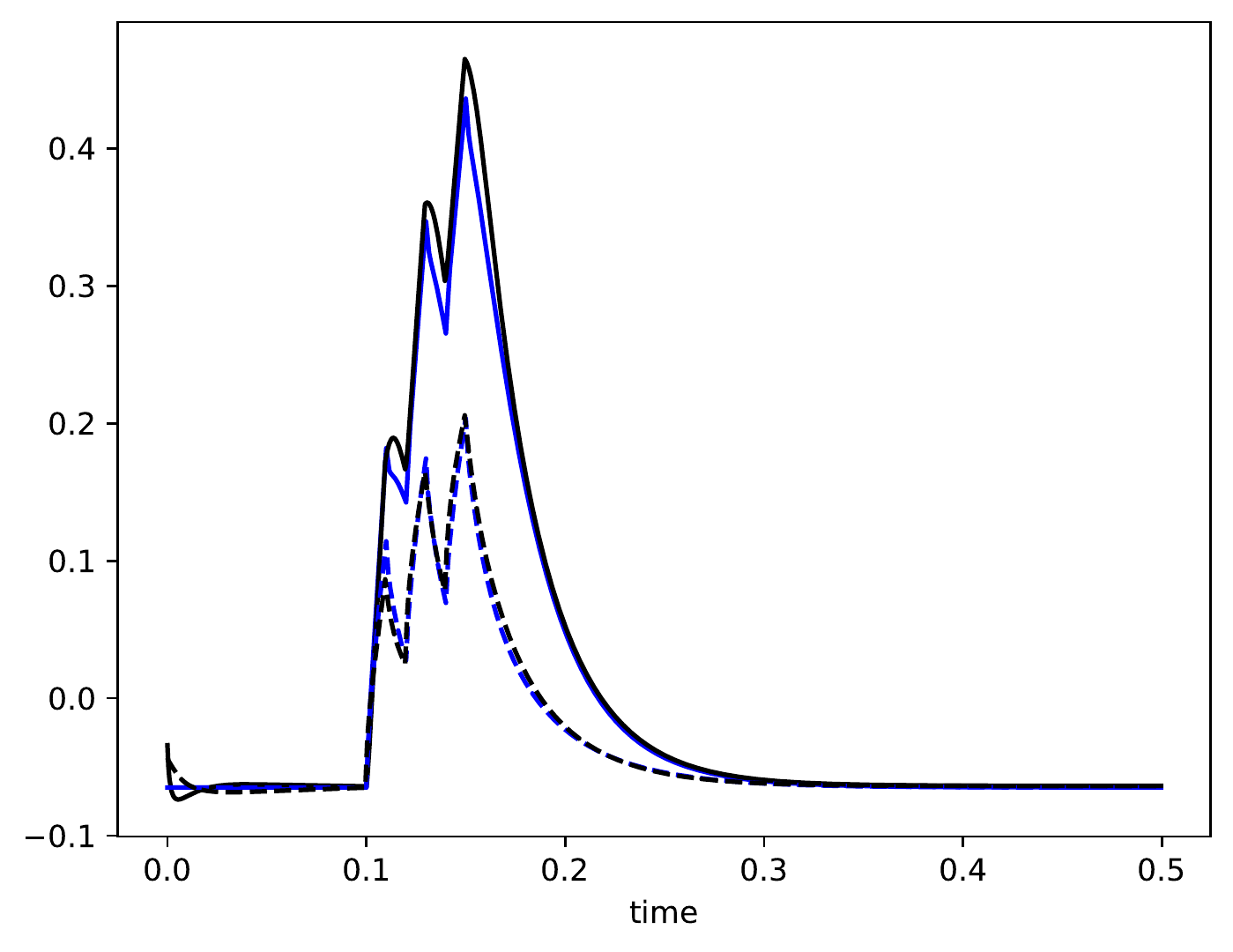} & 
		\includegraphics[width=0.16\textwidth]{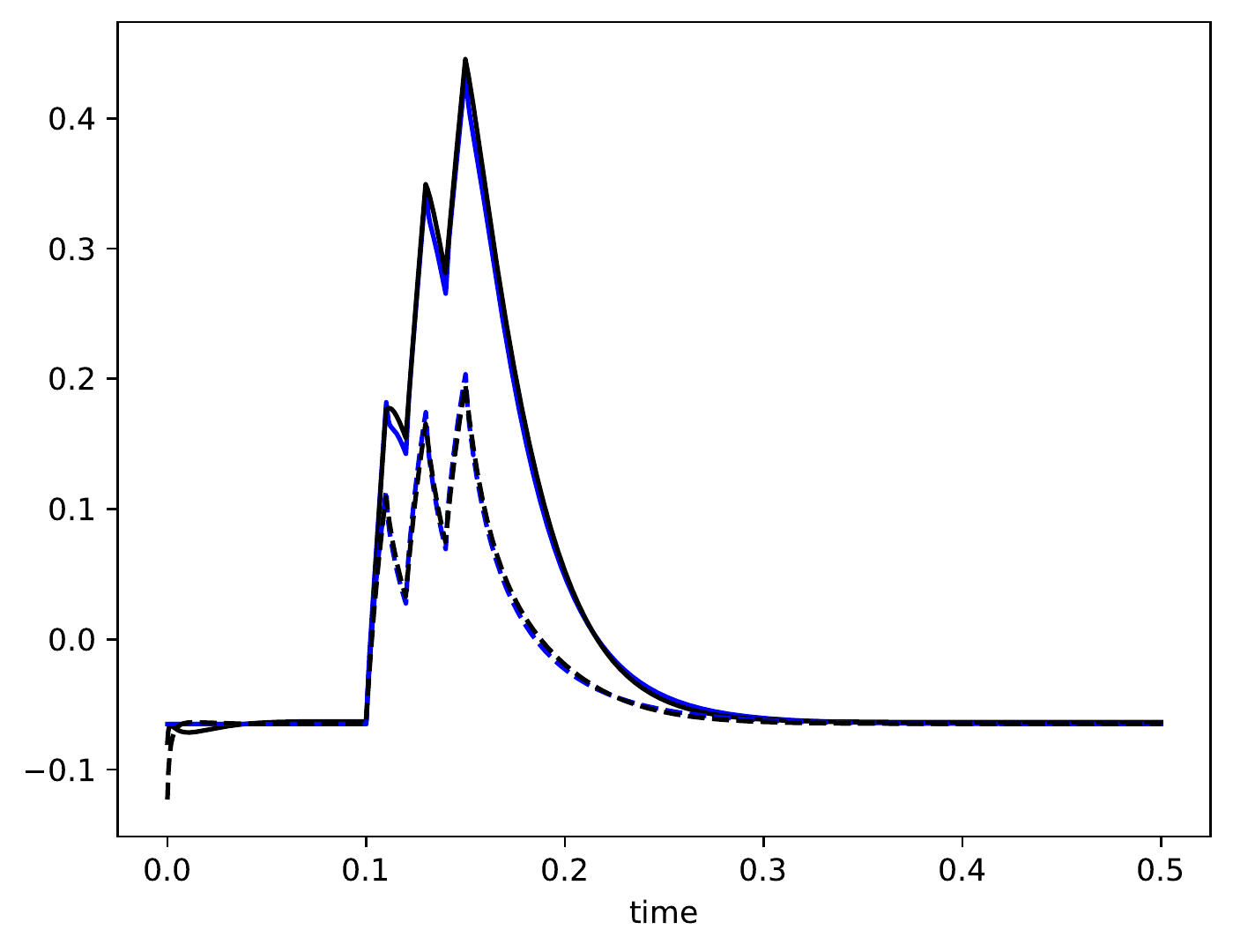} & 
	\\
		\includegraphics[width=0.16\textwidth]{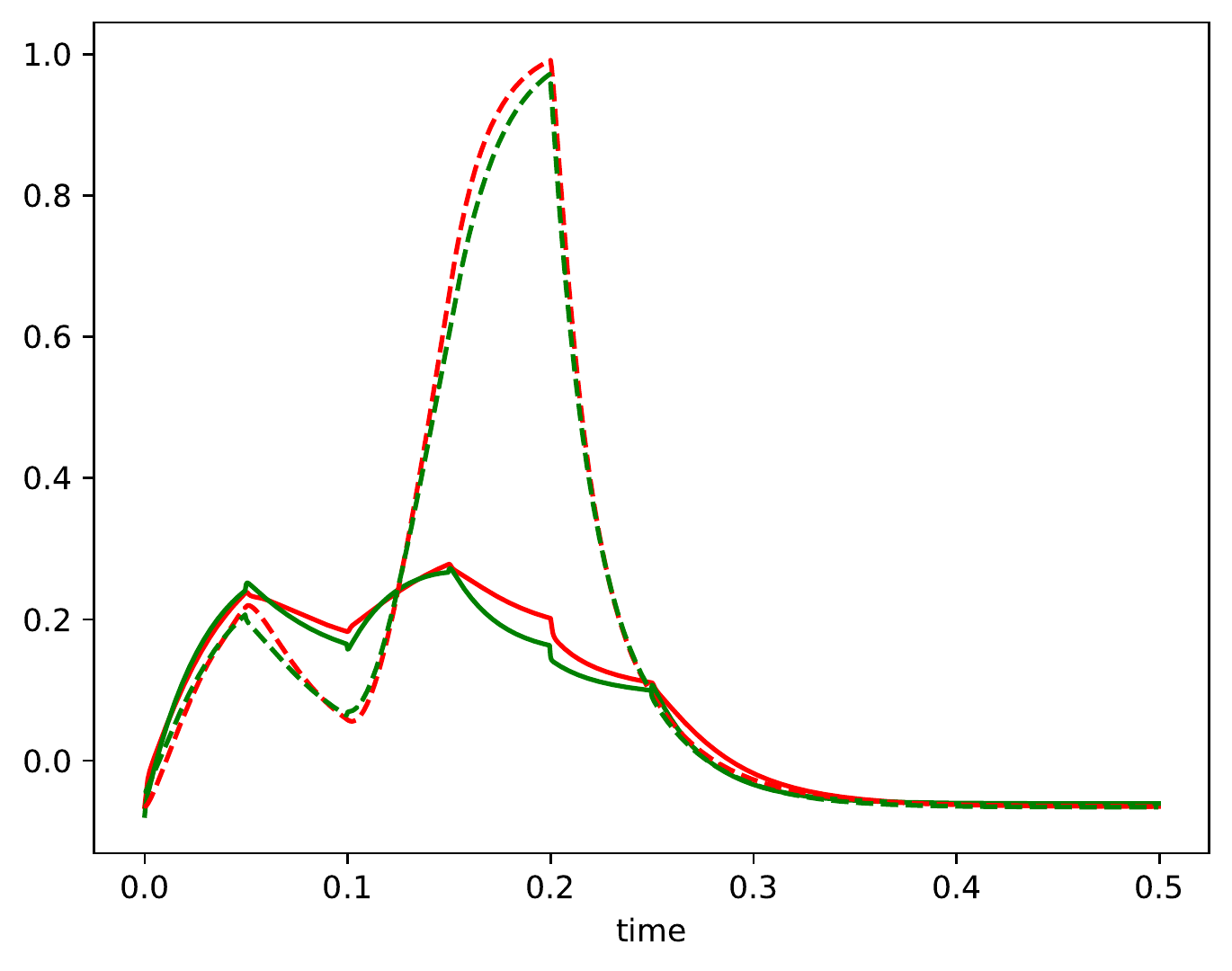} & 
		\includegraphics[width=0.16\textwidth]{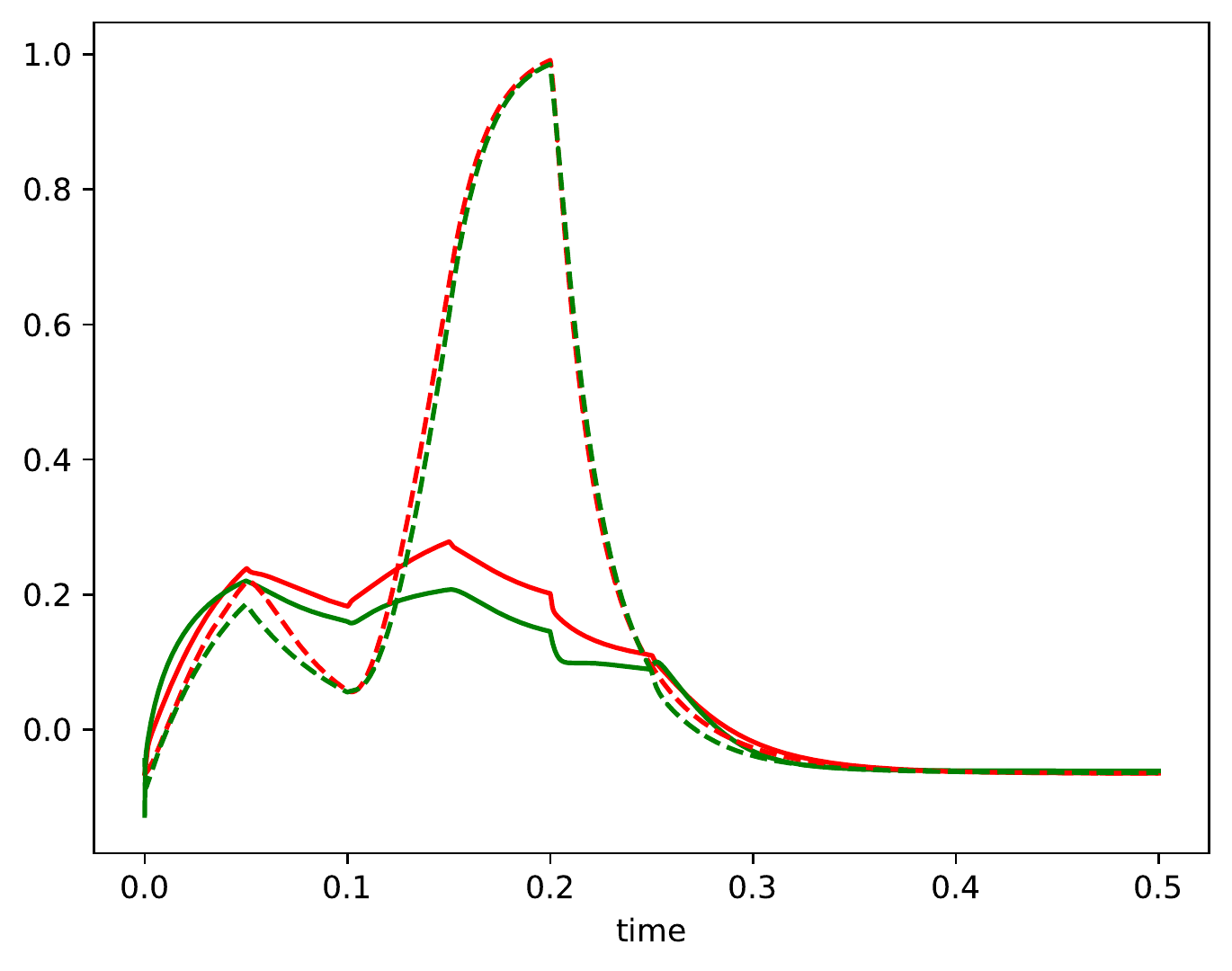} & 
		\includegraphics[width=0.16\textwidth]{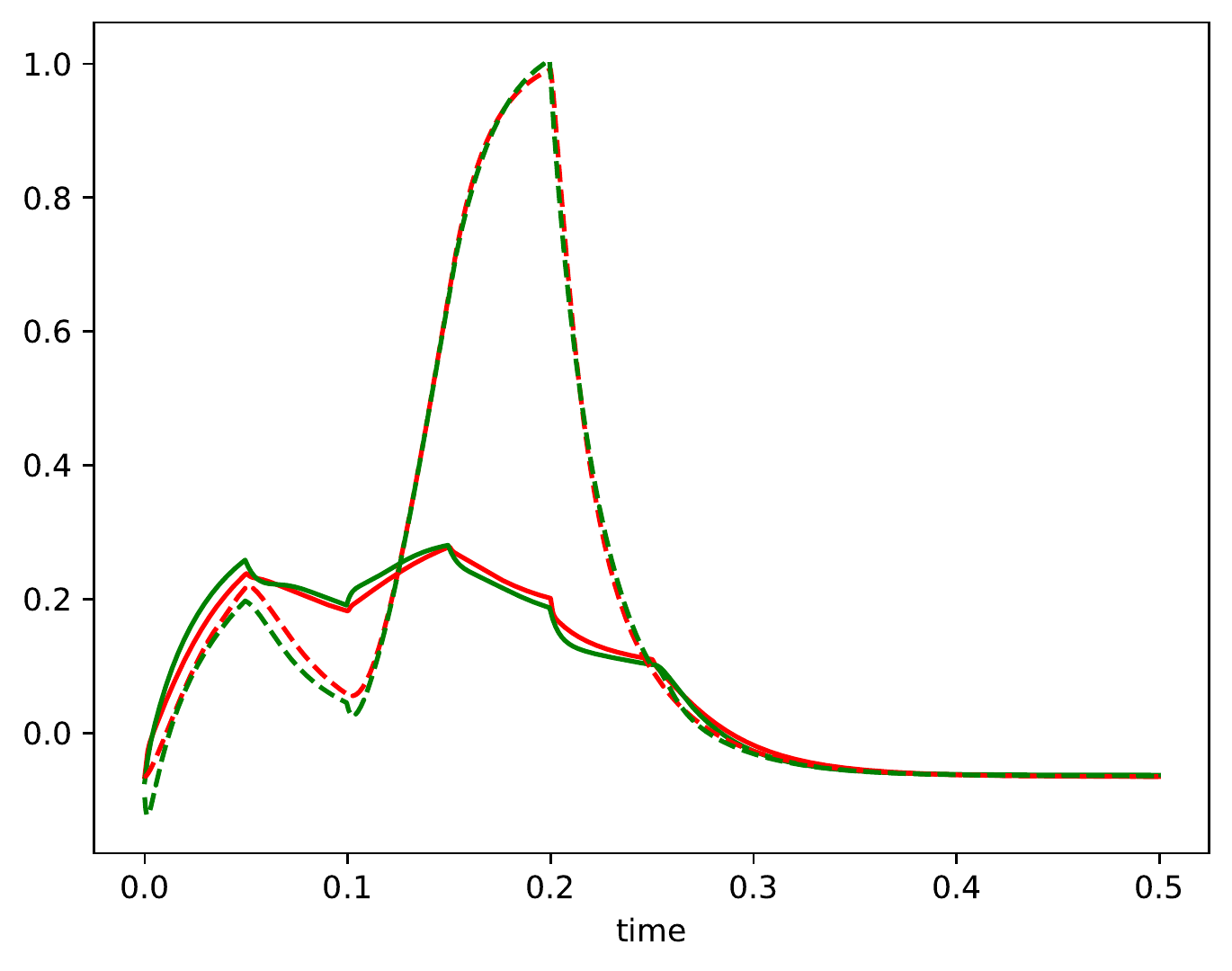} & 
		\includegraphics[width=0.16\textwidth]{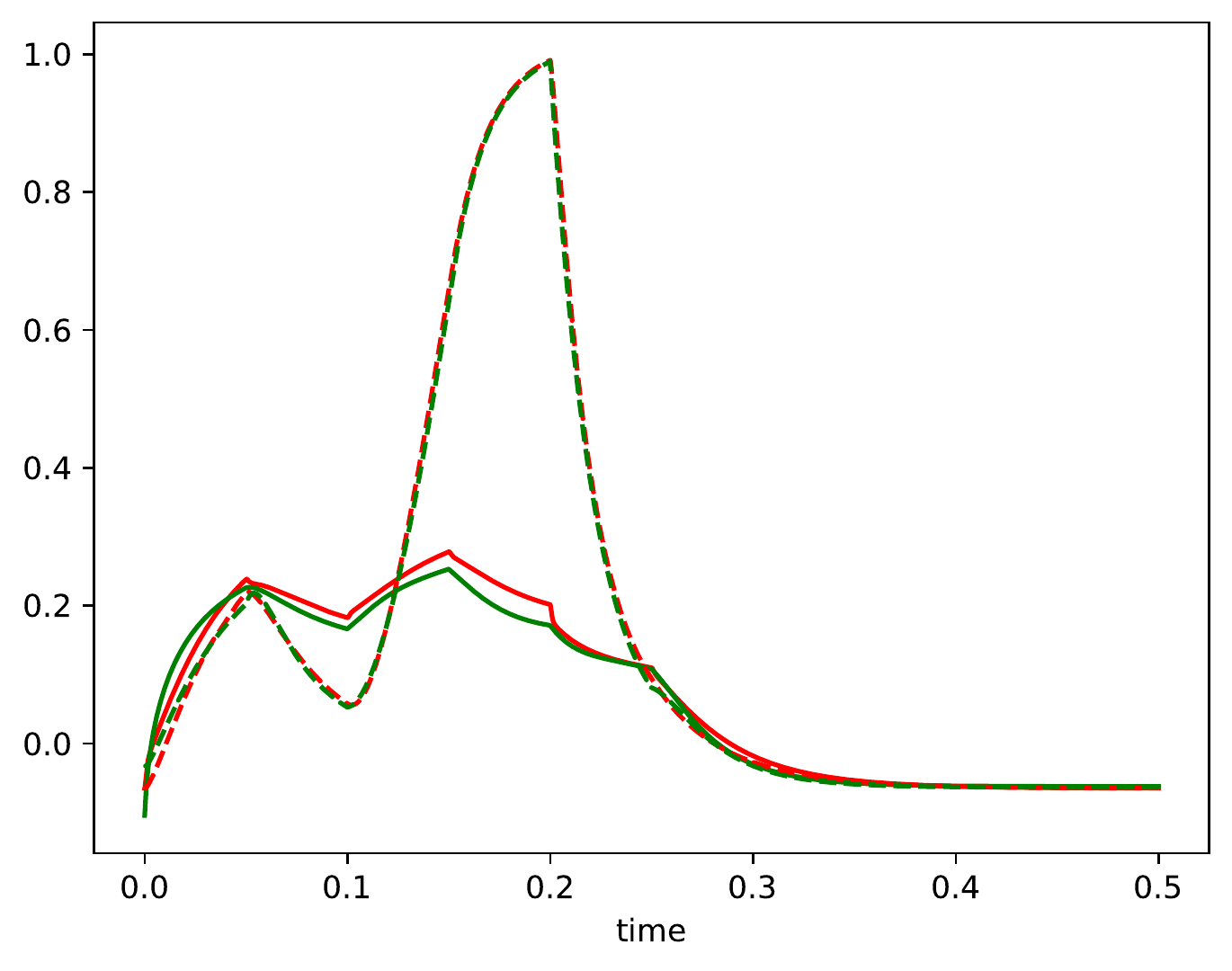} & 
	\\
		\includegraphics[width=0.16\textwidth]{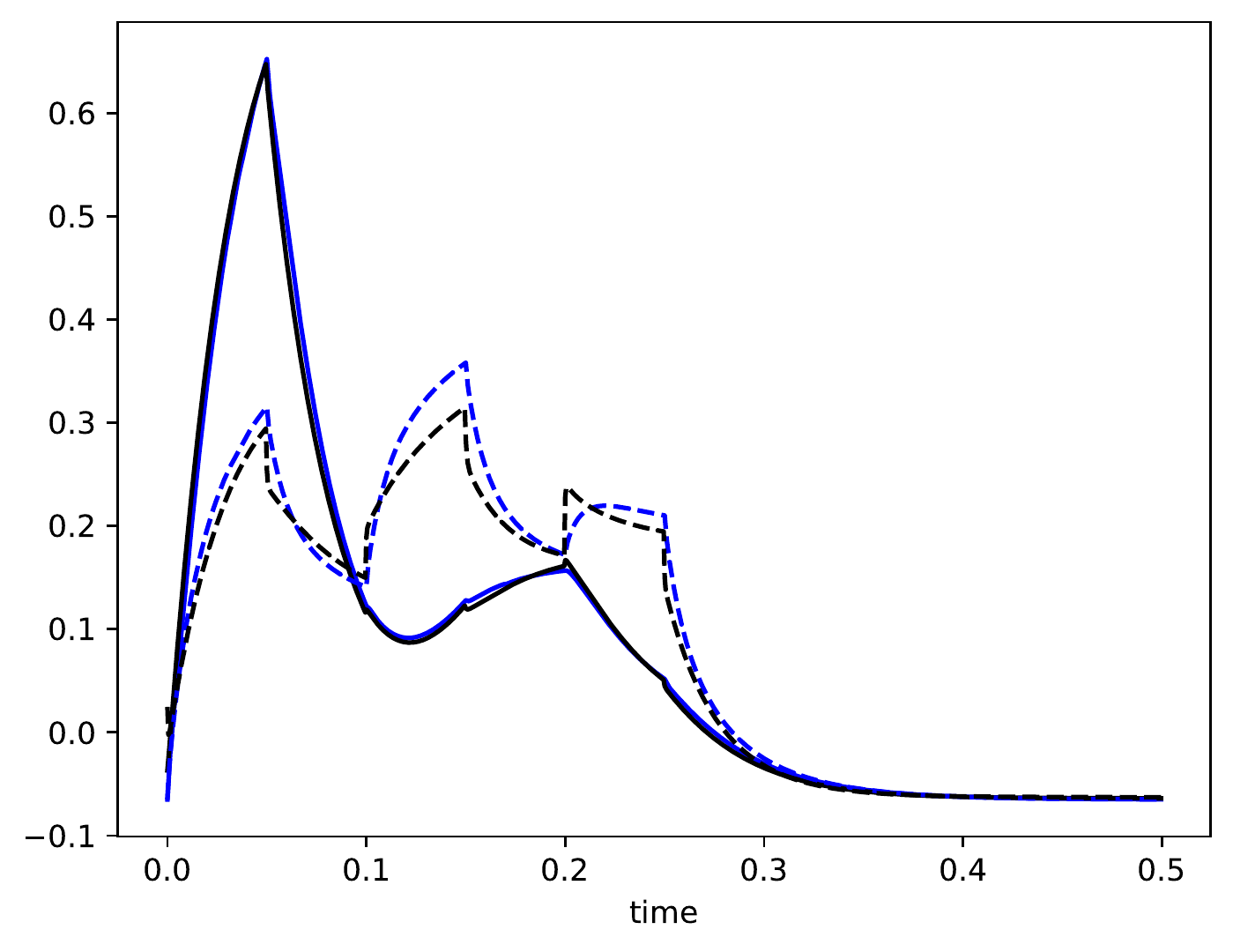} & 
		\includegraphics[width=0.16\textwidth]{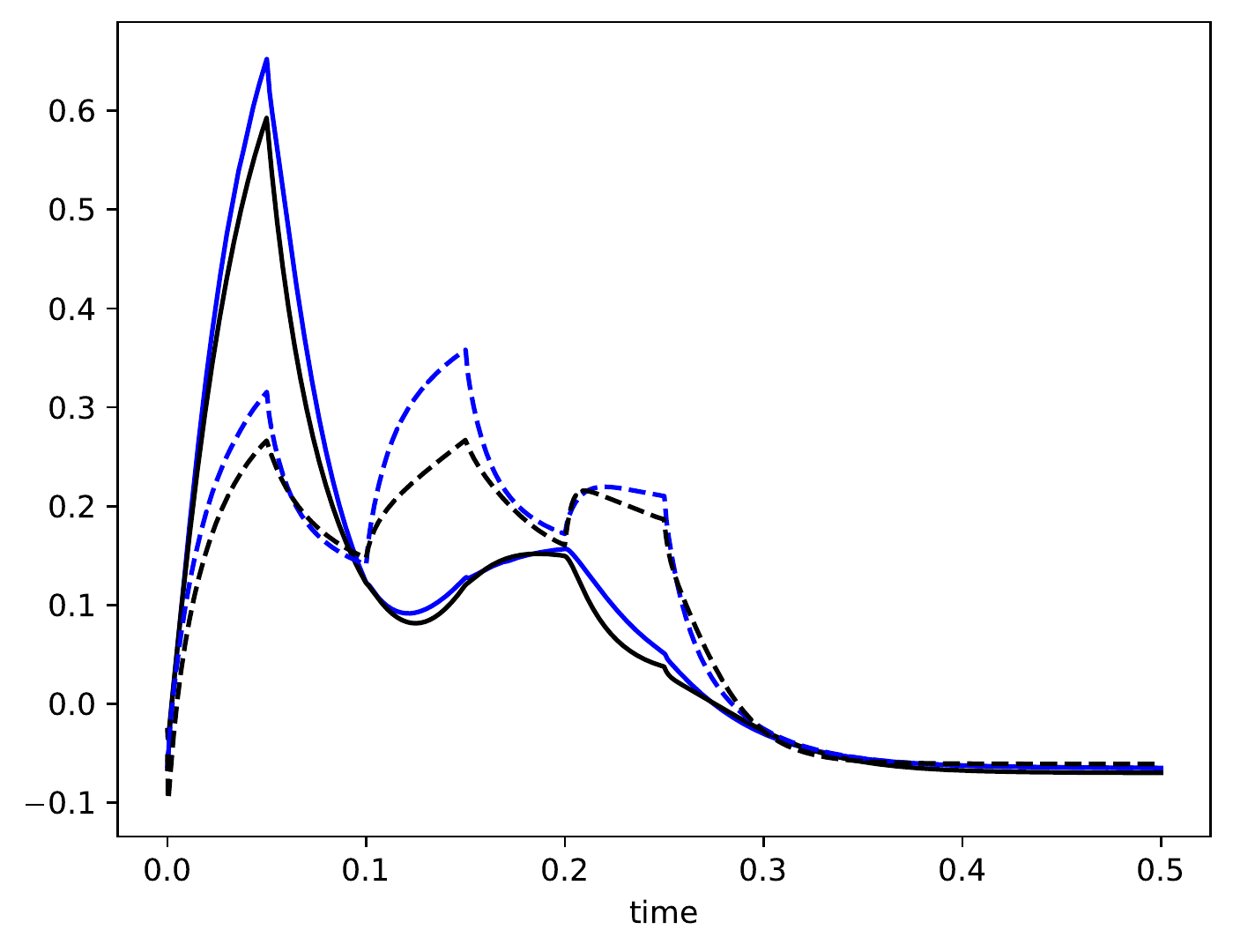} & 
		\includegraphics[width=0.16\textwidth]{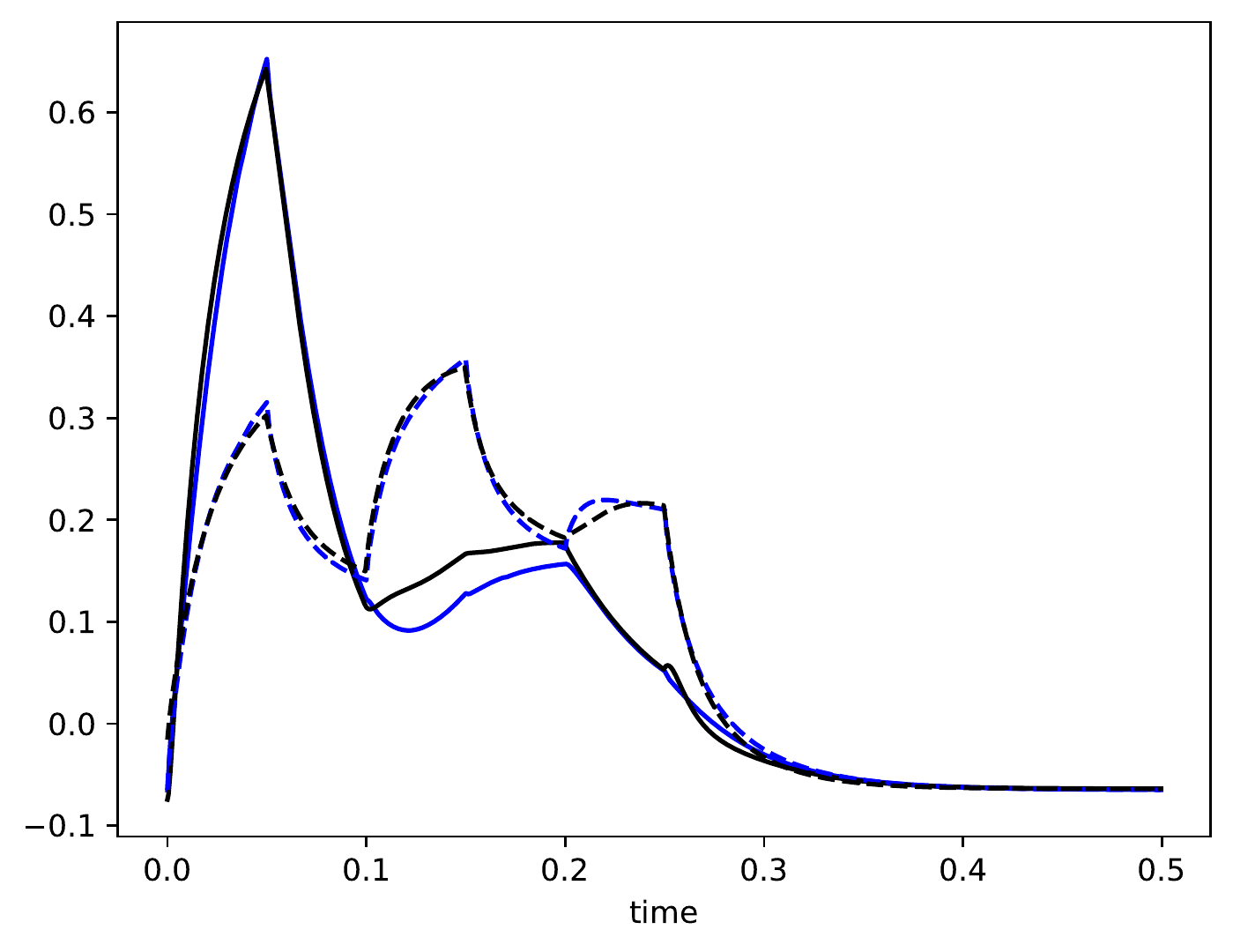} & 
		\includegraphics[width=0.16\textwidth]{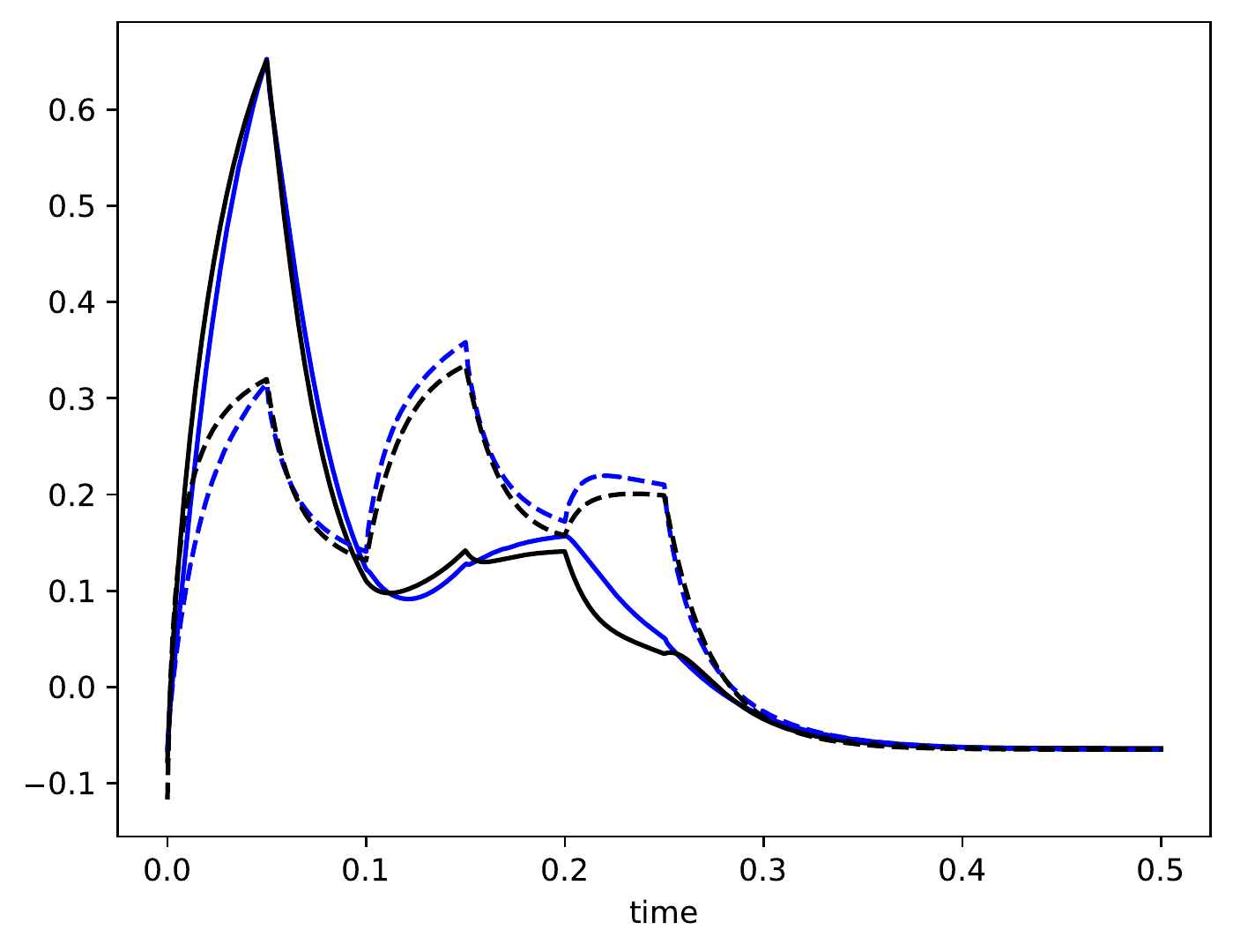} & 
	\\
  \end{tabular}
   \caption{Real (red \& blue) and predicted (green \& black) sequences on Exp. 3 for DB1 and LUAL ($1^{st}$ and $3^{rd}$ rows) and PVR and VB1 ($2^{nd}$ and $4^{th}$ rows) for two input sequences of the test set.}
  \label{fig:experiment3grid}
\end{figure}


\begin{table}[h]
  \caption{The losses obtained for the GRU model, for the two datasets and two sizes of the recurrent layer (4 and 8 hidden units), for the iteration with the smallest validation loss.}
  \label{tab:experiment3}
  \centering
  \begin{tabular}{lllll}
    \toprule
& \textbf{Dataset1 - 4} & \textbf{Dataset2 -  4} & \textbf{Dataset1 - 8} & \textbf{Dataset2 - 8} \\ \midrule
Training    & 1.3813e-04    & 2.0611e-04    & 7.9803e-05    & 7.9383e-05            \\ \midrule
Validation & 1.3098e-03    & 1.2156e-03    & 1.5419e-03    & 1.2928e-03            \\ \midrule
Test       & 1.9200e-04    & 3.9695e-04    & 1.8009e-04    & 1.7164e-04            \\
    \bottomrule
  \end{tabular}
\end{table}


From Figure~\ref{fig:exp3loss} it is clear that even though the model takes more time to converge in the finer time step case, it ends up stabilizing with a loss of the same order in both cases, as it can be seen in Table~\ref{tab:experiment3}. 
Figure~\ref{fig:experiment3grid} further strengthens this conclusion as the plots show the model fits the test data well.

\section{Conclusion}


In this paper we create models for the \textit{C. Elegans} nervous system with three different recurrent neural networks architectures: simple RNNs, LSTMs and GRUs. The objective is to generate a low-order description to replace the high-fidelity model in the NEURON simulator. 
To achieve this goal we seek a model as simple as possible and therefore the ideal unit would appear to be the simple RNN. However, this unit does not perform sufficiently well compared to the other two architectures. The LSTM and GRU give comparable results in terms of accuracy for different sizes of the recurrent layer. Due to its simplicity, GRU is preferable and with a hidden size of 4 units, is able to reproduce with high accuracy the high-fidelity model's responses to different types of stimuli.
Further work will concentrate on improving the automation in choosing appropriate stimuli for the training, validation and test sets as well as optimal parameter selection.
This will require a systematic analysis of compression possibilities of the learning-based models with error control.

These results nonetheless show that it could be feasible to develop recurrent neural network models able to infer input-output behaviours of real biological systems, enabling researchers to advance their understanding of these systems even in the absence of detailed level of connectivity.

%

\section*{Acknowledgements}

This work was partially supported by Portuguese national funds through FCT, Funda\c{c}\~{a}o para a Ci\^{e}ncia e a Tecnologia, under project UIDB/50021/2020 as well as project PTDC/EEI-EEE/31140/2017.

%

\medskip

{
\small

\bibliography{authorref}

\begin{thebibliography}{10}
\providecommand{\url}[1]{#1}
\csname url@samestyle\endcsname
\providecommand{\newblock}{\relax}
\providecommand{\bibinfo}[2]{#2}
\providecommand{\BIBentrySTDinterwordspacing}{\spaceskip=0pt\relax}
\providecommand{\BIBentryALTinterwordstretchfactor}{4}
\providecommand{\BIBentryALTinterwordspacing}{\spaceskip=\fontdimen2\font plus
\BIBentryALTinterwordstretchfactor\fontdimen3\font minus
  \fontdimen4\font\relax}
\providecommand{\BIBforeignlanguage}[2]{{%
\expandafter\ifx\csname l@#1\endcsname\relax
\typeout{** WARNING: IEEEtran.bst: No hyphenation pattern has been}%
\typeout{** loaded for the language `#1'. Using the pattern for}%
\typeout{** the default language instead.}%
\else
\language=\csname l@#1\endcsname
\fi
#2}}
\providecommand{\BIBdecl}{\relax}
\BIBdecl

\bibitem{cook2019whole}
S.~J. Cook, T.~A. Jarrell, C.~A. Brittin, Y.~Wang, A.~E. Bloniarz, M.~A.
  Yakovlev, K.~C. Nguyen, L.~T.-H. Tang, E.~A. Bayer, J.~S. Duerr
  \emph{et~al.}, ``Whole-animal connectomes of both {C}aenorhabditis elegans
  sexes,'' \emph{Nature}, vol. 571, no. 7763, pp. 63--71, 2019.

\bibitem{brittin2020beyond}
C.~A. Brittin, S.~J. Cook, D.~H. Hall, S.~W. Emmons, and N.~Cohen, ``Beyond the
  connectome: A map of a brain architecture derived from whole-brain volumetric
  reconstructions,'' \emph{bioRxiv}, 2020.

\bibitem{wormbase}
``Worm{B}ase, ws280,'' \url{https://www.wormbase.org/}, accessed: 2021-05-21.

\bibitem{wormgeneexp}
``Gene {E}xpression {D}atabase,'' \url{https://www.gfpworm.org/}, accessed:
  2021-05-21.

\bibitem{hunt2007high}
R.~Hunt-Newbury, R.~Viveiros, R.~Johnsen, A.~Mah, D.~Anastas, L.~Fang,
  E.~Halfnight, D.~Lee, J.~Lin, A.~Lorch \emph{et~al.}, ``High-throughput in
  vivo analysis of gene expression in {C}aenorhabditis elegans,'' \emph{PLoS
  Biol}, vol.~5, no.~9, p. e237, 2007.

\bibitem{wormimage}
A.~E. C. o.~M. {D}epartment~of {N}euroscience, ``The {W}orm{I}mage
  {D}atabase,'' \url{https://www.wormimage.org/}, accessed: 2021-05-21.

\bibitem{wormbook}
``Worm{B}ook, the online review of {C}. elegans biology,''
  \url{https://www.wormbook.org/}, accessed: 2021-05-21.

\bibitem{jackson2014use}
B.~M. Jackson, P.~Abete-Luzi, M.~W. Krause, and D.~M. Eisenmann, ``Use of an
  activated beta-catenin to identify {W}nt pathway target genes in
  {C}aenorhabditis elegans, including a subset of collagen genes expressed in
  late larval development,'' \emph{G3: Genes, Genomes, Genetics}, vol.~4,
  no.~4, pp. 733--747, 2014.

\bibitem{wormatlas}
A.~E. C. o.~M. {D}epartment~of {N}euroscience, ``Worm{A}tlas, {A} database
  featuring behavioral and structural anatomy of {C}aenorhabditis {E}legans,''
  \url{https://www.wormatlas.org/}, accessed: 2021-05-21.

\bibitem{opensourcebrain}
``Open {S}ource {B}rain,'' \url{https://www.opensourcebrain.org/}, accessed:
  2021-05-21.

\bibitem{gleeson2019open}
P.~Gleeson, M.~Cantarelli, B.~Marin, A.~Quintana, M.~Earnshaw, S.~Sadeh,
  E.~Piasini, J.~Birgiolas, R.~C. Cannon, N.~A. Cayco-Gajic \emph{et~al.},
  ``Open source brain: a collaborative resource for visualizing, analyzing,
  simulating, and developing standardized models of neurons and circuits,''
  \emph{Neuron}, vol. 103, no.~3, pp. 395--411, 2019.

\bibitem{openworm}
``Open{W}orm,'' \url{http://openworm.org/index.html}, accessed: 2021-05-21.

\bibitem{szigeti2014openworm}
B.~Szigeti, P.~Gleeson, M.~Vella, S.~Khayrulin, A.~Palyanov, J.~Hokanson,
  M.~Currie, M.~Cantarelli, G.~Idili, and S.~Larson, ``Open{W}orm: an
  open-science approach to modeling {C}aenorhabditis elegans,'' \emph{Frontiers
  in computational neuroscience}, vol.~8, p. 137, 2014.

\bibitem{geppetto}
``Geppetto. {B}uild robust neuroscience applications.''
  \url{http://www.geppetto.org/}, accessed: 2021-05-21.

\bibitem{cantarelli2018geppetto}
M.~Cantarelli, B.~Marin, A.~Quintana, M.~Earnshaw, R.~Court, P.~Gleeson,
  S.~Dura-Bernal, R.~A. Silver, and G.~Idili, ``Geppetto: a reusable modular
  open platform for exploring neuroscience data and models,''
  \emph{Philosophical Transactions of the Royal Society B: Biological
  Sciences}, vol. 373, no. 1758, p. 20170380, 2018.

\bibitem{varshney2011structural}
L.~R. Varshney, B.~L. Chen, E.~Paniagua, D.~H. Hall, and D.~B. Chklovskii,
  ``Structural properties of the {C}aenorhabditis elegans neuronal network,''
  \emph{PLoS Comput Biol}, vol.~7, no.~2, p. e1001066, 2011.

\bibitem{wormwiring}
A.~E. C. o.~M. {E}mmons {L}ab, ``Worm{W}iring, {N}ematode {C}onnectomics,''
  \url{https://www.wormwiring.org/}, accessed: 2021-05-21.

\bibitem{jarrell2012connectome}
T.~A. Jarrell, Y.~Wang, A.~E. Bloniarz, C.~A. Brittin, M.~Xu, J.~N. Thomson,
  D.~G. Albertson, D.~H. Hall, and S.~W. Emmons, ``The connectome of a
  decision-making neural network,'' \emph{Science}, vol. 337, no. 6093, pp.
  437--444, 2012.

\bibitem{gleeson2018c302}
P.~Gleeson, D.~Lung, R.~Grosu, R.~Hasani, and S.~D. Larson, ``c302: a
  multiscale framework for modelling the nervous system of caenorhabditis
  elegans,'' \emph{Philosophical {T}ransactions of the {R}oyal {S}ociety {B}:
  {B}iological {S}ciences}, vol. 373, no. 1758, p. 20170379, 2018.

\bibitem{karasozen2020model}
B.~Karas{\"o}zen, ``Model {O}rder {R}eduction in {N}euroscience,'' \emph{arXiv
  preprint arXiv:2003.05133}, 2020.

\bibitem{brunton2016extracting}
B.~W. Brunton, L.~A. Johnson, J.~G. Ojemann, and J.~N. Kutz, ``Extracting
  spatial--temporal coherent patterns in large-scale neural recordings using
  dynamic mode decomposition,'' \emph{Journal of {N}euroscience {M}ethods},
  vol. 258, pp. 1--15, 2016.

\bibitem{kellems2009low}
A.~R. Kellems, D.~Roos, N.~Xiao, and S.~J. Cox, ``Low-dimensional,
  morphologically accurate models of subthreshold membrane potential,''
  \emph{Journal of {C}omputational {N}euroscience}, vol.~27, no.~2, p. 161,
  2009.

\bibitem{lehtimaki2019projection}
M.~Lehtim{\"a}ki, L.~Paunonen, and M.-L. Linne, ``Projection-based order
  reduction of a nonlinear biophysical neuronal network model,'' in \emph{58th
  Conference on Decision and Control (CDC)}.\hskip 1em plus 0.5em minus
  0.4em\relax IEEE, 2019, pp. 1--6.

\bibitem{regazzoni2019machine}
F.~Regazzoni, L.~Dede, and A.~Quarteroni, ``Machine learning for fast and
  reliable solution of time-dependent differential equations,'' \emph{Journal
  of Computational physics}, vol. 397, p. 108852, 2019.

\bibitem{sun2020neupde}
Y.~Sun, L.~Zhang, and H.~Schaeffer, ``Neu{PDE}: {N}eural network based ordinary
  and partial differential equations for modeling time-dependent data,'' in
  \emph{Mathematical and Scientific Machine Learning}.\hskip 1em plus 0.5em
  minus 0.4em\relax PMLR, 2020, pp. 352--372.

\bibitem{lai2018modeling}
G.~Lai, W.-C. Chang, Y.~Yang, and H.~Liu, ``Modeling long-and short-term
  temporal patterns with deep neural networks,'' in \emph{The 41st
  International ACM SIGIR Conference on Research \& Development in Information
  Retrieval}, 2018, pp. 95--104.

\bibitem{jin2020prediction}
X.~Jin, X.~Yu, X.~Wang, Y.~Bai, T.~Su, and J.~Kong, ``Prediction for time
  series with {CNN} and {LSTM},'' in \emph{Proceedings of the 11th
  International Conference on Modelling, Identification and Control
  (ICMIC2019)}.\hskip 1em plus 0.5em minus 0.4em\relax Springer, 2020, pp.
  631--641.

\bibitem{massaoudi2019novel}
M.~Massaoudi, I.~Chihi, L.~Sidhom, M.~Trabelsi, S.~S. Refaat, and F.~S.
  Oueslati, ``A novel approach based deep {RNN} using hybrid {NARX}-{LSTM}
  model for solar power forecasting,'' \emph{arXiv preprint arXiv:1910.10064},
  2019.

\bibitem{gallicchio2018comparison}
C.~Gallicchio, A.~Micheli, and L.~Pedrelli, ``Comparison between {D}eep{ESN}s
  and gated {RNN}s on multivariate time-series prediction,'' \emph{arXiv
  preprint arXiv:1812.11527}, 2018.

\bibitem{yuan2020using}
Y.~Yuan, L.~Lin, L.-Z. Huo, Y.-L. Kong, Z.-G. Zhou, B.~Wu, and Y.~Jia, ``Using
  an attention-based {LSTM} encoder--decoder network for near real-time
  disturbance detection,'' \emph{IEEE Journal of Selected Topics in Applied
  Earth Observations and Remote Sensing}, vol.~13, pp. 1819--1832, 2020.

\bibitem{filonov2016multivariate}
P.~Filonov, A.~Lavrentyev, and A.~Vorontsov, ``Multivariate industrial time
  series with cyber-attack simulation: Fault detection using an lstm-based
  predictive data model,'' \emph{arXiv preprint arXiv:1612.06676}, 2016.

\bibitem{tavakoli2019modeling}
N.~Tavakoli, ``Modeling genome data using bidirectional {LSTM},'' in \emph{2019
  IEEE 43rd Annual Computer Software and Applications Conference (COMPSAC)},
  vol.~2.\hskip 1em plus 0.5em minus 0.4em\relax IEEE, 2019, pp. 183--188.

\bibitem{xu2020one}
G.~Xu, T.~Ren, Y.~Chen, and W.~Che, ``A one-dimensional {CNN}-{LSTM} model for
  epileptic seizure recognition using {EEG} signal analysis,'' \emph{Frontiers
  in Neuroscience}, vol.~14, p. 1253, 2020.

\bibitem{gucclu2017modeling}
U.~G{\"u}{\c{c}}l{\"u} and M.~A. van Gerven, ``Modeling the dynamics of human
  brain activity with recurrent neural networks,'' \emph{Frontiers in
  computational neuroscience}, vol.~11, p.~7, 2017.

\bibitem{rumelhart1986}
D.~E. Rumelhart, G.~E. Hinton, and R.~J. Williams, ``Learning representations
  by back-propagating errors,'' \emph{Nature}, vol. 323, pp. 533--536, 1986.

\bibitem{werbos1988}
\BIBentryALTinterwordspacing
P.~J. Werbos, ``Generalization of backpropagation with application to a
  recurrent gas market model,'' \emph{Neural Networks}, vol.~1, no.~4, pp.
  339--356, 1988. [Online]. Available:
  \url{https://www.sciencedirect.com/science/article/pii/089360808890007X}
\BIBentrySTDinterwordspacing

\bibitem{elman1990}
\BIBentryALTinterwordspacing
J.~L. Elman, ``Finding structure in time,'' \emph{Cognitive Science}, vol.~14,
  no.~2, pp. 179--211, 1990. [Online]. Available:
  \url{https://onlinelibrary.wiley.com/doi/abs/10.1207/s15516709cog1402_1}
\BIBentrySTDinterwordspacing

\bibitem{lstms1997original}
\BIBentryALTinterwordspacing
S.~Hochreiter and J.~Schmidhuber, ``{Long {S}hort-{T}erm {M}emory},''
  \emph{Neural Computation}, vol.~9, no.~8, pp. 1735--1780, 11 1997. [Online].
  Available: \url{https://doi.org/10.1162/neco.1997.9.8.1735}
\BIBentrySTDinterwordspacing

\bibitem{gers1999}
F.~Gers, J.~Schmidhuber, and F.~Cummins, ``Learning to forget: continual
  prediction with lstm,'' in \emph{1999 Ninth International Conference on
  Artificial Neural Networks ICANN 99. (Conf. Publ. No. 470)}, vol.~2, 1999,
  pp. 850--855 vol.2.

\bibitem{cho2014gru}
\BIBentryALTinterwordspacing
K.~Cho, B.~van Merrienboer, {\c{C}}.~G{\"{u}}l{\c{c}}ehre, F.~Bougares,
  H.~Schwenk, and Y.~Bengio, ``Learning phrase representations using {RNN}
  encoder-decoder for statistical machine translation,'' \emph{CoRR}, vol.
  abs/1406.1078, 2014. [Online]. Available:
  \url{http://arxiv.org/abs/1406.1078}
\BIBentrySTDinterwordspacing

\bibitem{werbos1990}
P.~Werbos, ``Backpropagation through time: what it does and how to do it,''
  \emph{Proceedings of the IEEE}, vol.~78, pp. 1550 -- 1560, 11 1990.

\bibitem{bengio1994}
Y.~Bengio, P.~Simard, and P.~Frasconi, ``Learning long-term dependencies with
  gradient descent is difficult,'' \emph{IEEE Transactions on Neural Networks},
  vol.~5, no.~2, pp. 157--166, 1994.

\bibitem{pascanu2013}
\BIBentryALTinterwordspacing
R.~Pascanu, T.~Mikolov, and Y.~Bengio, ``Understanding the exploding gradient
  problem,'' \emph{CoRR}, vol. abs/1211.5063, 2012. [Online]. Available:
  \url{http://arxiv.org/abs/1211.5063}
\BIBentrySTDinterwordspacing

\bibitem{gers2000}
\BIBentryALTinterwordspacing
F.~A. Gers, J.~Schmidhuber, and F.~Cummins, ``{Learning to Forget: Continual
  Prediction with {LSTM}},'' \emph{Neural Computation}, vol.~12, no.~10, pp.
  2451--2471, 10 2000. [Online]. Available:
  \url{https://doi.org/10.1162/089976600300015015}
\BIBentrySTDinterwordspacing

\bibitem{carnevale2006neuron}
N.~T. Carnevale and M.~L. Hines, \emph{The {NEURON} book}.\hskip 1em plus 0.5em
  minus 0.4em\relax Cambridge {U}niversity {P}ress, 2006.

\bibitem{python}
G.~Van~Rossum and F.~L. Drake, \emph{Python 3 Reference Manual}.\hskip 1em plus
  0.5em minus 0.4em\relax Scotts Valley, CA: CreateSpace, 2009.

\bibitem{keras}
F.~Chollet \emph{et~al.}, ``Keras,'' \url{https://keras.io}, 2015.

\bibitem{tensorflow}
\BIBentryALTinterwordspacing
M.~Abadi, A.~Agarwal, P.~Barham, E.~Brevdo, Z.~Chen, C.~Citro, G.~S. Corrado,
  A.~Davis, J.~Dean, M.~Devin, S.~Ghemawat, I.~Goodfellow, A.~Harp, G.~Irving,
  M.~Isard, Y.~Jia, R.~Jozefowicz, L.~Kaiser, M.~Kudlur, J.~Levenberg,
  D.~Man\'{e}, R.~Monga, S.~Moore, D.~Murray, C.~Olah, M.~Schuster, J.~Shlens,
  B.~Steiner, I.~Sutskever, K.~Talwar, P.~Tucker, V.~Vanhoucke, V.~Vasudevan,
  F.~Vi\'{e}gas, O.~Vinyals, P.~Warden, M.~Wattenberg, M.~Wicke, Y.~Yu, and
  X.~Zheng, ``{TensorFlow}: Large-scale machine learning on heterogeneous
  systems,'' 2015, software available from tensorflow.org. [Online]. Available:
  \url{https://www.tensorflow.org/}
\BIBentrySTDinterwordspacing

\bibitem{adam}
\BIBentryALTinterwordspacing
D.~P. Kingma and J.~Ba, ``Adam: {A} method for stochastic optimization,'' in
  \emph{3rd International Conference on Learning Representations, {ICLR} 2015,
  San Diego, CA, USA, May 7-9, 2015, Conference Track Proceedings}, Y.~Bengio
  and Y.~LeCun, Eds., 2015. [Online]. Available:
  \url{http://arxiv.org/abs/1412.6980}
\BIBentrySTDinterwordspacing

\end{thebibliography}
\bibliographystyle{IEEEtran}

\end{document}